\documentclass[eqsecnum,secnumarabic,nofootinbib,prd,11pt,tightenlines,showkeys]{revtex4}
\usepackage{amsfonts}
\usepackage{amssymb}
\usepackage{amsmath}
\usepackage{graphicx}
\usepackage[utf8]{inputenc}
\usepackage{fancyhdr}
\usepackage{slashed}
\usepackage{hyperref}

\begin{document}

\title{\textbf{Symmetries in one loop solutions: The AV, AVV, and AVVV
diagrams, from 2D, 4D, and 6D dimensions and the role of breaking
integration linearity.}}
\author{L. Ebani$^1$\footnote{luci.ebani@gmail.com}, T. J. Girardi$^1$\footnote{thalisjg@gmail.com}, 
J. F. Thuorst$^1$\footnote{jfernando.th@gmail.com, corresponding author}}
\affiliation{$^1$CBPF-Centro Brasileiro de Pesquisas F\'{\i}sicas, 22290-180, Rio de
Janeiro, RJ, Brazil}

\begin{abstract}
We investigated relations among green functions defined in the context of an
alternative strategy for coping with the divergences, also called Implicit
Regularization. Our targets are fermionic amplitudes in even space-time
dimensions, where anomalous tensors connect to finite amplitudes. Those
tensors depend on surface terms, whose non-zero values arise from finite
amplitudes as requirements of consistency with the linearity of integration
and uniqueness. Maintaining these terms implies breaking momentum-space
homogeneity and in a later step the Ward identities. Meanwhile, eliminating
them allows more than one mathematical expression for the same amplitude.
That is a consequence of choices related to the involved Dirac traces.
Independently of divergences, it is impossible to satisfy all symmetry
implications that require the vanishing of surface terms and linearity
simultaneously. Nonetheless, the symmetry violations are globally
independent of divergences and can be allocated appropriately. From this
perspective, we cast all the choices involved and the different meanings,
whose implications go beyond the scenario described.
\end{abstract}


\maketitle

\section{Introduction}

Since their inception, anomalies have been part of the culture of
practitioners of quantum field theory (QFT). The issue surges in multiple
avatars and is intricately linked to the conservation or not of the
classical currents in the operator framework of quantum theory. Met in the
end forties and, at the outset of the fifties by the authors \cite%
{Fukuda1949}, \cite{Steinberger1949}, \cite{Schwinger1951} and \cite%
{Rosenberg1963}. Rediscovered in two dimensions ($2D$) \cite{Johnson1963}
and in four dimensions ($4D$) by \cite{Adler1969} \cite{Bardeen1969} \cite%
{Jackiw1969}, the ABBJ anomaly of the triangle's graph. Later, many studies
considered perturbative and non-perturbative approaches to investigate these
phenomena, mainly the Fujikawa interpretation \cite{Bertlmann1996}, \cite%
{Bastianelli2006} of the path-integral's measure, and heat kernel
expansions, see \cite{Vassilevich2003} for a review, as well as
cohomological methods \cite{Bertlmann1996}. One of the reasons for its
importance comes from the fact that anomalies prevent the quantum
expectation value of Noether currents from satisfying their classical
symmetries. Hence, the anomaly manifestation breaks Ward-Takahashi
identities (WI) or Slavnov-Taylor identities to non-abelian gauge symmetry,
which are needed for the perturbative renormalizability of the gauge models,
even when spontaneous symmetry breaking is present as in the case of the
physical theory Standard Model. Another one the role in the phenomenological
description of particle decays, the neutral pion the most emblematic one.

Their general and most notable manifestation is in the perturbative scenario
where correlators of the axial and vector currents that are odd, linear
divergent tensors corresponding to the $AV^{n}$ amplitudes, in $d=2n$
dimensions, can not have all their WIs satisfied. All of them have three
structural properties.

First, in the last instance, they are $\left( n+1\right) $-rank tensor of
odd parity, functions of $n$ momenta variables, that for such features
posses, through contractions with their momenta variables, a set of low
energy theorems, such theorems come as a consequence of their WIs. One of
these identities, axial-WI, relates the same tensor, in some way, to the
amplitude $PV^{n}$, whose perturbative expression is a finite integral, and
this expression behaves in the point where the low energy theorem is stated
in such a way that either the $PV^{n}$ have a value that contradicts its
explicit form or some Ward identity is unavoidably violated. The anomaly.

Second, a property that is shared by those tensors, they have minimal
dimensionality such that they rise from Dirac traces that cast two more
gamma matrices than the dimension they are defined. This trace of $2n+2$
gamma matrices that are combinations of metric and Levi-Civita tensors are
known to exhibit a multitude of identical expressions, where they display
explicitly, different sets of indexes, signs, and number of monomials. All
of them differ by sums of totally antisymmetric tensors in $2n+1$ indexes,
therefore they are zero at $2n$ dimensions. The choice of one or another
form is arbitrary nonetheless.

Finally, their power counting leaves the possible presence of
non-identically vanishing surface terms opened, making these structures
depending on how the external momenta percolate the graphs. That means, the
final results generally show up many possibilities, since perturbative
solutions may unavoidably depend on arbitrary choices for routings and Dirac
traces.

This last proposition is inseparable from the fact that in perturbation
theory, divergences are the rule, in order to get some model predictions of
QFT, which means to acquire some information of the kinematical dependence,
symmetries, and so on, regularization techniques are adopted in an attempt
to circumvent these problems, as some examples, c\textrm{ut-off,
Pauli-Villars, analytic regularization, dimensional regularization (DR) \cite%
{Bollini1972} \cite{tHooft1972}, high covariant regularization, differential
renormalization}, and new methods have, until recently, been proposed to
deal with multi-loop calculation and aiming to an algorithmic implementation
for high precision numerical predictions \cite{Pittau2012} \cite%
{TodGnendiger2017}. Modifying, invariably, the amplitudes by making
divergent amplitudes finite, manipulations forbidden, or not guaranteed to
the original expressions become valid, such as shifts in the integration
variable. And determining how and which properties and parts of the algebras
present must be dealt with \cite{Breitenlohner1977}\cite{Jegerlehner2001}%
\cite{Tsai2011a}\cite{Tsai2011b}\cite{RFerrari2017}\cite{Bruque2018}, things
that are not inherent to the Feynman rules defining the objects under
investigation. As in the case of DR, this process is\textbf{\ }guided by
symmetry preservation and implies the elimination of surface terms. Then in
a later step, combining regularization and renormalization, predictions can
be established and compared to the experimental data resulting in the
success known, for example, in the Quantum Electrodynamics/Standard Model.

On the other hand, Feynman integrals of power counting linear or higher and
logarithmic, but tensors ones, own surface terms. For the linear ones, a
shift in the integration variable is acknowledged to require compensation
through non-zero surface terms \cite{Treiman1985} \cite{ChengLi1984} \cite%
{Bertlmann1996}, hence they can not be free-shifted, arbitrary routings are
the rule. It is a manifestation of the internal momenta arbitrariness,
notwithstanding they satisfy the energy-momentum conservation, as the
differences in the routings are functions of the physical momenta, by
themselves and the sums of these routings are arbitrary, and they may even
assume non-covariant expressions \cite{Sterman1993}. Given the fact that
non-zero surface terms imply the breaking of translational symmetry in the
momentum space, and this operation is thoroughly needed to prove WIs, then
it is reasonable that other violations of symmetries occur in such
calculations, and in fact, they do.

The question that is proposed and answered in this contribution is: What is
the extension of the consequences resulting from the three properties of the
tensors mentioned, over symmetries of the amplitudes and the mathematical
content of the diagrams, that means, the integration linearity,
translational symmetry, and uniqueness\footnote{%
To uniqueness, which needs a particular definition to work its consequences,
we provide it along the paper.}? That is independent of a particular set of
rules, even if some rule should be elected in some very final level of
analysis.

For such purpose, we use a general model where these aspects can be
generated: Spin-1/2 fermions coupled with boson fields of spin-zero and one,
of even and odd parity. The $n$-vertices polygon graphs of spin 1/2 internal
propagators are the center of the analysis, specifically the $2D$-$AV$ and $%
VA$ bubbles, $4D$-$AVV$, $VAV$, and $VVA$ triangles, and $6D$-$AVVV$ \ box.
The results are obtained within the context of an alternative procedure to
handle divergent and finite integrals also, that are independent of a
particular regularization, introduced in the Ph.D. thesis of O.A. Battistel 
\cite{PhdBattistel1999}. This strategy has been applied in a large number of
investigations, for example in two dimensions \cite{Battistel2012} \cite%
{Battistel2018} \cite{Ebani2018}, in four dimensions \cite{Battistel2002a} 
\cite{Battistel2002b}, in six dimensions \cite{Fonseca2014}, in odd
five-dimensions $\left( 5D\right) $ \cite{Fonseca2013}, and in even and odd,
from two to six, dimensions \cite{Battistel2014}. An incarnation of this
strategy has been known in the literature as Implicit Regularization (IREG),
and it has been applied in similar investigations \cite{Viglioni2016} \cite%
{Vieira2016} \cite{Ferreira2012}.

The idea is very simple, the divergences are isolated by means of an
identity of universal applicability that does not interfere with the Feynman
rules used. Insofar as diverging integrals are not evaluated explicitly and
the amplitudes are not modified at early steps of the calculations.
Arbitrary routings are adopted for the momenta of internal lines to preserve
their intrinsic arbitrariness which is a feature of such perturbative
calculations. Furthermore, in the strategy, we devise a notational scheme to
systematize the finite integrals and their divergent parts that appear in
this work, based on previous works about the subject,\textrm{\ }ref. \cite%
{Battistel2006} for single-mass integrals, and ref.\cite{Dallabona2012}\cite%
{SunYi2012} for multiples masses case. There are three ingredients for such:
irreducible divergent objects, tensor surface terms, and finite functions.
It is precisely this organization that allows a clear view of the relevant
points of our discussion.

The unique assumption made is that the linearity applies to Feynman
integrals, typical quantities of the perturbative calculations. At the
amplitude's level, the linearity of integration arises in the relations
among green functions (RAGFs) derived for the well-defined integrands, that
when integrated, the critical step of computations, embodies the referred
property. This aspect is one of the main points of the investigation since
if it was present automatically, it would be possible to prove all of Ward's
identities by methods blessed by translational symmetry as the DR.

Our stance on the undetermined character of perturbation theory quantities
enables a clear view of the active elements that produce the variety of
results, notably of how and where the presence of surface terms in
amplitudes is connected with ambiguities and symmetry violations. As a
result, in the first place, if one asks: Is there a unique tensor of the
external momenta? The answer will be, to even amplitudes, yes, any
interpretation of the divergences that make zero the surface terms render
the results unique and, as a by-product, symmetric. But if one asks soon
after: Does such action make all the amplitudes, including the odd ones,
unique functions of the external momenta? The answer is no, in other words,
there is more than one answer to be obtained, even if the divergences are
not touched at all and are taken consistently with the even amplitudes, such
attitude wreck the integration linearity and arbitrary combinations of equal
integrands give rise to arbitrary distinct polynomials in the integrals,
that means, once you lose uniqueness an uncountable (literally) number of
tensors can be reached from the same expression.

The other side is deep as well, once the value, the unique one, that saves
linearity and uniqueness is adopted, no matter what manipulation is used in
the traces, even if bilinears are not reduced in the splitting of divergent
parts, it provides one and only one tensor, here is the catch, of the
routing variables, implying that physical interpretation asks for arbitrary
parameters to account for the realization of the symmetries. In the odd
amplitudes, such freedom will, as in any other arbitrariness situation,
enable one to fix the known and desired content of the results. And the
striking consequence is that if universality is asked to play a role, even
amplitudes will always violate their WI.

The lack of momentum ambiguities leads to a lack of uniqueness in a
sub-class of pseudo-tensor amplitudes, where belongs precisely the anomalous
ones in even space-time dimensions, among them the $AV^{n}$\ amplitudes.
Moreover, in dimensions equal to four or higher, more amplitudes show the
behavior described here,\textrm{\ }$AV^{r}$, for $n\leq r<2n$. As an
example, the $4D$-$AVVV$ box, will show a dependence on surface terms, but
not on internal momenta, and a triangle topology with a tensor vertex, $TAV$%
, suffers from the same properties we present in this work.

To appreciate these statements we organized the work as follows. In section (%
\ref{ModlDef}), we have the general model, definitions, and a preliminary
discussion. Section (\ref{IREG}) deals with the alternative to
regularizations strategy in handling the divergences, where we give the
general defintion of the irreducible objects and tensor surface terms that
appear everywhere in the work. A detailed compilation of\ the effects of
traces and surface terms in two dimensions appears in section (\ref{2Dim2Pt}%
), there, for the first time, and in a simplified situation the linearity of
integration and uniqueness are fully analyzed through complete and
independent computation of all quantities involved in the relation among
green functions, the consequences of adopting results saving linearity and
of saving translational symmetry are presented, and interpreted in light of
low-energy theorems. The most lengthy section (\ref{4Dim3Pt}), handles with
all odd triangles, their RAGFs, and the nature of uniqueness that is more
convoluted in that case, sub-sections (\ref{LE4D}) and (\ref{LED4DSTS}) deal
with the general properties of low-energy theorems and offer a theorem
connecting linearity, low-energy behavior of finite amplitudes in general,
and surface terms. The last section (\ref{6Dim4Pt}), extends seamlessly the
propositions put in the realm of six dimensions, \cite{Fonseca2014} has
already worked with one of the possibilities, and among all the other
possibilities we chose one more to illustrate the behavior we have presented
in two and four dimensions.

Finally in section (\ref{finalremarks}), Final Remarks and Perspectives, we
present comments on some fine points of the work and a timeline of the
arguments leading to our main results and the character of some of our
conceptual tools. Just as importantly, and integral part is the appendices,
in (\ref{Tr6G4D}) we show how the attitude present in the main body of the
work, in four dimensions, is enough to give account for any non-trivial
proposition. Appendices (\ref{AppInt2D}, \ref{AppInt4D}, and \ref{AppInt6D})
contain the divergent and finite parts, obtained through the strategy
delineated in section (\ref{IREG}) for any structure used in the paper, in
addition to all the reductions and identities needed. The last appendix (\ref%
{AppSub}) is a compilation of results required in section (\ref{4Dim3Pt})
not present in the text due to size reasons.

\section{Notation, Definitions, Model and Preliminaries}

\label{ModlDef}

The Feynman rules, vertexes, and propagators, employed in this investigation
come from a model where fermionic currents coupled to bosonic fields of even
and odd parity $\left\{ \Phi \left( x\right) ,V_{\mu }\left( x\right) ,\Pi
\left( x\right) ,A_{\mu }\left( x\right) \right\} $ through the general
interacting action%
\begin{equation}
\mathcal{S}_{I}=\int \mathrm{d}^{2n}x\left[ e_{S}S\left( x\right) \Phi
\left( x\right) +e_{\Pi }P\left( x\right) \Pi \left( x\right) +e_{V}J^{\mu
}\left( x\right) V_{\mu }\left( x\right) +e_{A}J_{\ast }^{\mu }\left(
x\right) A_{\mu }\left( x\right) \right] .  \label{Action}
\end{equation}%
The currents $\left\{ S,P,J_{\mu },J_{\ast \mu }\right\} $ are bilinears in
the fermionic fields, $J_{i}=\bar{\psi}\left( x\right) \Gamma _{i}\psi
\left( x\right) ,$ and they deliver the vertexes proportional to%
\begin{equation}
\Gamma _{i}\in \left( S,P,V,A\right) =\left( 1,\gamma _{\ast },\gamma _{\mu
},\gamma _{\ast }\gamma _{\mu }\right) ,  \label{SetofVertexes}
\end{equation}%
the proportionality comes from the coupling constants $\left\{ e_{S},e_{\Pi
},e,e_{A}\right\} $ that are taken as the unit for our purposes, as they can
be easily recovered if needed. The elements $\gamma _{\mu }$ are the
generators of the Clifford algebra of Dirac matrices satisfying $\left\{
\gamma ^{\mu _{1}},\gamma ^{\mu _{2}}\right\} =2g^{\mu _{1}\mu _{2}}$. The
highest-weight element of the algebra, in $d=2n,$ is the chiral matrix of
that dimension that satisfies $\left\{ \gamma _{\ast },\gamma ^{\mu
_{k}}\right\} =0$, explicitly 
\begin{equation}
\gamma _{\ast }=i^{n-1}\gamma _{0}\gamma _{1}\cdots \gamma
_{2n-1}=i^{n-1}\prod_{s=0}^{2n-1}\gamma _{s}=\frac{i^{n-1}}{\left( 2n\right)
!}\varepsilon _{\nu _{1}\cdots \nu _{2n}}\gamma ^{\nu _{1}\cdots \nu _{2n}}.
\end{equation}

We often will adopt a notation of merging to the product of matrices $\gamma
^{\nu _{1}\cdots \nu _{2n}}=\gamma ^{\nu _{1}}\gamma ^{\nu _{2}}\cdots
\gamma ^{\nu _{2n}}$ and will adapt to the Lorentz indexes, $\mu _{1}\mu
_{2}\cdots \mu _{s}=\mu _{12\cdots s}$ when convenient and clear by context.
The behavior under the permutation of the indexes is determined by the
objects: $g_{\mu _{1}\mu _{2}}=g_{\mu _{12}}=g_{\mu _{21}}$ or $\varepsilon
_{\mu _{1}\mu _{2}\cdots \mu _{2n}}=\varepsilon _{\mu _{12\cdots
2n}}=-\varepsilon _{\mu _{21}\cdots \mu _{2n}}$. For the $2n$-dimensional,
follow the normalization $\varepsilon ^{0123\cdots 2n-1}=1$.

The elements of the algebra are the antisymmetrized products of gamma
matrices%
\begin{equation}
\gamma _{\left[ \mu _{1}\cdots \mu _{r}\right] }=\frac{1}{r!}\sum_{\pi \in
S_{r}}\mathrm{sign}\left( \pi \right) \gamma _{\mu _{\pi \left( 1\right)
}\cdots \mu _{\pi \left( r\right) }},
\end{equation}%
that satisfies the general identities, see by example the appendix of the
ref.\cite{deWit1986}.%
\begin{equation}
\gamma _{\ast }\gamma _{\left[ \mu _{1}\cdots \mu _{r}\right] }=\frac{%
i^{n-1+r\left( r+1\right) }}{\left( 2n-s\right) !}\varepsilon _{\mu
_{1}\cdots \mu _{r}}^{\hspace{0.95cm}\nu _{r+1}\cdots \nu _{2n}}\gamma _{%
\left[ \nu _{r+1}\cdots \nu _{2n}\right] }.  \label{Chiral-Id}
\end{equation}%
These identities are needed when taking traces with the chiral matrix.

\begin{figure}[tbph]
\begin{equation*}
T^{\Gamma _{1}\Gamma _{2}\cdots \Gamma _{n_{1}}}=%
\begin{array}{c}
\includegraphics[scale=0.8]{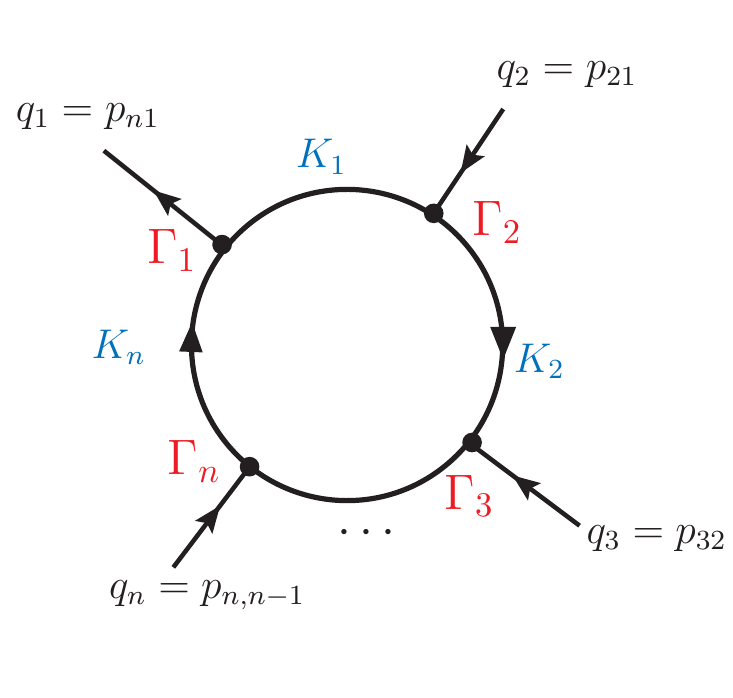}%
\end{array}%
\end{equation*}%
\caption{General diagram for the one-loop amplitudes of this work. }
\label{diag1}
\end{figure}

The spinorial Feynman propagators come naturally from the standard kinetic
term of Dirac fermions%
\begin{equation}
S_{F}\left( K_{i}\right) =\frac{1}{\left( \slashed{K}_{i}-m+i0^{+}\right) }=%
\frac{\left( \slashed{K}_{i}+m\right) }{D_{i}}  \label{Prop}
\end{equation}%
where $D_{i}=K_{i}^{2}-m^{2}$ and the momentum flowing through it%
\begin{equation}
K_{i}=k+k_{i},  \label{Ka}
\end{equation}%
where the $k$ is the unrestricted loop momentum, and $k_{i}$ are the
routings that keep tracking of the flux of external momenta through the
graph, see ref.\cite{Sterman1993}\footnote{%
Specifically in the section (4.1) for a rare comment on the level of
arbitrariness of these routings.}. They can not be written as a function of
the kinematical data in the divergent integrals. In our approach, they
codify the conditions of the satisfaction of symmetries or lack thereof.
Nonetheless, their differences are related to the external momenta through
the definition%
\begin{equation}
p_{ij}=k_{i}-k_{j},  \label{pij}
\end{equation}%
using the momenta conservation in the vertexes of the diagram in the fig. (%
\ref{diag1}) 

From the diagram and Feynman rules follow the integrand of the amplitude%
\begin{equation*}
t^{\Gamma _{1}\Gamma _{2}\cdots \Gamma _{n_{1}}}\left( k_{1},\cdots
,k_{n_{1}}\right) =\text{\textrm{tr}}\left[ \Gamma _{1}S_{F}\left(
K_{1}\right) \Gamma _{2}S_{F}\left( K_{2}\right) \cdots \Gamma
_{n_{1}}S_{F}\left( K_{n_{1}}\right) \right]
\end{equation*}%
that is a well-defined function of the external momenta as the undetermined
(by momentum conservation) sums%
\begin{equation}
P_{ij}=k_{i}+k_{j}.  \label{Pij}
\end{equation}%
Often we are going to adopt the simplification $S\left( i\right) \equiv
S_{F}\left( K_{i}\right) $, where numerical index $i$ represents all the
parameters of the corresponding line. The total amplitude comes from
integration in the loop momenta%
\begin{equation}
T^{\Gamma _{1}\Gamma _{2}\cdots \Gamma _{s}}\left( 1,\cdots ,s\right) =\int 
\frac{\mathrm{d}^{2n}k}{\left( 2\pi \right) ^{2n}}t^{\Gamma _{1}\Gamma
_{2}\cdots \Gamma _{s}}\left( 1,\cdots ,s\right) .
\end{equation}%
The vertexes $\Gamma _{i}$, when replaced by specific ones (\ref%
{SetofVertexes}), the notation accompanies the Lorentz indexes.

It is possible to establish identities among the Green functions exhibiting
Lorentz indexes from the vector and axial currents. They are commonly called
relations among green functions (RAGF), see some references about \cite%
{Battistel2002a}\cite{Battistel2012}\cite{Battistel2014}. The application of
the relations has been used in a wide range of perturbative investigations
in the scenario of IREG. Without claiming as such, they are often used in WI
investigations. However, they can be considered as conditions on the
linearity of integration before any symmetry problems arise. They function
as constraints or guides even before symmetry-specific WIs are asked to play
a consistent judgment role in Feynman perturbation diagrams.

As a working example let us take the amplitude $AV^{r-1}$,%
\begin{equation}
t_{\mu _{1}\mu _{2}\cdots \mu _{r}}^{AV\cdots V}=\text{\textrm{tr}}[\gamma
_{\ast }\gamma _{\mu _{1}}S\left( 1\right) \gamma _{\mu _{2}}S\left(
2\right) \cdots \gamma _{\mu _{r}}S\left( r\right) ]
\end{equation}%
when contracted with $p_{21}^{\mu _{2}}$ in the vectorial vertex $\gamma
_{\mu _{2}}$ it can have one of its propagators removed using (\ref{Ka}) and
(\ref{Prop}) in the standard manipulation $\slashed{p}_{21}=\slashed{K}_{2}-%
\slashed{K}_{1}=S^{-1}\left( 2\right) -S^{-1}\left( 1\right) $. The result
is direct the VRAGF 
\begin{equation*}
p_{21}^{\mu _{2}}t_{\mu _{1}\mu _{2}\cdots \mu _{r}}^{AV\cdots V}=\text{%
\textrm{tr}}[\gamma _{\ast }\gamma _{\mu _{1}}S\left( 1\right) \gamma _{\mu
_{3}}S\left( 3\right) \cdots \gamma _{\mu _{r}}S\left( r\right) ]-\text{%
\textrm{tr}}[\gamma _{\ast }\gamma _{\mu _{1}}S\left( 2\right) \cdots \gamma
_{\mu _{r}}S\left( r\right) ].
\end{equation*}%
The result is again an amplitude built out of the same rules. In this case,
a difference between two such amplitudes%
\begin{equation}
p_{21}^{\mu _{2}}t_{\mu _{1}\mu _{2}\cdots \mu _{r}}^{AV\cdots V}=t_{\mu _{1}%
\hat{\mu}_{2}\cdots \mu _{r}}^{AV\cdots V}(1,\hat{2},\cdots ,r)-t_{\mu _{1}%
\hat{\mu}_{2}\cdots \mu _{r}}^{AV\cdots V}(\hat{1},2,\cdots ,r).
\end{equation}%
The hats mean the omission of the propagator corresponding to that routing
and vertexes corresponding to the Lorentz index as well. In other words, the
RHS contains lower point functions that in general (but not always) are
singular under integration.

For the axial vertex 
\begin{eqnarray*}
p_{r1}^{\mu _{1}}t_{\mu _{12}\cdots \mu _{r}}^{AV\cdots V} &=&\text{\textrm{%
tr}}[S\left( r\right) \gamma _{\ast }S^{-1}\left( r\right) S\left( 1\right)
\gamma _{\mu _{2}}S\left( 2\right) \cdots \gamma _{\mu _{r-1}}S\left(
r-1\right) \gamma _{\mu _{r}}] \\
&&-\text{\textrm{tr}}[\gamma _{\ast }\gamma _{\mu _{2}}S\left( 2\right)
\cdots \gamma _{\mu _{r}}S\left( r\right) ].
\end{eqnarray*}%
We must use $S\left( r\right) \gamma _{\ast }S^{-1}\left( r\right) =\left(
-\gamma _{\ast }-2mS\left( r\right) \gamma _{\ast }\right) ,$ and the
commutation product of the chiral and Dirac matrices. Thus follow the ARAGF%
\begin{equation}
p_{r1}^{\mu _{1}}t_{\mu _{12}\cdots \mu _{r}}^{AV\cdots V}=t_{\mu _{r}\hat{%
\mu}_{1}\mu _{2}\cdots \mu _{r-1}}^{AVV\cdots V}\left( 1,2,\cdots ,\hat{r}%
\right) -t_{\hat{\mu}_{1}\mu _{2}\cdots \mu _{r}}^{AVV\cdots V}(\hat{1}%
,2,\cdots ,r)-2mt_{\mu _{2}\cdots \mu _{r}}^{PV\cdots V}.
\end{equation}

After integration they become%
\begin{eqnarray*}
p_{r1}^{\mu _{1}}T_{\mu _{12}\cdots \mu _{r}}^{AV\cdots V} &=&T_{\mu _{r}%
\hat{\mu}_{1}\cdots \mu _{r-1}}^{AVV\cdots V}\left( 1,2,\cdots ,\hat{r}%
\right) -T_{\hat{\mu}_{1}\cdots \mu _{r}}^{AVV\cdots V}(\hat{1},2,\cdots
,r)-2mT_{\mu _{2}\cdots \mu _{r}}^{PV\cdots V},\text{ and} \\
p_{21}^{\mu _{2}}T_{\mu _{12}\cdots \mu _{r}}^{AV\cdots V} &=&T_{\mu _{1}%
\hat{\mu}_{2}\cdots \mu _{r}}^{AV\cdots V}(1,\hat{2},\cdots ,r)-T_{\mu _{1}%
\hat{\mu}_{2}\cdots \mu _{r}}^{AV\cdots V}(\hat{1},2,\cdots ,r).
\end{eqnarray*}%
From these equations, it is clear that they embody the assumptions of
linearity of integration in perturbative computations. This characteristic
is not guaranteed for divergent amplitudes. We will expose this scenario
through the full calculations of the amplitudes and their relations.
Although these equations can be an important structural property, they are
not a priori linked to the particularities of the model and its symmetries.
However, after summing up all the contributions from the cross diagram (if
applicable), the contraction with momenta, assuming some properties for the
lower-point green functions, must correspond to the WIs.

The WIs are equations satisfied by the Green functions as a consequence of
the continuous symmetries of the action. They are valid in the perturbative
approximations built on Feynman's rules unless they are inevitably
anomalous. They arise from the joint application of the algebra of the
quantized currents and the motion's equations to the currents: $\partial
_{\mu }J^{\mu }=0$ and $\partial _{\mu }J_{\ast }^{\mu }=-2miP$, remember
the currents are bilinears in the fermion. Their expression in the position
space of the AWI%
\begin{equation}
\frac{\partial }{\partial x_{1}^{\mu _{1}}}\left\langle J_{\ast }^{\mu
_{1}}\left( x_{1}\right) J_{\mu _{2}}\left( x_{2}\right) \cdots J_{\mu
_{n}}\left( x_{n}\right) \right\rangle =-2mi\left\langle P\left(
x_{1}\right) J_{\mu _{2}}\left( x_{2}\right) \cdots J_{\mu _{n}}\left(
x_{n}\right) \right\rangle ,
\end{equation}%
where $\left\langle \cdots \right\rangle =\left\langle 0\left\vert T\left[
\cdots \right] \right\vert 0\right\rangle ,$ and of the VWI to the same
correlator%
\begin{equation}
\frac{\partial }{\partial x_{2}^{\mu _{1}}}\left\langle J_{\ast \mu
_{1}}\left( x_{1}\right) J^{\mu _{2}}\left( x_{2}\right) \cdots J_{\mu
_{n}}\left( x_{n}\right) \right\rangle =0.
\end{equation}

In our notation for the perturbative amplitudes, we must have an analogous
equation%
\begin{equation*}
q_{1}^{\mu _{1}}T_{\mu _{12}\cdots \mu _{r}}^{A\rightarrow V\cdots
V}=-2mT_{\mu _{2}\cdots \mu _{r}}^{PV\cdots V};\quad q_{2}^{\mu _{2}}T_{\mu
_{12}\cdots \mu _{r}}^{A\rightarrow V\cdots V}=0;\cdots \quad q_{r}^{\mu
_{r}}T_{\mu _{12}\cdots \mu _{r}}^{A\rightarrow V\cdots V}=0.
\end{equation*}%
The connection between the RAGF and the WIs is direct. The violation of the
RAGF implies violations in the WIs. In this way, the satisfaction of the
total set of WI will be conditional on satisfying all RAGFs plus
translational invariant amplitudes in momentum space, a requirement that we
will show to be impossible in general.

The last point in this section is related to the calculations themselves. To
compute the amplitudes, we have got to take the Dirac traces using
commutation relations of the algebra, after that, any amplitude is always
expressed as linear combinations of bare Feynman integrals to which we adopt
a definition%
\begin{equation}
\bar{J}_{n_{2}}^{\mu _{1}\mu _{2}\cdots \mu _{n_{1}}}\left( 1,2,\cdots
,n_{2}\right) =\int \frac{\mathrm{d}^{2n}k}{\left( 2\pi \right) ^{2n}}\frac{%
K_{i}^{\mu _{1}}\cdots K_{i}^{\mu _{n_{1}}}}{D_{1}D_{2}\cdots D_{n_{2}}},
\label{JInt}
\end{equation}%
simplifying the dependence of the functions on their arguments $f\left(
k_{1},k_{2},\cdots \right) =f\left( 1,2,\cdots \right) $ and when convenient
omitting them at all. The combination $K_{i}=k+k_{i}$ in the definition is
only a convenience. To change from a reference routing $k_{j}$ to $k_{i}$ it
is just a matter of recognizing the definition of $p_{ij}$ in (\ref{pij})
and writing $K_{i}=K_{j}+p_{ij}$. These integrals have power-counting given
by%
\begin{equation}
\omega =2n+n_{1}-2n_{2},
\end{equation}%
being $n_{1}$ the tensor rank and $n_{2}$ the number of denominators
present. Just a set of five types of integral will arise in each amplitude
that we will investigate in this work, they are the theme of the section (%
\ref{FinFcts}).

The amplitudes whose properties will be detailed are:

\begin{itemize}
\item The $2D$ bubbles: $T_{\mu _{1}\mu _{2}}^{AV};$ $T_{\mu _{1}\mu
_{2}}^{VA}$

\item The $4D$ triangles: $T_{\mu _{1}\mu _{2}\mu _{3}}^{AVV};$ $T_{\mu
_{1}\mu _{2}\mu _{3}}^{VAV};$ $T_{\mu _{1}\mu _{2}\mu _{3}}^{VVA};$ $T_{\mu
_{1}\mu _{2}\mu _{3}}^{AAA}$

\item The $6D$ box: $T_{\mu _{1}\mu _{2}\mu _{3}\mu _{4}}^{AVVV}$
\end{itemize}

All of them are divergent odd tensors of rank at least two, in two
dimensions a logarithmic power counting and in four and six dimensions a
linear power counting after taking the Dirac traces. Because of their power
counting, to determine the validity of linearity and the symmetries required
for such tensor we will devise the procedure to handle the divergent
integrals in the next section.

\section{Procedure to Handle the Divergences and the Finite Integrals}

\label{IREG}Before presenting the strategy used to solve the divergent
amplitudes, let us make a digression about the divergent-integrals issue
present in QFT.

It is known that the products of propagators, that are not regular
distribution, are in general ill-defined, as a good example is the equation%
\begin{equation}
I=\int \frac{\mathrm{d}^{4}k}{\left( 2\pi \right) ^{4}}\mathrm{tr}\left[
S_{F}\left( k-p\right) S_{F}\left( k\right) \right] =\int \mathrm{d}^{4}x%
\mathrm{tr}\left[ S_{F}^{2}\left( x\right) \right] \mathrm{e}^{-ip\cdot x},
\end{equation}%
where in the LHS\textit{\ }lays a divergent convolution integral of two
Fourier transformed Feynman propagators in momentum space, as the RHS is the
Fourier transform of the square of the propagator in position space. Both
sides are not defined as distributions for the fact when the point-wise
product of distributions does not exist the convolution product of their
Fourier transform doesn't as well and vice versa. These are the
short-distance UV singularities that are manifested in the divergences of
the loop momentum integrals.

Their origins can be rigorously traced back to multiplications of singular
distributions by discontinuous step function in the chronological ordering
of operators in the interaction picture that leads, through the Wick
theorem, to the Feynman rules, e.g., G.Scharf \cite{Scharf2014,Scharf2010},
originally in Epstein and Glaser \cite{EpsteinGlaser1973}. Although, the
undefined Feynman diagrams can be averted by carefully studying the
splitting of distributions with causal support in the setting of causal
perturbation theory \cite{Aste1997}\cite{Aste2003}, where no divergent
integral appears at all, we are going to keep working with the Feynman rules
in the context of regularizations.

However, the systematic procedure to handle the divergences we will employ
is slightly different from the usual regularizations. It is the framework
known as Implicit Regularization (IREG) a method that was introduced and
developed at the end of 1990's in the Ph.D. thesis of O.A. Battistel \cite%
{PhdBattistel1999}, whose the first references about the subject are \cite%
{Battistel1997,BattistelNemes1999}.

It has the objective to keep the connection at all moments with the
expression of the bare Feynman rules removing all the parameters such as
routings and masses from divergent integrals and putting them in, strictly,
finite integrals that are integrated without restriction, while the
divergent ones do not suffer any modification besides an organization in
surface terms and irreducible scalar integrals.

This objective is realized by noticing that all Feynman integrals possess
propagators-like functions, remember that $D_{i}=\left( k+k_{i}\right)
^{2}-m^{2}$ in eq. (\ref{Prop}), that can be rewritten as%
\begin{equation}
\frac{1}{D_{i}}=\frac{1}{D_{\lambda }+A_{i}}=\frac{1}{D_{\lambda }}\frac{1}{%
\left[ 1-\left( -A_{i}/D_{\lambda }\right) \right] },  \label{DecompDi}
\end{equation}%
where $D_{\lambda }=k^{2}-\lambda ^{2}$ and $A_{i}=2k\cdot k_{i}+\left(
k_{i}^{2}+\lambda ^{2}-m^{2}\right) $. It is worth noting that in this
decomposition, the dependence of the parameters that are not integrated is
contained only at $A_{i}$.

To motivate the first element used in this strategy, which means a general
identity capable of realizing the aforementioned objectives, we shall
digress shortly about the sum of the geometric progression of order $N$ and
ratio $x$ and its relation with the behavior of the propagator, namely the
sum $s=1+x+\cdots +x^{N}$, that is easily expressed in closed form, and then
we can write succinctly%
\begin{equation}
\left( 1-x\right) ^{-1}=\sum_{r=0}^{N}x^{r}+x^{N+1}\left( 1-x\right) ^{-1},
\label{IdentGeom}
\end{equation}%
where immediately it is possible to determine the asymptotic behavior of the
terms in this progression when is identified $x$ in our organization of the
propagator%
\begin{equation}
x=-\frac{A_{i}}{D_{\lambda }}=\frac{2k\cdot k_{i}+\left( k_{i}^{2}+\lambda
^{2}-m^{2}\right) }{\left( k^{2}-\lambda ^{2}\right) },
\end{equation}%
we see that the asymptotic behavior at infinity is $\left\Vert k\right\Vert
^{-1}$. Thus follow that the sequence $\sum_{r=0}^{N}x^{r}$ depends on the
routings only in the numerator, and as a polynomial, decays stronger as
bigger it is $N$.

Being valid for arbitrary $N$ and an identity, to any power counting, in a
product of propagators, is always possible to obtain the external momenta,
through the definition (\ref{pij}), in finite integrals because with the
help of eq.'s (\ref{IdentGeom}) and (\ref{DecompDi}), we get%
\begin{equation}
\frac{1}{D_{i}}=\sum_{r=0}^{N}\left( -1\right) ^{r}\frac{A_{i}^{r}}{%
D_{\lambda }^{r+1}}+\left( -1\right) ^{N+1}\frac{A_{i}^{N+1}}{D_{\lambda
}^{N+1}D_{i}}.  \label{id}
\end{equation}

Making $N$ equals the power counting $\omega $, the last integral is finite
and regularization independent. After this first step, we must mention that
for any $N$ the identity is independent of $\lambda ^{2}$, as can be
verified by taking the derivative in relation to this parameter, in the end,
this generates a connection among the divergent and finite parts of the
integrals and/or amplitudes. It implies specific behavior to the basic
divergent scalar integrals and it the straightforwardly satisfied, thus we
will adopt the mass of the propagator $\lambda ^{2}=m^{2}$ as such a scale
in this work.

Here again, to modularize the analysis and make clear the effects of
divergent and finite parts over the definition of the amplitudes, we adopted
this systematic: the finite integrals will be solved without restrictions,
and the divergences will be kept without any further modification. They will
not be resolved just standardized and basic properties for them will be
established, as we will see next.

\subsection{Divergent Terms\label{DivTerms}}

After applying conveniently the identity (\ref{id}), the content of the
Feynman integrals is going to be specified through, surface terms,
irreducible divergent objects, and finite functions. To clarify, the
divergent terms will appear as a set of pure integration-momentum integrals%
\begin{equation}
\int \frac{\mathrm{d}^{2n}k}{\left( 2\pi \right) ^{2n}}\frac{1}{D_{\lambda
}^{a}},\quad \int \frac{\mathrm{d}^{2n}k}{\left( 2\pi \right) ^{2n}}\frac{%
k_{\mu _{1}}k_{\mu _{2}}}{D_{\lambda }^{a+1}},\quad \int \frac{\mathrm{d}%
^{2n}k}{\left( 2\pi \right) ^{2n}}\frac{k_{\mu _{1}}k_{\mu _{2}}\cdots
k_{\mu _{2b-1}}k_{\mu _{2b}}}{D_{\lambda }^{a+b}},
\end{equation}%
in which $n\geq a$. A convenient systematization emerges naturally. Since
they have the same power counting, it is always possible to combine the
integrals above as surface terms noticing that%
\begin{equation}
-\frac{\partial }{\partial k^{\mu _{1}}}\frac{k_{\mu _{2}}\cdots k_{\mu
_{2n}}}{D_{\lambda }^{a}}=2a\frac{k_{\mu _{1}}k_{\mu _{2}}\cdots k_{\mu
_{2n}}}{D_{\lambda }^{a+1}}-g_{\mu _{1}\mu _{2}}\frac{k^{\mu _{3}}\cdots
k^{\mu _{2n}}}{D_{\lambda }^{a}}-\text{permutations.}
\end{equation}%
As can be seen, all the terms of the RHS have equal power counting and the
iterative use of this observation allow to recombine all of the integrals
above as surface terms.

In short, the highest-rank surface term generates a tower through linear
combinations of lower-rank surface terms up to scalar integrals that encode
exactly the divergent content of the original expression. They will keep the
possibility or not of shifting the integration variable, which means we are
trading the freedom of the operation of translation in the momentum-space
for the arbitrary choice of the routings in these perturbative corrections.

For linear and higher divergent or logarithmic-divergent tensor integrals,
these surface terms are always present, nonetheless, to the former, they
have some of its coefficients the ambiguous momenta defined in eq. (\ref{Pij}%
) as for logarithmic power counting, the coefficients are the external
momenta defined (\ref{pij}).

For our purposes, in this work we need to define the following combinations 
\begin{equation}
\Delta _{\left( n+1\right) ;\mu _{1}\mu _{2}}^{2n}(\lambda ^{2})=\int \frac{%
\mathrm{d}^{2n}k}{\left( 2\pi \right) ^{2n}}\left( \frac{2nk_{\mu
_{1}}k_{\mu _{2}}}{D_{\lambda }^{n+1}}-g_{\mu _{1}\mu _{2}}\frac{1}{%
D_{\lambda }^{n}}\right) =-\int \frac{\mathrm{d}^{2n}k}{\left( 2\pi \right)
^{2n}}\frac{\partial }{\partial k^{\mu _{1}}}\frac{k_{\mu _{2}}}{D_{\lambda
}^{n}},  \label{delta2n}
\end{equation}%
in two, four, and six dimensions, $n=1$, $n=2$, and $n=3$ respectively
indicated by the superscript.

The irreducible scalar is defined also by dimension and will be%
\begin{equation}
I_{\log }^{\left( 2n\right) }\left( \lambda ^{2}\right) =\int \frac{\mathrm{d%
}^{2n}k}{\left( 2\pi \right) ^{2n}}\frac{1}{D_{\lambda }^{n}}.
\end{equation}

As we have seen, the separation highlights diverging structures that
organize them without performing any analytic operation. Moreover, it makes
clear that the divergent content is a local polynomial in the ambiguous and
physical momenta which is obtained without expansions or limits. The finite
parts are regularization free and obtained by integrating just finite
integrals that will be explained in the section (\ref{FinFcts}).

\subsection{Finite Functions}

\label{FinFcts}

After separating the finite part using the identity (\ref{id}), we solve the
integrals using the usual techniques of perturbative calculations. It is
possible to project the results into a family of functions. For the
two-point basic functions they are given by%
\begin{equation}
Z_{n_{1}}^{\left( -1\right) }=\int_{0}^{1}\mathrm{d}x\frac{x^{n_{1}}}{Q}%
;\quad Z_{n_{1}}^{\left( 0\right) }=\int_{0}^{1}\mathrm{d}xx^{n_{1}}\log 
\frac{Q}{-m^{2}}
\end{equation}%
where $Q$ is a polynomial given by%
\begin{equation}
Q=p^{2}x\left( 1-x\right) -m^{2}
\end{equation}

For the three-point basic we have\footnote{%
Thes polynomials can be written as a quadric form with the Symanzik
polynomials constructed using the spanning trees and two-forests of the
graph.} 
\begin{equation}
Q=p^{2}x_{1}\left( 1-x_{1}\right) +q^{2}x_{2}\left( 1-x_{2}\right) -2\left(
p\cdot q\right) x_{1}x_{2}-m^{2}
\end{equation}%
and the basic functions%
\begin{eqnarray}
Z_{n_{1}n_{2}}^{\left( -1\right) } &=&\int_{0}^{1}\mathrm{d}%
x_{1}\int_{0}^{1-x_{1}}\mathrm{d}x_{2}\frac{x_{1}^{n_{1}}x_{2}^{n_{2}}}{Q}
\label{Znm(-1)} \\
Z_{n_{1}n_{2}}^{\left( 0\right) } &=&\int_{0}^{1}\mathrm{d}%
x_{1}\int_{0}^{1-x_{1}}\mathrm{d}x_{2}x_{1}^{n_{1}}x_{2}^{n_{2}}\log \frac{Q%
}{-m^{2}}
\end{eqnarray}%
And finally, for the four-point is%
\begin{eqnarray}
Q &=&p^{2}x_{1}\left( 1-x_{1}\right) +q^{2}x_{2}\left( 1-x_{2}\right)
+r^{2}x_{3}\left( 1-x_{3}\right)  \notag \\
&&-2\left( p\cdot q\right) x_{1}x_{2}-2\left( p\cdot r\right)
x_{1}x_{3}-2\left( q\cdot r\right) x_{2}x_{3}-m^{2}
\end{eqnarray}%
and the basic functions%
\begin{eqnarray}
Z_{n_{1}n_{2}n_{3}}^{\left( -1\right) } &=&\int_{0}^{1}\mathrm{d}%
x_{1}\int_{0}^{1-x_{1}}\mathrm{d}x_{2}\int_{0}^{1-x_{1}-x_{2}}\mathrm{d}x_{3}%
\frac{x_{1}^{n_{1}}x_{2}^{n_{2}}x_{3}^{n_{3}}}{Q}  \label{Zmnp(-1)} \\
Z_{n_{1}n_{2}n_{3}}^{\left( 0\right) } &=&\int_{0}^{1}\mathrm{d}%
x_{1}\int_{0}^{1-x_{1}}\mathrm{d}x_{2}\int_{0}^{1-x_{1}-x_{2}}\mathrm{d}%
x_{3}x_{1}^{n_{1}}x_{2}^{n_{2}}x_{3}^{n_{3}}\log \frac{Q}{-m^{2}}
\end{eqnarray}

It is possible to write the parameters in terms of derivatives of the
polynomials and, then, use partial integration to obtain relations among
these functions. More precisely, they are reductions of the parameter powers 
$n_{1}+n_{2}$, for the equation (\ref{Znm(-1)}) and $n_{1}+n_{2}+n_{3},$ for
the equation (\ref{Zmnp(-1)}) (see appendices \ref{AppInt4D}, and \ref%
{AppInt6D}), they were approched in the papers \cite{Battistel2006}, \cite%
{Dallabona2012}, and \cite{SunYi2012}. This resource is necessary to perform
the operations displayed in this paper.

\subsection{Basis of Feynman Integrals\label{BasisFI}}

From the general definition presented in eq. (\ref{JInt}) for the integrals
that appear soon after taking Dirac traces, we will describe in a nutshell
the ones that arise in the $AV^{n}$ amplitudes in $d=2n$, more generally,
any amplitude of $n+1$ points of odd parity. They are%
\begin{equation}
\left( \bar{J}_{n};\bar{J}_{n}^{\mu _{1}}\right) =\int \frac{\mathrm{d}^{2n}k%
}{\left( 2\pi \right) ^{2n}}\frac{(1;\ K_{1}^{\mu _{1}})}{D_{12\cdots n}}
\end{equation}%
and%
\begin{equation}
\left( \bar{J}_{n+1};\bar{J}_{n+1}^{\mu _{1}};\bar{J}_{n+1}^{\mu
_{12}}\right) =\int \frac{\mathrm{d}^{2n}k}{\left( 2\pi \right) ^{2n}}\frac{%
(1;\ K_{1}^{\mu _{1}};\ K_{1}^{\mu _{12}})}{D_{12\cdots n+1}},
\end{equation}%
the power counting of $n$-point integrals are $\omega (J_{n}^{\mu _{1}})=1,$ 
$\omega \left( J_{n}\right) =0$ and of the ($n+1$)-point integrals are $%
\omega (J_{n+1}^{\mu _{12}})=0$, $\omega (J_{n+1}^{\mu _{1}})=-1$, and $%
\omega \left( J_{n+1}\right) =-2$. Observe \ our notational conventions $%
K_{12\cdots i}^{\nu _{12}\cdots \nu _{i}}=K_{1}^{\nu _{1}}K_{2}^{\nu
_{2}}\cdots K_{i}^{\nu _{i}}$ and $D_{12\cdots i}=D_{1}D_{2}\cdots D_{i}$.

Therefore, as anticipated by the power counting, some of these integrals
contain finite and divergent parts, as is the case with $\bar{J}_{n+1}^{\mu
_{12}},$ $\bar{J}_{n}^{\mu _{1}},$ and $\bar{J}_{n}$. The integrals $%
J_{n+1}^{\mu _{1}}$ and $J_{n+1}$ are finite and then will not get an
overbar. All the time, we are working with the strictly finite part of the
divergent integrals, they will come free of the overbar.

As a quite important example, we chose to compute the highest power-counting
integral in $d=4$ to illustrate some of the features of our treatment, the
vector two-point integral%
\begin{equation}
\bar{J}_{2}^{\mu _{1}}=\int \frac{\mathrm{d}^{4}k}{\left( 2\pi \right) ^{4}}%
\frac{K_{1}^{\mu _{1}}}{D_{12}}
\end{equation}%
using the identity (\ref{id}) with $N=1$%
\begin{equation}
\frac{1}{D_{i}}=\frac{1}{D_{\lambda }}-\frac{A_{i}}{D_{\lambda }^{2}}+\frac{%
A_{i}^{2}}{D_{\lambda }^{2}D_{i}}
\end{equation}%
corresponding to its linear divergence, we get%
\begin{eqnarray}
\frac{K_{1}^{\mu _{1}}}{D_{12}} &=&\frac{K_{1}^{\mu _{1}}}{D_{\lambda }^{2}}-%
\frac{\left( A_{1}+A_{2}\right) }{D_{\lambda }^{3}}K_{1}^{\mu _{1}}+\frac{%
A_{1}A_{2}}{D_{\lambda }^{4}}K_{1}^{\mu _{1}}  \notag \\
&&+\left[ \frac{A_{1}^{2}}{D_{\lambda }^{3}D_{1}}+\frac{A_{2}^{2}}{%
D_{\lambda }^{3}D_{2}}-\frac{A_{1}A_{2}^{2}}{D_{\lambda }^{4}D_{2}}-\frac{%
A_{2}A_{1}^{2}}{D_{\lambda }^{4}D_{1}}+\frac{A_{1}^{2}A_{2}^{2}}{D_{\lambda
}^{4}D_{12}}\right] K_{1}^{\mu _{1}},
\end{eqnarray}%
collecting the purely divergent integrals and integrating the remaining
finite integrals comes%
\begin{equation}
\bar{J}_{2}^{\mu _{1}}=J_{2}^{\mu _{1}}\left( p_{21}\right) -\frac{1}{2}%
\left[ P_{21}^{\nu _{1}}\Delta _{3\nu _{1}}^{\left( 4\right) \mu
_{1}}+p_{21}^{\mu _{1}}I_{\log }^{\left( 4\right) }\right]  \label{J2bar4D}
\end{equation}%
where%
\begin{equation}
J_{2}^{\mu _{1}}\left( p_{21}\right) =\frac{i}{\left( 4\pi \right) ^{2}}%
p_{21}^{\mu _{1}}Z_{1}^{\left( 0\right) }\left( p_{21}\right)
\end{equation}%
showing all the elements we have presented before, the local divergent terms
organized, and the finite part integrated without restrictions.

The same type of steps leads, in $d=6$ dimensions, to%
\begin{equation}
\bar{J}_{3}^{\mu _{1}}=\int \frac{\mathrm{d}^{6}k}{\left( 2\pi \right) ^{6}}%
\frac{K_{1}^{\mu _{1}}}{D_{123}}
\end{equation}%
\begin{equation}
\bar{J}_{3}^{\mu _{1}}=-\frac{1}{3}\left( k_{1}^{\nu _{1}}+k_{2}^{\nu
_{1}}+k_{3}^{\nu _{1}}\right) \Delta _{4\nu _{1}}^{\left( 6\right) \mu _{1}}-%
\frac{1}{3}\left( p_{21}^{\mu _{1}}+p_{31}^{\mu _{1}}\right) I_{\log
}^{\left( 6\right) }+J_{3}^{\mu _{1}},
\end{equation}%
where the finite part is simply%
\begin{equation}
J_{3}^{\mu _{1}}\left( p_{21},p_{31}\right) =\frac{i}{\left( 4\pi \right)
^{3}}\left[ p_{21}^{\mu _{1}}Z_{10}^{\left( 0\right) }\left(
p_{21},p_{31}\right) +p_{31}^{\mu _{1}}Z_{01}^{\left( 0\right) }\left(
p_{21},p_{31}\right) \right] ,
\end{equation}%
as in two dimensions, we have%
\begin{equation}
\bar{J}_{1\mu _{1}}=\int \frac{\mathrm{d}^{2}k}{\left( 2\pi \right) ^{2}}%
\frac{K_{1\mu _{1}}}{D_{1}}=-k_{1}^{\nu _{1}}\Delta _{2\nu _{1}\mu
_{1}}^{\left( 2\right) },
\end{equation}%
that is a pure surface term, and this is the reason to illustrate for $d=4$
dimensions first.

For all explicit results used see the appendices (\ref{AppInt4D}) and (\ref%
{AppInt6D}).

\section{Two Dimensional $AV$ and $VA$ Two-Point Functions}

\label{2Dim2Pt}To establish the connection among linearity, uniqueness, and
WIs we study the two Lorentz indexes amplitudes $AV$, and $VA$. Amid this
process, the relation among all two-point functions will emerge. As the
power counting is zero, it is not expected they depend on the routings
through their sums besides they are a function of only two routings, hence
we will adopt the simplification $q=p_{21}=k_{2}-k_{1}$, when seen as suit.

Therefore, to start with, we have the RAGFs coming from the vector vertex%
\begin{eqnarray}
q^{\mu _{2}}t_{\mu _{12}}^{AV} &=&t_{\mu _{1}}^{A}\left( 1\right) -t_{\mu
_{1}}^{A}\left( 2\right) \\
q^{\mu _{1}}t_{\mu _{12}}^{VA} &=&t_{\mu _{2}}^{A}\left( 1\right) -t_{\mu
_{2}}^{A}\left( 2\right) ,  \label{p1VA}
\end{eqnarray}%
and from the axial vertex%
\begin{eqnarray}
q^{\mu _{1}}t_{\mu _{12}}^{AV} &=&t_{\mu _{2}}^{A}\left( 1\right) -t_{\mu
_{2}}^{A}\left( 2\right) -2mt_{\mu _{2}}^{PV} \\
q^{\mu _{2}}t_{\mu _{12}}^{VA} &=&t_{\mu _{1}}^{A}\left( 1\right) -t_{\mu
_{1}}^{A}\left( 2\right) +2mt_{\mu _{1}}^{VP},  \label{p2VA}
\end{eqnarray}%
obtained using the procedure delineated in the section (\ref{ModlDef}).
Taking their integrals, we should have%
\begin{eqnarray}
q^{\mu _{2}}T_{\mu _{12}}^{AV} &=&T_{\mu _{1}}^{A}\left( 1\right) -T_{\mu
_{1}}^{A}\left( 2\right) \\
q^{\mu _{1}}T_{\mu _{12}}^{AV} &=&T_{\mu _{2}}^{A}\left( 1\right) -T_{\mu
_{2}}^{A}\left( 2\right) -2mT_{\mu _{2}}^{PV},
\end{eqnarray}%
only by the linearity of integration. Similarly to $VA$ amplitudes.

On the other hand, the WIs will require%
\begin{equation}
q^{\mu _{2}}T_{\mu _{12}}^{AV}=0;\quad q^{\mu _{1}}T_{\mu
_{12}}^{AV}=-2mT_{\mu _{2}}^{PV},
\end{equation}%
that in turn, through the general tensor structure of these amplitudes,
imply kinematic properties to the scalar invariants of these tensors with
the same status as the symmetry properties.

As an example, to the $AV$ amplitude, we can write%
\begin{equation}
T_{\mu _{12}}^{AV}=\varepsilon _{\mu _{1}\mu _{2}}F_{1}+\varepsilon _{\mu
_{1}\nu }q^{\nu }q_{\mu _{2}}F_{2}+\varepsilon _{\mu _{2}\nu }q^{\nu }q_{\mu
_{1}}F_{3},  \label{AVForm}
\end{equation}%
being the $F_{i}$ the scalar invariants, then by contracting with the
external momenta in the respective indexes, we get%
\begin{eqnarray}
q^{\mu _{2}}T_{\mu _{12}}^{AV} &=&\varepsilon _{\mu _{1}\nu }q^{\nu }\left(
q^{2}F_{2}+F_{1}\right) \text{ and}  \label{q2ContAV} \\
q^{\mu _{1}}T_{\mu _{12}}^{AV} &=&\varepsilon _{\mu _{2}\nu }q^{\nu }\left(
q^{2}F_{3}-F_{1}\right) ,  \label{q1ContAV}
\end{eqnarray}%
by vector conservation, in the first equation, we trade $F_{1}=-q^{2}F_{2}$
in the second equation to obtain 
\begin{equation}
q^{\mu _{1}}T_{\mu _{12}}^{AV}=\varepsilon _{\mu _{2}\nu }q^{\nu
}q^{2}\left( F_{3}+F_{2}\right) ,
\end{equation}%
and finally, under the hypothesis of regularity, we have the low energy
theorem for the contraction with the index of the axial current $\left.
q^{\mu _{1}}T_{\mu _{12}}^{AV}\right\vert _{q^{2}=0}=0$. However, if the WI
that relates this contraction to the $PV$ function is satisfied, comes the
consequence%
\begin{equation}
\left. q^{\mu _{1}}T_{\mu _{12}}^{AV}\right\vert _{0}=\left. -2mT_{\mu
_{2}}^{PV}\right\vert _{0}=\varepsilon _{\mu _{2}\nu }q^{\nu }\Omega
^{PV}\left( q^{2}=0\right) =0.
\end{equation}%
Therefore, if the symmetries were respected and the hypothesis of regular
form factors is met, such behavior must be attained and in this sense, we
have said that it has the status of symmetry.

All these constraints must be seen in the light of explicit computations
that will be unfolded and analyzed in the sequel. Following the definitions
of the previous section, after integration the amplitude becomes%
\begin{equation*}
T^{\Gamma _{1}\Gamma _{2}}\left( 1,2\right) =\int \frac{\mathrm{d}^{2}k}{%
\left( 2\pi \right) ^{2}}t^{\Gamma _{1}\Gamma _{2}}\left( 1,2\right) ,
\end{equation*}%
expanding the terms of mass and momentum, 
\begin{eqnarray}
t^{\Gamma _{1}\Gamma _{2}} &=&+K_{12}^{\nu _{12}}\text{\textrm{tr}}\left[
\Gamma _{1}\gamma _{\nu _{1}}\Gamma _{2}\gamma _{\nu _{2}}\right] \frac{1}{%
D_{12}}  \label{2ptexp} \\
&&+mK_{1}^{\nu _{1}}\text{\textrm{tr}}\left[ \Gamma _{1}\gamma _{\nu
_{1}}\Gamma _{2}\right] \frac{1}{D_{12}}  \notag \\
&&+mK_{2}^{\nu _{1}}\text{\textrm{tr}}\left[ \Gamma _{1}\Gamma _{2}\gamma
_{\nu _{1}}\right] \frac{1}{D_{12}}  \notag \\
&&+m^{2}\text{\textrm{tr}}\left[ \Gamma _{1}\Gamma _{2}\right] \frac{1}{%
D_{12}},  \notag
\end{eqnarray}%
from the formula above, choosing appropriately the vertexes and keeping the
non-zero traces, we have%
\begin{eqnarray*}
t_{\mu _{12}}^{AV} &=&K_{12}^{\nu _{12}}\mathrm{tr}(\gamma _{\ast }\gamma
_{\mu _{1}\nu _{1}\mu _{2}\nu _{2}})\frac{1}{D_{12}}+m^{2}\mathrm{tr}(\gamma
_{\ast }\gamma _{\mu _{1}\mu _{2}})\frac{1}{D_{12}}, \\
t_{\mu _{12}}^{VA} &=&K_{12}^{\nu _{12}}\mathrm{tr}(\gamma _{\ast }\gamma
_{\mu _{1}\nu _{1}\mu _{2}\nu _{2}})\frac{1}{D_{12}}-m^{2}\mathrm{tr}(\gamma
_{\ast }\gamma _{\mu _{1}\mu _{2}})\frac{1}{D_{12}}.
\end{eqnarray*}

The main point is that the trace of four gamma matrices, which is a linear
combination of the metric and epsilon tensor, can be expressed in a variety
of forms, obtained using the substitutions in the identity (\ref{Chiral-Id}%
), that means%
\begin{equation*}
\gamma _{\ast }=\varepsilon _{\nu _{12}}\gamma ^{\nu _{12}}/2;\text{\quad }%
\gamma _{\ast }\gamma _{\mu _{1}}=-\varepsilon _{\mu _{1}\nu _{1}}\gamma
^{\nu _{1}};\text{\quad }\gamma _{\ast }\gamma _{\left[ \mu _{1}\mu _{2}%
\right] }=-\varepsilon _{\mu _{1}\mu _{2}}.
\end{equation*}%
They lead to expressions that are not automatically equal after integration.
Effectively they turn the RAGFs into equations among functions. To unfold
the rationale, let us apply the definition of the chiral matrix in the form%
\begin{eqnarray*}
\mathrm{tr}\left( \gamma _{\ast }\gamma _{abcd}\right) &=&\varepsilon
_{\alpha _{12}}\mathrm{tr}\left( \gamma ^{\alpha _{12}}\gamma _{abcd}\right)
/2 \\
&=&2\left[ -g_{ab}\varepsilon _{cd}+g_{ac}\varepsilon
_{bd}-g_{ad}\varepsilon _{bc}-g_{bc}\varepsilon _{ad}+g_{bd}\varepsilon
_{ac}-g_{cd}\varepsilon _{ab}\right] ,
\end{eqnarray*}%
where the Latin indexes make it simple to perform substitutions to obtain
the trace when the definition of the chiral matrix is deployed adjacent to
the first or the second vertex, by example $\left( a,b,c,d\right) =\left(
\mu _{1},\nu _{1},\mu _{2},\nu _{2}\right) $ and $\left( a,b,c,d\right)
=\left( \mu _{2},\nu _{2},\mu _{1},\nu _{1}\right) $. These traces will
differ by the signs of the terms only and have all the indexes of the trace
explicitly present although distinctly displayed, these seemingly innocuous
observations have far-reaching implications. With these aspects in mind, let
us call these two expressions version one and two of the traces and carry
over to the amplitudes the same nomenclature, as we will demonstrate they
are enough to reach any other expression.

\textbf{First Version:}%
\begin{eqnarray*}
K_{12}^{\nu _{12}}\mathrm{tr}\left( \gamma _{\ast }\gamma _{\mu _{1}}\gamma
_{\nu _{1}}\gamma _{\mu _{2}}\gamma _{\nu _{2}}\right) &=&-2\varepsilon
_{\mu _{1}\nu _{1}}\left( K_{1\mu _{2}}K_{2}^{\nu _{1}}+K_{2\mu
_{2}}K_{1}^{\nu _{1}}\right) -2\varepsilon _{\mu _{2}\nu _{1}}\left( K_{1\mu
_{1}}K_{2}^{\nu _{1}}-K_{2\mu _{1}}K_{1}^{\nu _{1}}\right) \\
&&+2\varepsilon _{\mu _{1}\mu _{2}}\left( K_{1}\cdot K_{2}\right) +2g_{\mu
_{1}\mu _{2}}\varepsilon _{\nu _{1}\nu _{2}}K_{12}^{\nu _{12}}
\end{eqnarray*}

\textbf{Second Version}:%
\begin{eqnarray*}
K_{12}^{\nu _{12}}\mathrm{tr}(\gamma _{\ast }\gamma _{\mu _{2}}\gamma _{\nu
_{2}}\gamma _{\mu _{1}}\gamma _{\nu _{1}}) &=&+2\varepsilon _{\mu _{1}\nu
_{1}}\left( K_{1\mu _{2}}K_{2}^{\nu _{1}}-K_{2\mu _{2}}K_{1}^{\nu
_{1}}\right) -2\varepsilon _{\mu _{2}\nu _{1}}\left( K_{1\mu _{1}}K_{2}^{\nu
_{1}}+K_{2\mu _{1}}K_{1}^{\nu _{1}}\right) \\
&&-2\varepsilon _{\mu _{1}\mu _{2}}\left( K_{1}\cdot K_{2}\right) -2g_{\mu
_{1}\mu _{2}}\varepsilon _{\nu _{1}\nu _{2}}K_{12}^{\nu _{12}}.
\end{eqnarray*}%
Here, we have already contracted with $K_{12}^{\nu _{12}}$. Now we note that
in the first row of each version, it is possible to identify, as will be
done in four and six dimensions, a common tensor%
\begin{equation}
t_{\mu _{2}}^{\left( s\right) \nu _{1}}=\left( K_{1\mu _{2}}K_{2}^{\nu
_{1}}+sK_{2\mu _{2}}K_{1}^{\nu _{1}}\right) \frac{1}{D_{12}},  \label{t(s)}
\end{equation}%
where $s=\pm 1$. And in the other rows appear the amplitude $SP$ obtained
when substituting the respective vertexes in (\ref{2ptexp}), $%
t^{SP}=2\varepsilon _{\nu _{12}}K_{12}^{\nu _{12}}/D_{12}=-t^{PS}.$

Completing the two amplitudes summing the mass terms using $\mathrm{tr}%
(\gamma _{\ast }\gamma _{\mu _{12}})=-2\varepsilon _{\mu _{12}}$, we get the
first and the second versions of the $AV$%
\begin{eqnarray*}
\left( t_{\mu _{12}}^{AV}\right) _{1} &=&-2\varepsilon _{\mu _{1}\nu
_{1}}t_{\mu _{2}}^{\left( +\right) \nu _{1}}-\varepsilon _{\mu _{1}\mu
_{2}}t^{PP}-2\varepsilon _{\mu _{2}\nu _{1}}t_{\mu _{1}}^{\left( -\right)
\nu _{1}}+g_{\mu _{1}\mu _{2}}t^{SP} \\
\left( t_{\mu _{12}}^{AV}\right) _{2} &=&-2\varepsilon _{\mu _{2}\nu
_{1}}t_{\mu _{1}}^{\left( +\right) \nu _{1}}-\varepsilon _{\mu _{1}\mu
_{2}}t^{SS}+2\varepsilon _{\mu _{1}\nu _{1}}t_{\mu _{2}}^{\left( -\right)
\nu _{1}}-g_{\mu _{1}\mu _{2}}t^{SP},
\end{eqnarray*}%
similar expressions can be obtained to the $VA$ amplitude.

In the above relations, we have identified the scalar two-point amplitudes%
\begin{eqnarray}
t^{PP} &=&q^{2}\frac{1}{D_{12}}-\frac{1}{D_{1}}-\frac{1}{D_{2}}  \label{PP}
\\
t^{SS} &=&(4m^{2}-q^{2})\frac{1}{D_{12}}+\frac{1}{D_{1}}+\frac{1}{D_{2}},
\label{SS}
\end{eqnarray}%
to obtain these amplitudes we have used%
\begin{equation}
S_{ij}=K_{i}\cdot K_{j}-m^{2}=\frac{1}{2}(D_{i}+D_{j}-p_{ij}^{2}),
\label{Sij}
\end{equation}%
to reduce the bilinears that appear in their definitions in the eq.(\ref%
{2ptexp}). So it is straightforward to identify in the middle of our
expressions,%
\begin{equation*}
t_{\mu _{1}\mu _{2}}^{VV}=2t_{\mu _{1}\mu _{2}}^{\left( +\right) }+g_{\mu
_{1}\mu _{2}}t^{PP};\text{ and\quad }t_{\mu _{1}\mu _{2}}^{AA}=2t_{\mu
_{1}\mu _{2}}^{\left( +\right) }-g_{\mu _{1}\mu _{2}}t^{SS},
\end{equation*}%
they follow immediately after the traces are taken in the respective
definitions and clearly will relate the odd amplitudes to the even ones.

When integrated, it is always possible to see that some terms are finite and
vanishing. Namely $\varepsilon _{\mu _{i}\nu }T_{\mu _{j}}^{\left( -\right)
\nu }=0$ and $T^{SP}=0$, due of the properties of the finite vector integral
that is proportional to momentum $q=p_{21}$ and the scalar integral $J_{2}$
see in (\ref{2DJ2}). Therefore, the integrals will provide the basic
relations%
\begin{eqnarray*}
\left( T_{\mu _{12}}^{AV}\right) _{1} &=&-\varepsilon _{\mu _{1}}^{\quad \nu
_{1}}T_{\nu _{1}\mu _{2}}^{VV};\text{\qquad }\left( T_{\mu
_{12}}^{AV}\right) _{2}=-\varepsilon _{\mu _{2}}^{\quad \nu _{1}}T_{\mu
_{1}\nu _{1}}^{AA} \\
\left( T_{\mu _{12}}^{VA}\right) _{1} &=&-\varepsilon _{\mu _{1}}^{\quad \nu
_{1}}T_{\nu _{1}\mu _{2}}^{AA};\text{\qquad }\left( T_{\mu
_{12}}^{VA}\right) _{2}=-\varepsilon _{\mu _{2}}^{\quad \nu _{1}}T_{\mu
_{1}\nu _{1}}^{VV}.
\end{eqnarray*}

However, if one applies the second relation $\gamma _{\ast }\gamma _{\mu
_{1}}=-\varepsilon _{\mu _{1}\nu _{1}}\gamma ^{\nu _{1}}$ around the same
vertexes, we get directly%
\begin{equation*}
\left( t_{\mu _{12}}^{AV}\right) _{1}\tilde{=}-\varepsilon _{\mu
_{1}}^{\quad \nu _{1}}t_{\nu _{1}\mu _{2}}^{VV};\qquad \left( t_{\mu
_{12}}^{AV}\right) _{2}\tilde{=}-\varepsilon _{\mu _{2}}^{\quad \nu
_{1}}t_{\mu _{1}\nu _{1}}^{AA}.
\end{equation*}%
The sign $\tilde{=}$ means they are equal up to terms that are finite and
vanish under integration. The other independent version is obtained using
the third relation $\gamma _{\ast }\gamma _{\left[ \mu _{1}\mu _{2}\right]
}=-\varepsilon _{\mu _{1}\mu _{2}},$ in the form $\gamma _{\ast }\gamma
_{\mu _{1}}\gamma _{\nu _{1}}=-\varepsilon _{\mu _{1}\nu _{1}}+g_{\mu
_{1}\nu _{1}}\gamma _{\ast }$, given in the trace the expression $-2\left(
\varepsilon _{\mu _{1}\nu _{1}}g_{\mu _{2}\nu _{2}}+\varepsilon _{\mu
_{2}\nu _{2}}g_{\mu _{1}\nu _{1}}\right) $, that in the amplitudes enable us
to arrange the result%
\begin{eqnarray*}
\left( t_{\mu _{12}}^{AV}\right) _{3} &=&-\frac{1}{2}\left[ \varepsilon
_{\mu _{1}}^{\hspace{7pt}\nu _{1}}\left( t_{\nu _{1}\mu _{2}}^{VV}\right)
+\varepsilon _{\mu _{2}}^{\hspace{7pt}\nu _{1}}\left( t_{\mu _{1}\nu
_{1}}^{AA}\right) \right] -\varepsilon _{\mu _{2}\nu _{1}}t_{\mu
_{1}}^{\left( -\right) \nu _{1}}+\varepsilon _{\mu _{1}\nu _{1}}t_{\mu
_{2}}^{\left( -\right) \nu _{1}}, \\
\left( t_{\mu _{12}}^{VA}\right) _{3} &=&-\frac{1}{2}\left[ \varepsilon
_{\mu _{1}}^{\hspace{7pt}\nu _{1}}\left( t_{\nu _{1}\mu _{2}}^{AA}\right)
+\varepsilon _{\mu _{2}}^{\hspace{7pt}\nu _{1}}\left( t_{\mu _{1}\nu
_{1}}^{VV}\right) \right] -\varepsilon _{\mu _{2}\nu _{1}}t_{\mu
_{1}}^{\left( -\right) \nu _{1}}+\varepsilon _{\mu _{1}\nu _{1}}t_{\mu
_{2}}^{\left( -\right) \nu _{1}},
\end{eqnarray*}%
thereby their integrals provide us with the following%
\begin{equation}
\left( T_{\mu _{12}}^{AV}\right) _{3}=\frac{1}{2}\left[ \left( T_{\mu
_{12}}^{AV}\right) _{1}+\left( T_{\mu _{12}}^{AV}\right) _{2}\right] ;\quad
\left( T_{\mu _{12}}^{VA}\right) _{3}=\frac{1}{2}\left[ \left( T_{\mu
_{12}}^{VA}\right) _{1}+\left( T_{\mu _{12}}^{VA}\right) _{2}\right] .
\label{AV3}
\end{equation}%
This form is present in the equation (85) of the paper \cite{Battistel2004},
for example. This last form, obtained a linear combination of the other two
is a particular aspect that is made clear in the section (\ref{4Dim3Pt}),
here is present because it comes from the identites we introduced and we
wish to make a pedestrian approach in two dimensions.

To put the consequences of versions for amplitudes into perspective, we need
integrated expressions. It will be possible to see that the sampling of the
indices that appear between the finite and divergent parts makes the
expressions not automatically equal when integrated. However, the versions
are related through linearity violations reflected in the RAGFs. Using the
explicit results found in the appendix (\ref{AppInt2D}), we will have, from
the expressions (\ref{PP}) and (\ref{SS}), 
\begin{eqnarray}
T^{PP} &=&q^{2}J_{2}-2I_{\log } \\
T^{SS} &=&\left( 4m^{2}-p^{2}\right) J_{2}+2I_{\log },
\end{eqnarray}%
and for the sign tensor (\ref{t(s)}) 
\begin{eqnarray}
T_{\mu _{1}\mu _{2}}^{\left( +\right) } &=&2\left( \bar{J}_{2\mu _{1}\mu
_{2}}+q_{\mu _{1}}J_{2\mu _{2}}\right) =2\theta _{\mu _{1}\mu _{2}}\left(
q\right) \left( i/4\pi +m^{2}J_{2}\right) \\
&&-\frac{1}{2}g^{\mu _{1}\mu _{2}}q^{2}J_{2}+\left( \Delta _{2\mu _{1}\mu
_{2}}+g_{\mu _{1}\mu _{2}}I_{\log }\right) ,
\end{eqnarray}%
to the two-rank, two-point, even amplitudes%
\begin{eqnarray}
T_{\mu _{1}\mu _{2}}^{VV} &=&2\Delta _{2\mu _{1}\mu _{2}}+4\theta _{\mu
_{1}\mu _{2}}\left( m^{2}J_{2}+i/4\pi \right) , \\
T_{\mu _{1}\mu _{2}}^{AA} &=&2\Delta _{2\mu _{1}\mu _{2}}+4\theta _{\mu
_{1}\mu _{2}}\left( m^{2}J_{2}+i/4\pi \right) -g_{\mu _{1}\mu _{2}}\left(
4m^{2}J_{2}\right) ,
\end{eqnarray}%
where $\theta _{\alpha \lambda }\left( q\right) =\left( g_{\alpha \lambda
}q^{2}-q_{\alpha }q_{\lambda }\right) /q^{2}$ is the transversal projector.
And in this way, the odd amplitude $AV$ will turn up with the explicitly
expressions%
\begin{eqnarray}
\left( T_{\mu _{12}}^{AV}\right) _{1} &=&-2\varepsilon _{\mu _{1}}^{\quad
\nu }\Delta _{2\mu _{2}\nu }-4\varepsilon _{\mu _{1}\nu }\theta _{\mu
_{2}}^{\nu }\left( m^{2}J_{2}+i/4\pi \right)  \label{AV1} \\
\left( T_{\mu _{12}}^{AV}\right) _{2} &=&-2\varepsilon _{\mu _{2}}^{\quad
\nu }\Delta _{2\mu _{1}\nu }-4\varepsilon _{\mu _{2}\nu }\theta _{\mu
_{1}}^{\nu }\left( m^{2}J_{2}+i/4\pi \right) -\varepsilon _{\mu _{1}\mu
_{2}}\left( 4m^{2}J_{2}\right) .  \label{AV2}
\end{eqnarray}%
The expression for (\ref{AV3}) is obtained as a combination of the
expressions above.

The two-point functions, that are finite, and appear in the RAGFs 
\begin{eqnarray}
T_{\mu }^{PV} &=&-T_{\mu }^{VP}=\varepsilon _{\mu \nu }q^{\nu }\left[
-2mJ_{2}\left( q\right) \right] .  \label{PV} \\
T_{\mu }^{PA} &=&-T_{\mu }^{AP}=q_{\mu }\left[ +2mJ_{2}\left( q\right) %
\right] .  \notag
\end{eqnarray}%
And the one-point ones arising in that relations as well%
\begin{eqnarray*}
T_{\mu }^{V}\left( i\right) &=&-2k_{i}^{\nu }\Delta _{2\mu \nu } \\
T_{\mu }^{A}\left( i\right) &=&-2\varepsilon _{\mu \nu _{1}}k_{i}^{\nu
_{2}}\Delta _{2\nu _{2}}^{\nu _{1}}.
\end{eqnarray*}

Now it is possible to state that same if all integrands are the same, the
same is not true for the integrals. In the case of even amplitudes $VV$ and $%
AA$, the expressions depend in a unique way on their divergences once a
method is chosen. In the case of odd amplitudes, the expressions depend on
the version of the trace used. We will show that this dependence manifests
itself in a predictable way in all dimensions in the corresponding
amplitudes.

Where is the reason for such discrepancies? Even if it has been applied to
just identities, deploying the definition of the chiral matrix around the
first or the second vertexes has sampled that the indexes among the finite
and divergent parts are not automatically equal after integration. This
statement becomes clear when we subtract one expression from the other%
\begin{eqnarray*}
\left( T_{\mu _{12}}^{AV}\right) _{1}-\left( T_{\mu _{12}}^{AV}\right) _{2}
&=&-2\left[ \varepsilon _{\mu _{1}\nu }\Delta _{2\mu _{2}}^{\nu
}-\varepsilon _{\mu _{2}\nu }\Delta _{2\mu _{1}}^{\nu }\right] -4\left[
\varepsilon _{\mu _{1}\nu _{1}}\theta _{\mu _{2}}^{\nu }-\varepsilon _{\mu
_{2}\nu _{1}}\theta _{\mu _{1}}^{\nu }\right] \left( m^{2}J_{2}+i/4\pi
\right) \\
&&+4\varepsilon _{\mu _{1}\mu _{2}}m^{2}J_{2}.
\end{eqnarray*}%
For the rearranging of the indexes in the finite part and in the object $%
\Delta _{2\mu \nu }$ we use the Schouten identity\footnote{%
The notation to antisymmetrization of indexes adopted by us is 
\begin{equation*}
A_{[\alpha _{1}\cdots \alpha _{r}}B_{\alpha _{r+1}\cdots \alpha _{s}]}=\frac{%
1}{s!}\sum_{\pi \in S_{s}}\mathrm{sign}(\pi )A_{\alpha _{\pi \left( 1\right)
}\cdots \alpha _{\pi \left( r\right) }}B_{\alpha _{\pi \left( r+1\right)
}\cdots \alpha _{\pi \left( s\right) }}
\end{equation*}%
the normalizing factor is irrelevant to all the identities used. Through the
antisymmetry of the Levi-Civita tensor, follows%
\begin{eqnarray}
\varepsilon _{\mu _{1}\nu }\Delta _{2\mu _{2}}^{\nu }+\varepsilon _{\mu
_{2}\mu _{1}}\Delta _{2\nu }^{\nu }+\varepsilon _{\nu \mu _{2}}\Delta _{2\mu
_{1}}^{\nu } &=&0  \label{SchDiv} \\
\varepsilon _{\mu _{1}\nu }\theta _{\mu _{2}}^{\nu }+\varepsilon _{\mu
_{2}\mu _{1}}\theta _{\nu }^{\nu }+\varepsilon _{\nu \mu _{2}}\theta _{\mu
_{1}}^{\nu } &=&0.  \label{SchTeta}
\end{eqnarray}%
} in two dimensions $\varepsilon _{\lbrack \mu _{1}\nu }\Delta _{2\mu
_{2}]}^{\nu }=0$ and $\varepsilon _{\lbrack \mu _{1}\nu }\theta _{\mu
_{2}]}^{\nu }=0$, the difference between the versions is reduced to 
\begin{equation}
\left( T_{\mu _{12}}^{AV}\right) _{1}-\left( T_{\mu _{12}}^{AV}\right)
_{2}=-\varepsilon _{\mu _{1}\mu _{2}}\left( 2\Delta _{2\alpha }^{\alpha
}+i/\pi \right) .  \label{Uni-2D}
\end{equation}

The linearity can be translated as, if over the true equation $\left( t_{\mu
_{12}}^{AV}\right) _{1}-\left( t_{\mu _{12}}^{AV}\right) _{2}=0$ we apply an
operation to replace the integral, that is defined to be a linear operation,
then one should have the equation above identically vanishing, meantime, for
this to be satisfied we would have a condition about the value of the object 
$\Delta _{2\alpha }^{\alpha },$ determined by the unique relation (\ref%
{Uni-2D}). Before delving into this issue, let us analyze how this condition
is manifested in the RAGFs.

\subsection{Verification and Consequences of the RAGFs}

The RAGFs are given by the procedure illustrated in the section (\ref%
{ModlDef}). The even ones readily comply with their relations determined to
their integrands, which means%
\begin{eqnarray*}
q^{\mu _{1}}T_{\mu _{12}}^{VV} &=&T_{\mu _{2}}^{V}\left( 1\right) -T_{\mu
_{2}}^{V}\left( 2\right) =2q^{\nu _{1}}\Delta _{2\mu _{2}\nu _{1}} \\
q^{\mu _{1}}T_{\mu _{12}}^{AA}+2mT_{\mu _{2}}^{PA} &=&T_{\mu _{2}}^{V}\left(
1\right) -T_{\mu _{2}}^{V}\left( 2\right) =2q^{\nu _{1}}\Delta _{2\mu
_{2}\nu _{1}},
\end{eqnarray*}%
where we have used the following result $T_{\mu }^{V}\left( i\right)
=-2k_{i}^{\nu }\Delta _{2\mu \nu }$ and the presence of the projector to
eliminate some finite parts by $q^{\mu _{2}}\theta _{\mu _{2}}^{\nu }=0$.
Exactly the same results with the RAGF to the index $\mu _{2}$.

On the other hand, the odd amplitudes have a substantially more subtle
behavior. Contracting the first version of $AV$, eq.(\ref{AV1}), in the
vector vertex we get%
\begin{equation*}
q^{\mu _{2}}\left( T_{\mu _{12}}^{AV}\right) _{1}=-2\varepsilon _{\mu
_{1}\nu _{1}}q^{\nu _{2}}\Delta _{2\nu _{2}}^{\nu _{1}}=\left[ T_{\mu
_{1}}^{A}\left( 1\right) -T_{\mu _{1}}^{A}\left( 2\right) \right] ,
\end{equation*}%
where $T_{\mu }^{A}\left( i\right) =-2\varepsilon _{\mu \nu _{1}}k_{i}^{\nu
_{2}}\Delta _{2\nu _{2}}^{\nu _{1}}$. Note this happen without any
restriction.

Quite a different situation occurs when contracting with the axial vertex,
there we have%
\begin{equation*}
q^{\mu _{1}}\left( T_{\mu _{12}}^{AV}\right) _{1}=-2q^{\mu _{1}}\varepsilon
_{\mu _{1}\nu }\Delta _{2\mu _{2}}^{\nu }-4q^{\mu _{1}}\varepsilon _{\mu
_{1}\nu }\theta _{\mu _{2}}^{\nu }\left( m^{2}J_{2}+i/4\pi \right) .
\end{equation*}%
Using the Schouten identity as (\ref{SchDiv}) and (\ref{SchTeta}), and
projector properties, $\theta _{\nu }^{\nu }=1$ and transversality, it is
then obtained%
\begin{equation}
q^{\mu _{1}}\left( T_{\mu _{12}}^{AV}\right) _{1}=\left[ T_{\mu
_{2}}^{A}\left( 1\right) -T_{\mu _{2}}^{A}\left( 2\right) \right] -2mT_{\mu
_{2}}^{PV}+\varepsilon _{\mu _{2}\nu _{1}}q^{\nu _{1}}\left[ 2\Delta _{2\nu
_{2}}^{\nu _{2}}+\left( i/\pi \right) \right] ,  \label{pAV1}
\end{equation}%
where the $PV$ amplitude is given by (\ref{PV}). Note the last term in (\ref%
{pAV1}) spoils the automatic satisfaction of this RAGF.

Following the same reasoning for the expression (\ref{AV2}), we see the
opposite behavior,%
\begin{eqnarray}
q^{\mu _{2}}\left( T_{\mu _{12}}^{AV}\right) _{2} &=&\left[ T_{\mu
_{1}}^{A}\left( 1\right) -T_{\mu _{1}}^{A}\left( 2\right) \right]
+\varepsilon _{\mu _{1}\nu }q^{\nu }\left[ 2\Delta _{2\alpha }^{\alpha
}+\left( i/\pi \right) \right]  \label{pAV2} \\
q^{\mu _{1}}\left( T_{\mu _{12}}^{AV}\right) _{2} &=&\left[ T_{\mu
_{2}}^{A}\left( 1\right) -T_{\mu _{2}}^{A}\left( 2\right) \right] -2mT_{\mu
_{2}}^{PV},
\end{eqnarray}%
namely, the ARAGF coming from the contraction in the axial vertex is
satisfied without restrictions, however, the VRAGF arises conditioned by the
value of the object $\Delta _{2\mu \nu }$.

From the relations ARAGF in (\ref{pAV1}) and VRAGF in (\ref{pAV2}), it is
straightforward to see that for the expression (\ref{AV3}), of the third
version, both vertexes have potential violating terms, because it is given
by a combination of the other two. For that version, none of the RAGFs is
automatically satisfied.

The same happens to $VA$ amplitude. The vertex having the offending term
corresponds to the version in question. For the first version, the relation (%
\ref{p1VA}) is satisfied, while for the second version, the relation (\ref%
{p2VA}) is satisfied, and the possibly violating terms occur in 
\begin{eqnarray*}
q^{\mu _{1}}\left( T_{\mu _{12}}^{VA}\right) _{1} &=&\varepsilon _{\mu
_{2}\nu _{1}}q^{\nu _{1}}\left[ 2\Delta _{2\alpha }^{\alpha }+\left( i/\pi
\right) \right] +T_{\mu _{2}}^{A}\left( 1\right) -T_{\mu _{2}}^{A}\left(
2\right) \\
q^{\mu _{2}}\left( T_{\mu _{12}}^{VA}\right) _{2} &=&\varepsilon _{\mu
_{1}\nu _{1}}q^{\nu _{1}}\left[ 2\Delta _{2\alpha }^{\alpha }+\left( i/\pi
\right) \right] +T_{\mu _{1}}^{A}\left( 1\right) -T_{\mu _{1}}^{A}\left(
2\right) +2mT_{\mu _{1}}^{VP}.
\end{eqnarray*}

From the RAGFs verified above, we can conclude that even amplitudes have
their relations satisfied for any values of the surface term and therefore
do not violate linearity. On the other hand, odd amplitudes require the
condition%
\begin{equation}
\Delta _{2\alpha }^{\alpha }=-\frac{i}{2\pi }.  \label{finite1}
\end{equation}%
The location of this term is in the vertex where the version is defined.
Choosing a finite value for the surface term also requires that the
expression (\ref{Uni-2D}) is null, ensuring that the two versions have the
same content. However, this requirement implies that $\left[ T_{\mu
}^{V}\left( 1\right) -T_{\mu }^{V}\left( 2\right) \right] =-\left( i/2\pi
\right) q_{\mu }$ and for the axial function is the same because the
amplitudes are related through $T_{\mu }^{A}\left( k_{i}\right)
=-\varepsilon _{\mu }^{\text{\quad }\nu }\left[ T_{\nu }^{V}\left(
k_{i}\right) \right] $. The implications are obvious but we need to analyze
the consequences for the IWs.

\subsection{Ward Identities}

When a WI is unavoidably violated, we have an anomaly. With this simple
assertion in mind, we will establish from now on all possible scenarios for
the calculations outlined above and how the satisfaction or not of the RAGF
affects the WIs. The VWI, in turn, asks for the identical vanishing of the
one-point functions, for example, for the vector current conservation 
\begin{equation*}
q^{\mu _{1}}T_{\mu _{12}}^{VV}=\left[ T_{\mu _{2}}^{V}\left( 1\right)
-T_{\mu _{2}}^{V}\left( 2\right) \right] =2q^{\nu _{1}}\Delta _{2\mu _{2}\nu
_{1}}=0.
\end{equation*}%
As for the axial current, we have the partial conservation of the axial
current 
\begin{equation*}
q^{\mu _{1}}T_{\mu _{12}}^{AV}=\left[ T_{\mu _{2}}^{A}\left( 1\right)
-T_{\mu _{2}}^{A}\left( 2\right) \right] -2mT_{\mu _{2}}^{PV}=-2mT_{\mu
_{2}}^{PV}.
\end{equation*}

If this result was straightforward, it would be enough to enforce the
vanishing of $\Delta _{2\mu \nu }=0$. For the logarithmic divergences, it is
possible to see which coefficients are not ambiguous (as will see for linear
divergences as happens in $4D$-$AVV$) and depends only on the external
momentum (\ref{pij}). Nonetheless in this case linearity is violated in one
or the other version and the violating term ends up offending the WIs. The
condition is reflected in the WIs of the double vector function, as well,
see table (\ref{tab2d}) for global results. 
\begin{table}[h]
\caption{Violations for vanishing surface term in each version.}
\label{tab2d}\renewcommand{\baselinestretch}{1.1}{\normalsize \centering{\ }%
\ }$%
\begin{tabular}{|l|l|}
\hline
$q^{\mu _{1}}\left( T_{\mu _{12}}^{AV}\right) _{1}=-2mT_{\mu
_{2}}^{PV}+\left( i/\pi \right) \varepsilon _{\mu _{2}\nu _{1}}q^{\nu _{1}}$
& $q^{\mu _{2}}\left( T_{\mu _{12}}^{AV}\right) _{1}=0$ \\ \hline
$q^{\mu _{1}}\left( T_{\mu _{12}}^{AV}\right) _{2}=-2mT_{\mu _{2}}^{PV}$ & $%
q^{\mu _{2}}\left( T_{\mu _{12}}^{AV}\right) _{2}=\left( i/\pi \right)
\varepsilon _{\mu _{1}\nu _{1}}q^{\nu _{1}}$ \\ \hline
$q^{\mu _{1}}\left( T_{\mu _{12}}^{AV}\right) _{3}=-2mT_{\mu
_{2}}^{PV}+\left( i/2\pi \right) \varepsilon _{\mu _{2}\nu _{1}}q^{\nu _{1}}$
& $q^{\mu _{2}}\left( T_{\mu _{12}}^{AV}\right) _{3}=\left( i/2\pi \right)
\varepsilon _{\mu _{1}\nu _{1}}q^{\nu _{1}}$ \\ \hline
$q^{\mu _{1}}T_{\mu _{12}}^{VV}=0$ & $q^{\mu _{2}}T_{\mu _{12}}^{VV}=0$ \\ 
\hline
$q^{\mu _{1}}T_{\mu _{12}}^{AA}=-2mT_{\mu _{2}}^{PA}$ & $q^{\mu _{2}}T_{\mu
_{12}}^{AA}=2mT_{\mu _{2}}^{AP}$ \\ \hline
\end{tabular}%
$%
\end{table}

But if linearity is preserved and we assume the result (\ref{finite1}) as
valid, the amplitudes are unique but violate all WIs, see the table (\ref%
{tabuniq}), notice we do not need to index the version. Nonetheless, in this
scenario where all possible manipulations led to only one answer, one
consequence over the $VV$, as we saw through $\left[ T_{\mu }^{V}\left(
1\right) -T_{\mu }^{V}\left( 2\right) \right] =-\left( i/2\pi \right) q_{\mu
}$, is that its WIs are violated. 
\begin{table}[h]
\caption{Violations for unique amplitudes}
\label{tabuniq}\renewcommand{\baselinestretch}{1.1}{\normalsize \centering{\ 
}\ }$%
\begin{tabular}{|l|l|}
\hline
$q^{\mu _{1}}T_{\mu _{12}}^{AV}=-2mT_{\mu _{2}}^{PV}+\left( i/2\pi \right)
\varepsilon _{\mu _{2}\nu }q^{\nu }$ & $q^{\mu _{2}}T_{\mu
_{12}}^{AV}=\left( i/2\pi \right) \varepsilon _{\mu _{1}\nu }q^{\nu }$ \\ 
\hline
$q^{\mu _{1}}T_{\mu _{12}}^{VV}=-\left( i/2\pi \right) q_{\mu _{2}}$ & $%
q^{\mu _{2}}T_{\mu _{12}}^{VV}=-\left( i/2\pi \right) q_{\mu _{2}}$ \\ \hline
$q^{\mu _{1}}T_{\mu _{12}}^{AA}=-2mT_{\mu _{2}}^{PA}-\left( i/2\pi \right)
q_{\mu _{2}}$ & $q^{\mu _{2}}T_{\mu _{12}}^{AA}=2mT_{\mu _{2}}^{AP}-\left(
i/2\pi \right) q_{\mu _{2}}$ \\ \hline
\end{tabular}%
$%
\end{table}

The scenario is typical of anomalies where it is inevitable some sort of
violation in the WIs. The singularities of perturbation theory are the main
reason presented for such a state of affairs. However, we will also
establish a finite reason for such observations, through a low-energy
property of a finite function. At the beginning of this section we establish
that if the two WIs to $AV$ are satisfied by hypothesis, therefore we should
have%
\begin{equation*}
\left. q^{\mu _{1}}T_{\mu _{12}}^{AV}\right\vert _{q^{2}=0}=\varepsilon
_{\mu _{2}\nu }q^{\nu }\Omega ^{PV}\left( 0\right) =0,
\end{equation*}%
but if we look at the $PV$ amplitude, eq. (\ref{PV}), it is immediate to
note that it is finite with the form factor given by $\Omega ^{PV}=\left(
i/\pi \right) m^{2}Z_{0}^{\left( -1\right) }\left( q\right) $\footnote{%
\begin{equation*}
m^{2}Z_{0}^{\left( -1\right) }\left( q^{2}\right) =m^{2}\left. \int_{0}^{1}%
\mathrm{d}x\frac{1}{\left[ q^{2}x\left( 1-x\right) -m^{2}\right] }%
\right\vert _{q^{2}=0}=-1
\end{equation*}%
}, and in the point $q^{2}=0$, happens that $Q\left( q^{2}=0\right) =-m^{2}$%
, thereby we have%
\begin{equation}
\Omega ^{PV}(q^{2}=0)=-i/\pi ,
\end{equation}%
that it is the opposite of the result deduced in the equation above, to the
satisfaction of both WIs, hence there is no possibility of satisfaction of
both, even if all the elements involved were finite, as long as they are
connected to the finite $PV$ amplitude. The fact that remains is that if the
VWI is observed, the violation of the AWI is the amount corresponding to the
negative of the $\Omega ^{PV}(0)$, only due to reasons of tensor structure.

An analogous conclusion follows if in the contractions of the general tensor
representing the $AV$ structure, eq.(\ref{AVForm}), we adopt the axial WI as
the hypothesis, that means, in the eq.(\ref{q1ContAV}) we make $%
F_{1}=q^{2}F_{3}-\Omega ^{PV}$, and by substitution in the eq. (\ref%
{q2ContAV}), for the contraction in the vector vertex, one would get 
\begin{equation}
q^{\mu _{2}}T_{\mu _{12}}^{AV}=\varepsilon _{\mu _{1}\nu }q^{\nu }\left[
q^{2}\left( F_{2}+F_{3}\right) -\Omega ^{PV}\right] ,
\end{equation}%
and by the regularity of the form factors, we reach the conclusion%
\begin{equation}
\left. q^{\mu _{2}}T_{\mu _{12}}^{AV}\right\vert _{q^{2}=0}=-\varepsilon
_{\mu _{1}\nu }q^{\nu }\Omega ^{PV}\left( 0\right) =\left( \frac{i}{\pi }%
\right) \varepsilon _{\mu _{1}\nu }p^{\nu },
\end{equation}%
hence it is clear that the kinematical behavior of the $PV$ finite amplitude
is responsible for the series of violations observed in the odd correlators.
In four dimensions, we will show that, even if it is obtained violations in
all WIs, a specific combination is always dictated by the kinematical
properties of finite functions, and such result applies here, as well.

The conclusions drawn are immune to the divergent character of the $AV$ or $%
VA$ amplitude, and its eventual role in answering the question of inevitable
symmetry breaking is, in the limit we are treating, obfuscated by the finite
structure mentioned, notwithstanding it is possible to show, in general,
that if linearity is preserved as condition, we have a connection $\Omega
^{PV}\left( 0\right) =2\Delta _{2\alpha }^{\alpha }$, see the analogous
result fully demonstrated in $4D$ eq. (\ref{ME}).

From now on, we will enter into a more elaborated scenario, and we will show
that in the physical dimension the same conclusions can be drawn for the
case of odd triangles. The presence of the anomaly can be anticipated
through conclusions similar to those drawn for the two-dimensional case.

\section{Four Dimensions Three Point Functions}

\label{4Dim3Pt}In this dimension, the amplitudes that exhibit the claimed
behavior are the rank three triangles, $AVV$ its permutations $VAV$, $VVA$,
and $AAA$. The way they displace the violating terms, their uniqueness
properties, and violation of RAGF equations are only consequences of the
ambiguity of the integrated expression provoked by the traces of six gamma
matrices and an odd number of chiral matrices.

Their computations then, boil down to twenty-four triangles of rank one,
twelve parity-even triangles, $VPP$, $ASP$, and all their permutations, and
twelve parity-odd tensors $ASS$, $APP$, $VPS$, and all its permutations, in
addition to three tensors as in two dimensions that can't be written down as
other amplitudes and depend only on the leading odd trace of six gamma
matrices.

Given the multitude and ambiguities in the path to express all the results,
we will get excessively pedantic and detailed in some steps.

A general three-point function is obtained through the appropriate choice of
vertexes factors $\Gamma _{i}$ in 
\begin{equation}
t^{\Gamma _{1}\Gamma _{2}\Gamma _{3}}=\text{\textrm{tr}}\left[ \Gamma
_{1}S\left( 1\right) \Gamma _{2}S\left( 2\right) \Gamma _{3}S\left( 3\right) %
\right] ,
\end{equation}%
and as usual in this text, the integrated form gets a capital letter%
\begin{equation*}
T^{\Gamma _{1}\Gamma _{2}\Gamma _{3}}=\int \frac{\mathrm{d}^{4}k}{\left(
2\pi \right) ^{4}}t^{\Gamma _{1}\Gamma _{2}\Gamma _{3}},
\end{equation*}%
in the expansion of the terms of mass and momenta, it is possible to see
that in the rank zero and rank two triangles the non-zero terms are the odd
powers of the mass, and to the rank one and rank three the even power of
mass have non-zero traces.

The backbone of amplitudes we are interested in are the odd tensors%
\begin{eqnarray*}
t_{\mu _{123}}^{AVV} &=&\text{\textrm{tr}}\left[ \gamma _{\ast }\gamma _{\mu
_{1}}S\left( 1\right) \gamma _{\mu _{2}}S\left( 2\right) \gamma _{\mu
_{3}}S\left( 3\right) \right] \\
t_{\mu _{123}}^{VAV} &=&\text{\textrm{tr}}\left[ \gamma _{\mu _{1}}S\left(
1\right) \gamma _{\ast }\gamma _{\mu _{2}}S\left( 2\right) \gamma _{\mu
_{3}}S\left( 3\right) \right] \\
t_{\mu _{123}}^{VVA} &=&\text{\textrm{tr}}\left[ \gamma _{\mu _{1}}S\left(
1\right) \gamma _{\mu _{2}}S\left( 2\right) \gamma _{\ast }\gamma _{\mu
_{3}}S\left( 3\right) \right] \\
t_{\mu _{123}}^{AAA} &=&\text{\textrm{tr}}\left[ \gamma _{\ast }\gamma _{\mu
_{1}}S\left( 1\right) \gamma _{\ast }\gamma _{\mu _{2}}S\left( 2\right)
\gamma _{\ast }\gamma _{\mu _{3}}S\left( 3\right) \right] .
\end{eqnarray*}%
Keeping the non-vanishing traces they assume the form%
\begin{eqnarray}
t_{\mu _{123}}^{AVV} &=&K_{123}^{\nu _{123}}\mathrm{tr}(\gamma _{\ast \mu
_{1}\nu _{1}\mu _{2}\nu _{2}\mu _{3}\nu _{3}})\frac{1}{D_{123}}%
+m^{2}\varepsilon _{\mu _{1}\mu _{2}\mu _{3}\nu _{1}}\left( K_{1}^{\nu
_{1}}-K_{2}^{\nu _{1}}+K_{3}^{\nu _{1}}\right) \frac{1}{D_{123}}
\label{AVVexp} \\
t_{\mu _{123}}^{VAV} &=&K_{123}^{\nu _{123}}\mathrm{tr}(\gamma _{\ast \mu
_{1}\nu _{1}\mu _{2}\nu _{2}\mu _{3}\nu _{3}})\frac{1}{D_{123}}%
+m^{2}\varepsilon _{\mu _{1}\mu _{2}\mu _{3}\nu _{1}}\left( K_{1}^{\nu
_{1}}+K_{2}^{\nu _{1}}-K_{3}^{\nu _{1}}\right) \frac{1}{D_{123}}
\label{VAVexp} \\
t_{\mu _{123}}^{VVA} &=&K_{123}^{\nu _{123}}\mathrm{tr}(\gamma _{\ast \mu
_{1}\nu _{1}\mu _{2}\nu _{2}\mu _{3}\nu _{3}})\frac{1}{D_{123}}%
-m^{2}\varepsilon _{\mu _{1}\mu _{2}\mu _{3}\nu _{1}}\left( K_{1}^{\nu
_{1}}-K_{2}^{\nu _{1}}-K_{3}^{\nu _{1}}\right) \frac{1}{D_{123}}
\label{VVAexp} \\
t_{\mu _{123}}^{AAA} &=&K_{123}^{\nu _{123}}\mathrm{tr}(\gamma _{\ast \mu
_{1}\nu _{1}\mu _{2}\nu _{2}\mu _{3}\nu _{3}})\frac{1}{D_{123}}%
-m^{2}\varepsilon _{\mu _{1}\mu _{2}\mu _{3}\nu _{1}}\left( K_{1}^{\nu
_{1}}+K_{2}^{\nu _{1}}+K_{3}^{\nu _{1}}\right) \frac{1}{D_{123}}.
\label{AAAexp}
\end{eqnarray}

Apart from the other traces, the leading trace in all these pre-diagrams
containing six matrices can be computed in lots of ways. We will show after
how the protocol we are presenting is enough to achieve any possible result
of any identity in the appendix (\ref{Tr6G4D}).

Starting with the definition%
\begin{equation}
\gamma _{\ast }=\frac{i}{4!}\varepsilon _{\nu _{1234}}\gamma ^{\nu _{1234}}
\end{equation}%
the formula below gives all results to the trace of six matrices using the
definition anywhere in the string%
\begin{eqnarray}
\left( 4i\right) ^{-1}\mathrm{tr}\left( \gamma _{\ast }\gamma
_{abcdef}\right) &=&+g_{ab}\varepsilon _{cdef}+g_{ad}\varepsilon
_{bcef}+g_{af}\varepsilon _{bcde}  \label{Mold6} \\
&&+g_{bc}\varepsilon _{adef}+g_{cd}\varepsilon _{abef}+g_{cf}\varepsilon
_{abde}  \notag \\
&&+g_{be}\varepsilon _{acdf}+g_{de}\varepsilon _{abcf}+g_{ef}\varepsilon
_{abcd}  \notag \\
&&-g_{bd}\varepsilon _{acef}-g_{df}\varepsilon _{abce}-g_{bf}\varepsilon
_{acde}  \notag \\
&&-g_{ac}\varepsilon _{bdef}-g_{ce}\varepsilon _{abdf}-g_{ae}\varepsilon
_{bcdf}.  \notag
\end{eqnarray}

Although valid the identities%
\begin{equation}
\mathrm{tr}(\gamma _{\ast }\gamma _{\mu _{1}\nu _{1}\mu _{2}\nu _{2}\mu
_{3}\nu _{3}})=\mathrm{tr}(\gamma _{\ast }\gamma _{\mu _{2}\nu _{2}\mu
_{3}\nu _{3}\mu _{1}\nu _{1}})=\mathrm{tr}(\gamma _{\ast }\gamma _{\mu
_{3}\nu _{3}\mu _{1}\nu _{1}\mu _{2}\nu _{2}})
\end{equation}%
by using the eq. (\ref{Mold6}) above and adopting $\left( a,b,c,d,e,f\right)
=\left( \mu _{1},\nu _{1},\mu _{2},\nu _{2},\mu _{3},\nu _{3}\right) $, $%
\left( a,b,c,d,e,f\right) =\left( \mu _{2},\nu _{2},\mu _{3},\nu _{3},\mu
_{1},\nu _{1}\right) $ and $\left( a,b,c,d,e,f\right) =\left( \mu _{3},\nu
_{3},\mu _{1},\nu _{1},\mu _{2},\nu _{2}\right) $ the resulting expressions
differ only in a couple of signs leading, when integrated, to three not
automatically equivalent expressions to the odd rank-three amplitudes.

To these results for the traces and after contracting with factor $%
K_{123}^{\nu _{123}}$ we display in the equations below, because the place
we deploy the definition of the chiral matrix is related to the position of
the vertex we adopt a label of \ the version to these expressions, namely

\textbf{First Version}%
\begin{eqnarray}
K_{123}^{\nu _{123}}\mathrm{tr}\left( \gamma _{\ast }\gamma _{\mu _{1}\nu
_{1}\mu _{2}\nu _{2}\mu _{3}\nu _{3}}\right) &=&-\varepsilon _{\mu _{2}\mu
_{3}\nu _{1}\nu _{2}}\left[ K_{1\mu _{1}}K_{23}^{\nu _{12}}-K_{2\mu
_{1}}K_{13}^{\nu _{12}}+K_{3\mu _{1}}K_{12}^{\nu _{12}}\right]  \label{tr1}
\\
&&-\varepsilon _{\mu _{1}\mu _{3}\nu _{1}\nu _{2}}\left[ K_{1\mu
_{2}}K_{23}^{\nu _{12}}+K_{2\mu _{2}}K_{13}^{\nu _{12}}-K_{3\mu
_{2}}K_{12}^{\nu _{12}}\right]  \notag \\
&&+\varepsilon _{\mu _{1}\mu _{2}\nu _{1}\nu _{2}}\left[ K_{1\mu
_{3}}K_{23}^{\nu _{12}}-K_{2\mu _{3}}K_{13}^{\nu _{12}}-K_{3\mu
_{3}}K_{12}^{\nu _{12}}\right]  \notag \\
&&+\varepsilon _{\mu _{1}\mu _{2}\mu _{3}\nu _{1}}\left[ -K_{1}^{\nu
_{1}}\left( K_{2}\cdot K_{3}\right) +K_{2}^{\nu _{1}}\left( K_{1}\cdot
K_{3}\right) -K_{3}^{\nu _{1}}\left( K_{1}\cdot K_{2}\right) \right]  \notag
\\
&&+\left[ -g_{\mu _{1}\mu _{2}}\varepsilon _{\mu _{3}\nu _{1}\nu _{2}\nu
_{3}}-g_{\mu _{2}\mu _{3}}\varepsilon _{\mu _{1}\nu _{1}\nu _{2}\nu
_{3}}+g_{\mu _{1}\mu _{3}}\varepsilon _{\mu _{2}\nu _{1}\nu _{2}\nu _{3}}%
\right] K_{123}^{\nu _{123}}  \notag
\end{eqnarray}

\textbf{Second Version}%
\begin{eqnarray}
K_{123}^{\nu _{123}}\mathrm{tr}\left( \gamma _{\ast }\gamma _{\mu _{2}\nu
_{2}\mu _{3}\nu _{3}\mu _{1}\nu _{1}}\right) &=&+\varepsilon _{\mu _{1}\mu
_{3}\nu _{1}\nu _{2}}\left[ K_{1\mu _{2}}K_{23}^{\nu _{12}}-K_{2\mu
_{2}}K_{13}^{\nu _{12}}+K_{3\mu _{2}}K_{12}^{\nu _{12}}\right]  \label{tr2}
\\
&&-\varepsilon _{\mu _{1}\mu _{2}\nu _{1}\nu _{2}}\left[ K_{1\mu
_{3}}K_{23}^{\nu _{23}}+K_{2\mu _{3}}K_{13}^{\nu _{13}}+K_{3\mu
_{3}}K_{12}^{\nu _{12}}\right]  \notag \\
&&-\varepsilon _{\mu _{2}\mu _{3}\nu _{1}\nu _{2}}\left[ K_{1\mu
_{1}}K_{23}^{\nu _{23}}+K_{2\mu _{1}}K_{13}^{\nu _{13}}-K_{3\mu
_{1}}K_{12}^{\nu _{12}}\right]  \notag \\
&&+\varepsilon _{\mu _{1}\mu _{2}\mu _{3}\nu _{1}}\left[ -K_{1}^{\nu
_{1}}\left( K_{2}\cdot K_{3}\right) -K_{2}^{\nu _{1}}\left( K_{1}\cdot
K_{3}\right) +K_{3}^{\nu _{1}}\left( K_{1}\cdot K_{2}\right) \right]  \notag
\\
&&+\left[ g_{\mu _{1}\mu _{2}}\varepsilon _{\mu _{3}\nu _{1}\nu _{2}\nu
_{3}}-g_{\mu _{1}\mu _{3}}\varepsilon _{\mu _{2}\nu _{1}\nu _{2}\nu
_{3}}-g_{\mu _{2}\mu _{3}}\varepsilon _{\mu _{1}\nu _{1}\nu _{2}\nu _{3}}%
\right] K_{123}^{\nu _{123}}  \notag
\end{eqnarray}

\textbf{The Third Version}%
\begin{eqnarray}
K_{123}^{\nu _{123}}\mathrm{tr}\left( \gamma _{\ast }\gamma _{\mu _{3}\nu
_{3}\mu _{1}\nu _{1}\mu _{2}\nu _{2}}\right) &=&-\varepsilon _{\mu _{1}\mu
_{2}\nu _{1}\nu _{2}}\left[ K_{1\mu _{3}}K_{23}^{\nu _{12}}-K_{2\mu
_{3}}K_{13}^{\nu _{12}}+K_{3\mu _{3}}K_{12}^{\nu _{12}}\right]  \label{tr3}
\\
&&-\varepsilon _{\mu _{2}\mu _{3}\nu _{1}\nu _{2}}\left[ K_{1\mu
_{1}}K_{23}^{\nu _{12}}-K_{2\mu _{1}}K_{13}^{\nu _{12}}-K_{3\mu
_{1}}K_{12}^{\nu _{12}}\right]  \notag \\
&&-\varepsilon _{\mu _{1}\mu _{3}\nu _{1}\nu _{2}}\left[ K_{1\mu
_{2}}K_{23}^{\nu _{12}}+K_{2\mu _{2}}K_{13}^{\nu _{12}}+K_{3\mu
_{2}}K_{12}^{\nu _{12}}\right]  \notag \\
&&+\varepsilon _{\mu _{1}\mu _{2}\mu _{3}\nu _{1}}\left[ K_{1}^{\nu
_{1}}\left( K_{2}\cdot K_{3}\right) -K_{2}^{\nu _{1}}\left( K_{1}\cdot
K_{3}\right) -K_{3}^{\nu _{1}}\left( K_{1}\cdot K_{2}\right) \right]  \notag
\\
&&+\left[ -g_{\mu _{1}\mu _{2}}\varepsilon _{\mu _{3}\nu _{1}\nu _{2}\nu
_{3}}-g_{\mu _{1}\mu _{3}}\varepsilon _{\mu _{2}\nu _{1}\nu _{2}\nu
_{3}}+g_{\mu _{2}\mu _{3}}\varepsilon _{\mu _{1}\nu _{1}\nu _{2}\nu _{3}}%
\right] K_{123}^{\nu _{123}}  \notag
\end{eqnarray}

Preceding the discussion of any particular third rank amplitude, we shall
introduce here some definitions and general results which will be useful in
what follows. If one looks at the three first rows of (\ref{tr1}), (\ref{tr2}%
) and (\ref{tr3}), they all can be written in terms of tensors, which we
call sign tensors (equivalently to 2D..)%
\begin{equation}
\varepsilon _{\mu _{ab}\nu _{12}}t_{\mu _{c}}^{\nu _{12}\left(
s_{1}s_{2}\right) }=\varepsilon _{\mu _{ab}\nu _{12}}\left( K_{1\mu
_{c}}K_{23}^{\nu _{12}}+s_{1}K_{2\mu _{c}}K_{13}^{\nu _{12}}+s_{2}K_{3\mu
_{c}}K_{12}^{\nu _{12}}\right) \frac{1}{D_{123}}
\end{equation}%
where $s_{i}=\pm 1$, thus using $K_{i}=K_{j}+p_{ij}$ and $\varepsilon _{\mu
_{ab}\nu _{12}}K_{ij}^{\nu _{12}}=\varepsilon _{\mu _{ab}\nu
_{12}}p_{ji}^{\nu _{2}}K_{i}^{\nu _{1}}$ and writing $p_{32}=p_{31}-p_{21}$
when necessary, we have 
\begin{eqnarray}
\varepsilon _{\mu _{ab}\nu _{12}}t_{\mu _{c}}^{\nu _{12}\left(
s_{1}s_{2}\right) } &=&\varepsilon _{\mu _{ab}\nu _{12}}\left[ \left(
1+s_{1}\right) p_{31}^{\nu _{2}}-\left( 1-s_{2}\right) p_{21}^{\nu _{2}}%
\right] \frac{K_{1}^{\nu _{1}}K_{1\mu _{c}}}{D_{123}}  \notag \\
&&+\varepsilon _{\mu _{ab}\nu _{12}}\left[ p_{21}^{\nu _{1}}p_{32}^{\nu
_{2}}K_{1\mu _{c}}+\left( s_{1}p_{21\mu _{c}}p_{31}^{\nu _{2}}+s_{2}p_{31\mu
_{c}}p_{21}^{\nu _{2}}\right) K_{1}^{\nu _{1}}\right] \frac{1}{D_{123}}
\end{eqnarray}%
and will be clear that the integral of one of these tensors is finite, namely%
\begin{equation}
\varepsilon _{\mu _{ab}\nu _{12}}t_{\mu _{c}}^{\nu _{12}\left( -,+\right)
}=\varepsilon _{\mu _{ab}\nu _{12}}\left[ p_{21}^{\nu _{1}}p_{32}^{\nu
_{2}}K_{1\mu _{c}}+\left( -p_{21\mu _{c}}p_{31}^{\nu _{2}}+p_{31\mu
_{c}}p_{21}^{\nu _{2}}\right) K_{1}^{\nu _{1}}\right] \frac{1}{D_{123}},
\end{equation}%
because the presence of only one factor of the vector $K_{i}$ or in other
words only vector three-point integrals of power counting $\omega =-1$
appear here. After integration, and with the help of the result 
\begin{equation}
J_{3}^{\mu }=-p_{21}^{\mu }Z_{10}^{\left( -1\right) }-p_{31}^{\mu
}Z_{01}^{\left( -1\right) },  \label{4DJ3mu}
\end{equation}%
and the anti-symmetry of $\varepsilon $-tensor, we get the first relevant
result%
\begin{equation}
\varepsilon _{\mu _{ab}\nu _{12}}T_{\mu _{c}}^{\nu _{12}\left( -+\right)
}=\varepsilon _{\mu _{ab}\nu _{12}}\left[ p_{21}^{\nu _{1}}p_{32}^{\nu
_{2}}J_{3\mu _{c}}+\left( -p_{21\mu _{c}}p_{31}^{\nu _{2}}+p_{31\mu
_{c}}p_{21}^{\nu _{2}}\right) J_{3}^{\nu _{1}}\right] =0.  \label{T-+}
\end{equation}

The other three non-zero ones, containing the log-diverging integral $\bar{J}%
_{3\mu _{c}}^{\nu _{1}}$ responsible for the divergent content, we present
below%
\begin{eqnarray}
\varepsilon _{\mu _{ab}\nu _{12}}T_{\mu _{c}}^{\nu _{12}\left( +-\right) }
&=&2\varepsilon _{\mu _{ab}\nu _{12}}\left[ p_{31}^{\nu _{2}}\left( J_{3\mu
_{c}}^{\nu _{1}}+p_{21\mu _{c}}J_{3}^{\nu _{1}}\right) -p_{21}^{\nu
_{2}}\left( J_{3\mu _{c}}^{\nu _{1}}+p_{31\mu _{c}}J_{3}^{\nu _{1}}\right) %
\right]  \label{T+-} \\
&&+\frac{1}{2}\left( \varepsilon _{\mu _{ab}\nu _{12}}p_{32}^{\nu
_{2}}\Delta _{3\mu _{c}}^{\nu _{1}}+\varepsilon _{\mu _{abc}\nu
_{1}}p_{32}^{\nu _{1}}I_{\log }\right)  \notag
\end{eqnarray}%
\begin{equation}
\varepsilon _{\mu _{ab}\nu _{12}}T_{\mu _{c}}^{\nu _{12}\left( --\right)
}=-2\varepsilon _{\mu _{ab}\nu _{12}}p_{21}^{\nu _{2}}\left( J_{3\mu
_{c}}^{\nu _{1}}+p_{31\mu _{c}}J_{3}^{\nu _{1}}\right) -\frac{1}{2}\left(
\varepsilon _{\mu _{ab}\nu _{12}}p_{21}^{\nu _{2}}\Delta _{3\mu _{c}}^{\nu
_{1}}+\varepsilon _{\mu _{abc}\nu _{1}}p_{21}^{\nu _{1}}I_{\log }\right)
\label{T--}
\end{equation}%
\begin{equation}
\varepsilon _{\mu _{ab}\nu _{12}}T_{\mu _{c}}^{\nu _{12}\left( ++\right)
}=+2\varepsilon _{\mu _{ab}\nu _{12}}p_{31}^{\nu _{2}}\left( J_{3\mu
_{c}}^{\nu _{1}}+p_{21\mu _{c}}J_{3}^{\nu _{1}}\right) +\frac{1}{2}\left(
\varepsilon _{\mu _{ab}\nu _{12}}p_{31}^{\nu _{2}}\Delta _{3\mu _{c}}^{\nu
_{1}}+\varepsilon _{\mu _{abc}\nu _{1}}p_{31}^{\nu _{1}}I_{\log }\right)
\label{T++}
\end{equation}

Hence the three versions will turn up each with a tensor that is shared with
all rank-three pseudo triangles using that version of the trace we define
them as%
\begin{eqnarray}
c_{1\mu _{123}} &=&-\varepsilon _{\mu _{23}\nu _{12}}t_{\mu _{1}}^{\nu
_{12}\left( -+\right) }-\varepsilon _{\mu _{13}\nu _{12}}t_{\mu _{2}}^{\nu
_{12}\left( +-\right) }+\varepsilon _{\mu _{12}\nu _{12}}t_{\mu _{3}}^{\nu
_{12}\left( --\right) } \\
c_{2\mu _{123}} &=&+\varepsilon _{\mu _{13}\nu _{12}}t_{\mu _{2}}^{\nu
_{12}\left( -+\right) }-\varepsilon _{\mu _{12}\nu _{12}}t_{\mu _{3}}^{\nu
_{12}\left( ++\right) }-\varepsilon _{\mu _{23}\nu _{12}}t_{\mu _{1}}^{\nu
_{12}\left( +-\right) } \\
c_{3\mu _{123}} &=&-\varepsilon _{\mu _{12}\nu _{12}}t_{\mu _{3}}^{\nu
_{12}\left( -+\right) }-\varepsilon _{\mu _{23}\nu _{12}}t_{\mu _{1}}^{\nu
_{12}\left( --\right) }-\varepsilon _{\mu _{13}\nu _{12}}t_{\mu _{2}}^{\nu
_{12}\left( ++\right) }
\end{eqnarray}%
whose integrals in terms of the previous results eq.(\ref{T-+}) (\ref{T+-}) (%
\ref{T--}) (\ref{T++}) are%
\begin{eqnarray}
C_{1\mu _{123}} &=&-\varepsilon _{\mu _{13}\nu _{12}}T_{\mu _{2}}^{\nu
_{12}\left( +-\right) }+\varepsilon _{\mu _{12}\nu _{12}}T_{\mu _{3}}^{\nu
_{12}\left( --\right) }  \label{C1} \\
C_{2\mu _{123}} &=&-\varepsilon _{\mu _{12}\nu _{12}}T_{\mu _{3}}^{\nu
_{12}\left( ++\right) }-\varepsilon _{\mu _{23}\nu _{12}}T_{\mu _{1}}^{\nu
_{12}\left( +-\right) } \\
C_{3\mu _{123}} &=&-\varepsilon _{\mu _{23}\nu _{12}}T_{\mu _{1}}^{\nu
_{12}\left( --\right) }-\varepsilon _{\mu _{13}\nu _{12}}T_{\mu _{2}}^{\nu
_{12}\left( ++\right) }
\end{eqnarray}

Here another important point to notice is that the sampling of indexes
mentioned reflects in the absence of the index $\mu _{i}$ of the vertex $%
\Gamma _{i}$ in the $T^{\prime }$s of the $C_{i}$ because it is present
through the tensor $\varepsilon _{\mu _{ab}\nu _{12}}T_{\mu _{c}}^{\nu
_{12}\left( -+\right) }$ in the eq.(\ref{T-+}) that is finite and
identically zero. This will enable us to anticipate violations of either WIs
or RAGFs.

The next step is the vanishing of rank-one odd triangles, noticed that in
the last rows of eq.'s (\ref{tr1}), (\ref{tr2}) and (\ref{tr3}) they appear
as components. It happens they all differ only by a sign and turn out to be
finite and null, as an example%
\begin{equation}
t_{\mu _{i}}^{ASS}=4i\varepsilon _{\mu _{i}\nu _{123}}K_{123}^{\nu _{123}}%
\frac{1}{D_{123}}=4i\varepsilon _{\mu _{i}\nu _{123}}p_{21}^{\nu
_{2}}p_{31}^{\nu _{3}}\frac{K_{1}^{\nu _{1}}}{D_{123}}
\end{equation}%
due to the triple contraction, therefore using (\ref{4DJ3mu})%
\begin{equation}
T_{\mu _{i}}^{ASS}=4i\varepsilon _{\mu _{i}\nu _{123}}p_{21}^{\nu
_{2}}p_{31}^{\nu _{3}}J_{3}^{\nu _{1}}=0  \label{ASS}
\end{equation}%
and the same for%
\begin{eqnarray}
T_{\mu _{i}}^{ASS} &=&T_{\mu _{i}}^{SAS}=T_{\mu _{i}}^{SSA}=T_{\mu
_{i}}^{APP}=T_{\mu _{i}}^{PAP}=T_{\mu _{i}}^{PPA}=0 \\
T_{\mu _{i}}^{VSP} &=&T_{\mu _{i}}^{PVS}=T_{\mu _{i}}^{SPV}=T_{\mu
_{i}}^{VPS}=T_{\mu _{i}}^{SVP}=T_{\mu _{i}}^{PSV}=0
\end{eqnarray}%
disappearing then from the integrated expressions.

The last feature on this web of relations is the distinguishing rank-one
even triangles. Here we need a definite example, as the $VPP$ which is
defined by%
\begin{equation}
\left( t^{VPP}\right) ^{\nu _{1}}=\text{\textrm{tr}}\left[ \gamma ^{\nu
_{1}}S\left( 1\right) \gamma _{\ast }S\left( 2\right) \gamma _{\ast }S\left(
3\right) \right] =4\left( -K_{1}^{\nu _{1}}S_{23}+K_{2}^{\nu
_{1}}S_{13}-K_{3}^{\nu _{1}}S_{12}\right) \frac{1}{D_{123}},
\end{equation}%
as always, before the integration during the trace operation, remember the
combination $S_{ij}=K_{i}\cdot K_{j}-m^{2}$ was defined on eq. (\ref{Sij}).
Now is a direct task to see that combining the mass terms of the eq. (\ref%
{AVVexp}) to the $AVV$ integrand with the eq. (\ref{tr1}) to the first
version of the trace to get as a component sub-amplitude%
\begin{equation}
\mathrm{sub}\left( t_{\mu _{123}}^{AVV}\right) _{1}=4i\varepsilon _{\mu
_{1}\mu _{2}\mu _{3}\nu _{1}}\left( -K_{1}^{\nu _{1}}S_{23}+K_{2}^{\nu
_{1}}S_{13}-K_{3}^{\nu _{1}}S_{12}\right) \frac{1}{D_{123}}=i\varepsilon
_{\mu _{1}\mu _{2}\mu _{3}\nu _{1}}\left( t^{VPP}\right) ^{\nu _{1}}
\end{equation}%
where the pre-amplitude got the trace-version label. In the same vein then,
it is easy to verify the table (\ref{tabversions}) accounts for all other
possible combinations appearing in the explicit computation.

\begin{table}[tbph]
\caption{Even sub-amplitudes corresponding to the vertex configuration of
the rank-three odd ones and the version of the trace applied.}
\label{tabversions}\renewcommand{\baselinestretch}{1.4}{\normalsize %
\centering{\ }}$%
\begin{tabular}{|c|c|c|c|c|}
\hline
$\text{Version/Type}$ & $AVV$ & $VAV$ & $VVA$ & $AAA$ \\ \hline
1 & $+VPP$ & $+ASP$ & $-APS$ & $-VSS$ \\ \hline
2 & $-SAP$ & $+PVP$ & $+PAS$ & $-SVS$ \\ \hline
3 & $+SPA$ & $-PSA$ & $+PPV$ & $-SSV$ \\ \hline
\end{tabular}%
${\normalsize \ }
\end{table}

In other words, alongside the $C_{i}$ tensor of each of the versions to each
type of third-rank amplitude, there is one even and rank-one sub-amplitude.
Back to the $VPP$ example, its integrated form becomes%
\begin{eqnarray}
\left( T^{VPP}\right) ^{\nu _{1}} &=&-4\left( p_{21}\cdot p_{32}\right)
J_{3}^{\nu _{1}}  \label{TVPP} \\
&&+2\left[ \left( p_{31}^{\nu _{1}}p_{21}^{2}-p_{21}^{\nu
_{1}}p_{31}^{2}\right) J_{3}+p_{21}^{\nu _{1}}J_{2}\left( p_{21}\right)
-p_{32}^{\nu _{1}}J_{2}\left( p_{32}\right) \right]  \notag \\
&&+2\left[ P_{31}^{\nu _{2}}\Delta _{\nu _{2}}^{\nu _{1}}+\left( p_{21}^{\nu
_{1}}-p_{32}^{\nu _{1}}\right) I_{\log }\right]  \notag
\end{eqnarray}%
all the others are contemplated in the useful appendix (\ref{AppSub}), which
is essential to verify all the expressions presented in the paper.

For completeness, in the first version of $AVV$, we note that the divergent
part of the tensor $C_{1\mu _{1}\mu _{2}\mu _{3}}$, defined in eq. (\ref{C1}%
), can be read off from eq.'s (\ref{T+-}) and (\ref{T--}), for $\varepsilon
_{\mu _{13}\nu _{12}}T_{\mu _{2}}^{\nu _{12}\left( +-\right) }$ and $%
\varepsilon _{\mu _{12}\nu _{12}}T_{\mu _{3}}^{\nu _{12}\left( --\right) }$,
as 
\begin{equation*}
4iC_{1\mu _{1}\mu _{2}\mu _{3}}=-2i\left( \varepsilon _{\mu _{13}\nu
_{12}}p_{32}^{\nu _{2}}\Delta _{3\mu _{2}}^{\nu _{1}}+\varepsilon _{\mu
_{12}\nu _{12}}p_{21}^{\nu _{2}}\Delta _{3\mu _{3}}^{\nu _{1}}+\varepsilon
_{\mu _{123}\nu _{1}}\left( p_{21}^{\nu _{1}}-p_{32}^{\nu _{1}}\right)
I_{\log }\right) ,
\end{equation*}%
in such a way when combined with the divergent part of the sub-amplitude $%
VPP $ in the eq. (\ref{TVPP}) above, it exactly cancels the object $I_{\log
} $. Then the total divergent content will be expressed exactly and solely
in terms of surface terms%
\begin{equation}
S_{1\mu _{1}\mu _{2}\mu _{3}}=-2i\left( \varepsilon _{\mu _{13}\nu
_{12}}p_{32}^{\nu _{2}}\Delta _{3\mu _{2}}^{\nu _{1}}+\varepsilon _{\mu
_{12}\nu _{12}}p_{21}^{\nu _{2}}\Delta _{3\mu _{3}}^{\nu _{1}}\right)
+2i\varepsilon _{\mu _{123}\nu _{1}}P_{31}^{\nu _{2}}\Delta _{3\nu
_{2}}^{\nu _{1}}  \label{ST1}
\end{equation}%
hence, together with finite parts of the same eq.'s (\ref{T+-}) and (\ref%
{T--}), the complete result assumes the form 
\begin{eqnarray}
\left( T_{\mu _{123}}^{AVV}\right) _{1} &=&S_{1\mu _{1}\mu _{2}\mu _{3}}
\label{AVV1complete} \\
&&-8i\varepsilon _{\mu _{13}\nu _{12}}\left[ p_{31}^{\nu _{2}}\left( J_{3\mu
_{2}}^{\nu _{1}}+p_{21\mu _{2}}J_{3}^{\nu _{1}}\right) -p_{21}^{\nu
_{2}}\left( J_{3\mu _{2}}^{\nu _{1}}+p_{31\mu _{2}}J_{3}^{\nu _{1}}\right) %
\right]  \notag \\
&&-8i\varepsilon _{\mu _{12}\nu _{12}}p_{21}^{\nu _{2}}\left( J_{3\mu
_{3}}^{\nu _{1}}+p_{31\mu _{3}}J_{3}^{\nu _{1}}\right)  \notag \\
&&-4i\varepsilon _{\mu _{123}\nu _{1}}\left( p_{21}\cdot p_{32}\right)
J_{3}^{\nu _{1}}+2i\varepsilon _{\mu _{123}\nu _{1}}\left[ \left(
p_{31}^{\nu _{1}}p_{21}^{2}-p_{21}^{\nu _{1}}p_{31}^{2}\right) J_{3}\right]
\\
&&+2i\varepsilon _{\mu _{123}\nu _{1}}\left[ p_{21}^{\nu _{1}}J_{2}\left(
p_{21}\right) -p_{32}^{\nu _{1}}J_{2}\left( p_{32}\right) \right] .  \notag
\end{eqnarray}

Some statements are in order here, the exact cancellation among the
divergent objects $I_{\log }$ of the common tensors and sub-amplitudes,
separated from the finite integrals using the identity (\ref{id}), happens
for all versions and all amplitudes ($AVV$, $VAV$, $VVA$, $AAA$). And the
surface terms content, $S_{1\mu _{1}\mu _{2}\mu _{3}}$, of the combined
result is the same for all amplitudes, and only depends on the version in
question, this being one of the reasons we did not attach a super index to
these structures.

For later use, we must define the other two sets of surface terms appearing
in versions two and three of any amplitude%
\begin{eqnarray}
S_{2\mu _{123}} &=&-2i\left( \varepsilon _{\mu _{12}\nu _{12}}p_{31}^{\nu
_{2}}\Delta _{3\mu _{3}}^{\nu _{1}}+\varepsilon _{\mu _{23}\nu
_{12}}p_{32}^{\nu _{2}}\Delta _{3\mu _{1}}^{\nu _{1}}\right) +2i\varepsilon
_{\mu _{123}\nu _{1}}P_{21}^{\nu _{2}}\Delta _{3\nu _{2}}^{\nu _{1}}
\label{ST2} \\
S_{3\mu _{123}} &=&-2i\left( \varepsilon _{\mu _{13}\nu _{12}}p_{31}^{\nu
_{2}}\Delta _{3\mu _{2}}^{\nu _{1}}-\varepsilon _{\mu _{23}\nu
_{12}}p_{21}^{\nu _{2}}\Delta _{3\mu _{1}}^{\nu _{1}}\right) +2i\varepsilon
_{\mu _{123}\nu _{1}}P_{32}^{\nu _{2}}\Delta _{3\nu _{2}}^{\nu _{1}}
\label{ST3}
\end{eqnarray}

In summary, following strictly the same steps we have presented above, the
versions of $AVV$ are expressed as%
\begin{eqnarray}
\left( T_{\mu _{123}}^{AVV}\right) _{1} &=&4iC_{1\mu _{123}}+i\varepsilon
_{\mu _{123}\nu _{1}}\left( T^{VPP}\right) ^{\nu _{1}} \\
\left( T_{\mu _{123}}^{AVV}\right) _{2} &=&4iC_{2\mu _{123}}-i\varepsilon
_{\mu _{123}\nu _{1}}\left( T^{SAP}\right) ^{\nu _{1}} \\
\left( T_{\mu _{123}}^{AVV}\right) _{3} &=&4iC_{3\mu _{123}}+i\varepsilon
_{\mu _{123}\nu _{1}}\left( T^{SPA}\right) ^{\nu _{1}}
\end{eqnarray}%
as for $VAV$, $VVA$ and $AAA$ they have the same structure 
\begin{equation*}
\left( T_{\mu _{123}}^{\Gamma _{1}\Gamma _{2}\Gamma _{3}}\right)
_{i}=4iC_{i\mu _{123}}\pm i\varepsilon _{\mu _{123}\nu _{1}}\left( \text{%
Corresponding sub-amplitude}\right) ^{\nu _{1}}
\end{equation*}%
where the sub-amplitude is given by the table (\ref{tabversions}) above.

The set of surface terms $S_{i\mu _{1}\mu _{2}\mu _{3}}$ defined in eq.'s (%
\ref{ST1}), (\ref{ST2}) and (\ref{ST3}), pointing out again, are functions
only of the version and not of the vertex content of the diagram. Their full
verification is easily accomplished using the appendix (\ref{AppSub}) for
the sub-amplitudes, where both divergent and finite parts are expressed
explicitly. This is important because the mechanism of violation of RAGFs is
thoroughly dependent on the topology and the traces of six matrices in the
diagram.

The last element, that appear when investigating RAGFs, are the three-point
and rank-two amplitudes, they are finite after integrated and given by%
\begin{eqnarray}
-2mT_{\mu _{23}}^{PVV} &=&\varepsilon _{\mu _{23}\nu _{12}}p_{21}^{\nu
_{1}}p_{31}^{\nu _{2}}\left( 8im^{2}J_{3}\right) \\
2mT_{\mu _{13}}^{VPV} &=&\varepsilon _{\mu _{13}\nu _{12}}p_{21}^{\nu
_{1}}p_{31}^{\nu _{2}}\left( 8im^{2}J_{3}\right) \\
2mT_{\mu _{12}}^{VVP} &=&\varepsilon _{\mu _{12}\nu _{12}}p_{21}^{\nu
_{1}}p_{32}^{\nu _{2}}\left( -8im^{2}J_{3}\right) ,
\end{eqnarray}%
the signs and the mass factor are only a matter of convenience for they
appear in this form in the contractions with external momenta in the single
axial triangles. To the triple axial triangle, in its contractions with the
momenta, will appear the finite triangles below%
\begin{eqnarray}
-2mT_{\mu _{23}}^{PAA} &=&\varepsilon _{\mu _{23}\nu _{12}}p_{31}^{\nu _{2}} 
\left[ 8im^{2}\left( 2J_{3}^{\nu _{1}}+p_{21}^{\nu _{1}}J_{3}\right) \right]
\\
2mT_{\mu _{13}}^{APA} &=&\varepsilon _{\mu _{13}\nu _{12}}p_{21}^{\nu _{2}} 
\left[ -8im^{2}\left( 2J_{3}^{\nu _{1}}+p_{31}^{\nu _{1}}J_{3}\right) \right]
\\
2mT_{\mu _{12}}^{AAP} &=&\varepsilon _{\mu _{12}\nu _{12}}p_{32}^{\nu _{2}} 
\left[ 8im^{2}\left( 2J_{3}^{\nu _{1}}+p_{21}^{\nu _{1}}J_{3}\right) \right]
,
\end{eqnarray}%
the form factors of all these tensors can be obtaining retrieving the
definitions of the scalar and vector 3pt-integrals%
\begin{equation*}
J_{3}=\frac{i}{\left( 4\pi \right) ^{2}}Z_{00}^{\left( -1\right) };\quad
J_{3}^{\nu }=-\frac{i}{\left( 4\pi \right) ^{2}}\left[ p_{21}^{\nu
}Z_{10}^{\left( -1\right) }+p_{31}^{\nu }Z_{01}^{\left( -1\right) }\right]
\end{equation*}%
and are being explicited for future use, resorting to their value in the
point $q_{i}\cdot q_{j}=0$ 
\begin{equation*}
\left. Z_{rs}^{\left( -1\right) }\right\vert _{0}=-\frac{r!s!}{m^{2}\left[
\left( r+s+2\right) !\right] },
\end{equation*}%
where these momenta represent all the possible difference of routings, we
can display a precise low energy behavior to these tensors 
\begin{equation}
\left. -2mT_{\mu _{23}}^{PVV}\right\vert _{0}=\frac{\varepsilon _{\mu
_{23}\nu _{12}}p_{21}^{\nu _{1}}p_{32}^{\nu _{2}}}{\left( 2\pi \right) ^{2}}%
;\quad \left. 2mT_{\mu _{13}}^{VPV}\right\vert _{0}=\frac{\varepsilon _{\mu
_{13}\nu _{12}}p_{21}^{\nu _{1}}p_{32}^{\nu _{2}}}{\left( 2\pi \right) ^{2}}%
;\quad \left. 2mT_{\mu _{12}}^{VVP}\right\vert _{0}=\frac{\varepsilon _{\mu
_{12}\nu _{12}}p_{21}^{\nu _{1}}p_{32}^{\nu _{2}}}{\left( 2\pi \right) ^{2}}
\label{LEPVV}
\end{equation}%
and 
\begin{equation}
\left. -2mT_{\mu _{23}}^{PAA}\right\vert _{0}=\frac{\varepsilon _{\mu
_{23}\nu _{12}}p_{21}^{\nu _{1}}p_{32}^{\nu _{2}}}{3\left( 2\pi \right) ^{2}}%
;\quad \left. 2mT_{\mu _{13}}^{APA}\right\vert _{0}=\frac{\varepsilon _{\mu
_{13}\nu _{12}}p_{21}^{\nu _{1}}p_{32}^{\nu _{2}}}{3\left( 2\pi \right) ^{2}}%
;\quad \left. 2mT_{\mu _{12}}^{AAP}\right\vert _{0}=\frac{\varepsilon _{\mu
_{12}\nu _{12}}p_{21}^{\nu _{1}}p_{32}^{\nu _{2}}}{3\left( 2\pi \right) ^{2}}
\label{LEPAA}
\end{equation}%
the consequences of these values will be determined and examined in section (%
\ref{LE4D}).

\subsection{Relations Among Green Functions and Uniqueness}

\label{unique}Recalling the chapter two the standard procedure to obtain the
RAGFs, that allows to state to $t_{\mu _{123}}^{AVV}$ 
\begin{eqnarray*}
p_{31}^{\mu _{1}}t_{\mu _{123}}^{AVV} &=&t_{\mu _{32}}^{AV}\left( 1,2\right)
-t_{\mu _{23}}^{AV}\left( 2,3\right) -2mt_{\mu _{23}}^{PVV} \\
p_{21}^{\mu _{2}}t_{\mu _{123}}^{AVV} &=&t_{\mu _{13}}^{AV}\left( 1,3\right)
-t_{\mu _{13}}^{AV}\left( 2,3\right) \\
p_{32}^{\mu _{3}}t_{\mu _{123}}^{AVV} &=&t_{\mu _{12}}^{AV}\left( 1,2\right)
-t_{\mu _{12}}^{AV}\left( 1,3\right)
\end{eqnarray*}%
to $t_{\mu _{123}}^{VAV}$ 
\begin{eqnarray*}
p_{31}^{\mu _{1}}t_{\mu _{123}}^{VAV} &=&t_{\mu _{23}}^{AV}\left( 2,1\right)
-t_{\mu _{23}}^{AV}\left( 2,3\right) \\
p_{21}^{\mu _{2}}t_{\mu _{123}}^{VAV} &=&t_{\mu _{31}}^{AV}\left( 3,1\right)
-t_{\mu _{13}}^{AV}\left( 2,3\right) +2mt_{\mu _{13}}^{VPV} \\
p_{32}^{\mu _{3}}t_{\mu _{123}}^{VAV} &=&t_{\mu _{21}}^{AV}\left( 2,1\right)
-t_{\mu _{21}}^{AV}\left( 3,1\right)
\end{eqnarray*}%
to $t_{\mu _{123}}^{VVA}$%
\begin{eqnarray*}
p_{31}^{\mu _{1}}t_{\mu _{123}}^{VVA} &=&t_{\mu _{32}}^{AV}\left( 1,2\right)
-t_{\mu _{32}}^{AV}\left( 3,2\right) \\
p_{21}^{\mu _{2}}t_{\mu _{123}}^{VVA} &=&t_{\mu _{31}}^{AV}\left( 3,1\right)
-t_{\mu _{31}}^{AV}\left( 3,2\right) \\
p_{32}^{\mu _{3}}t_{\mu _{123}}^{VVA} &=&t_{\mu _{12}}^{AV}\left( 1,2\right)
-t_{\mu _{21}}^{AV}\left( 3,1\right) +2mt_{\mu _{12}}^{VVP}
\end{eqnarray*}%
to $t_{\mu _{123}}^{AAA}$%
\begin{eqnarray*}
p_{31}^{\mu _{1}}t_{\mu _{123}}^{AAA} &=&t_{\mu _{23}}^{AV}\left( 2,1\right)
-t_{\mu _{32}}^{AV}\left( 3,2\right) -2mt_{\mu _{23}}^{PAA} \\
p_{21}^{\mu _{2}}t_{\mu _{123}}^{AAA} &=&t_{\mu _{13}}^{AV}\left( 1,3\right)
-t_{\mu _{31}}^{AV}\left( 3,2\right) +2mt_{\mu _{13}}^{APA} \\
p_{32}^{\mu _{3}}t_{\mu _{123}}^{AAA} &=&t_{\mu _{21}}^{AV}\left( 2,1\right)
-t_{\mu _{12}}^{AV}\left( 1,3\right) +2mt_{\mu _{12}}^{AAP}.
\end{eqnarray*}

The RHS of these identities show two and three-point functions, the
two-point ones%
\begin{equation}
t_{\mu _{ij}}^{AV}\left( a,b\right) =\mathrm{tr}[\gamma _{\ast }\gamma _{\mu
_{i}}S\left( a\right) \gamma _{\mu _{j}}S\left( b\right) ]=-4i\varepsilon
_{\mu _{i}\mu _{j}\nu _{1}\nu _{2}}p_{ba}^{\nu _{2}}\frac{K_{a}^{\nu _{1}}}{%
D_{ab}}
\end{equation}%
are odd tensor functions of two variables, the external and ambiguous
momenta $\left( p_{ij},P_{ij}\right) $. After integration the result is
proportional to the vector two-point function that has a naturally linear
dependence on the ambiguous momenta, looking at the computed expression as
an example in the section (\ref{BasisFI}), eq. (\ref{J2bar4D}) it is
straightforward to see that any of these amplitudes is a pure surface 
\begin{equation}
T_{\mu _{ij}}^{AV}\left( a,b\right) =-4i\varepsilon _{\mu _{i}\mu _{j}\nu
_{1}\nu _{2}}p_{ba}^{\nu _{2}}\bar{J}_{2}^{\nu _{1}}\left( a,b\right)
=2i\varepsilon _{\mu _{i}\mu _{j}\nu _{1}\nu _{2}}p_{ba}^{\nu
_{2}}P_{ab}^{\nu _{3}}\Delta _{3\nu _{3}}^{\nu _{1}}  \label{AV4D}
\end{equation}%
with the help of the relations among the $\left( p_{ij},P_{ij}\right) $ to $%
AVV$ relations we obtain 
\begin{eqnarray}
T_{\mu _{32}}^{AV}\left( 1,2\right) -T_{\mu _{23}}^{AV}\left( 2,3\right)
&=&-2i\varepsilon _{\mu _{23}\nu _{12}}\left( p_{21}^{\nu _{2}}P_{12}^{\nu
_{3}}+p_{32}^{\nu _{2}}P_{32}^{\nu _{3}}\right) \Delta _{3\nu _{3}}^{\nu
_{1}}  \label{AV(-)1} \\
T_{\mu _{13}}^{AV}\left( 1,3\right) -T_{\mu _{13}}^{AV}\left( 2,3\right)
&=&-2i\varepsilon _{\mu _{13}\nu _{12}}\left( p_{32}^{\nu _{2}}P_{32}^{\nu
_{3}}-p_{31}^{\nu _{2}}P_{31}^{\nu _{3}}\right) \Delta _{3\nu _{3}}^{\nu
_{1}}  \label{AV(-)2} \\
T_{\mu _{12}}^{AV}\left( 1,2\right) -T_{\mu _{12}}^{AV}\left( 1,3\right)
&=&-2i\varepsilon _{\mu _{12}\nu _{12}}\left( p_{31}^{\nu _{2}}P_{31}^{\nu
_{3}}-p_{21}^{\nu _{2}}P_{21}^{\nu _{3}}\right) \Delta _{3\nu _{3}}^{\nu
_{1}}  \label{AV(-)3}
\end{eqnarray}

Such results could be different for the relations to the $VAV$, $VVA$, and $%
AAA$ diagrams, but turn out they are not, and in truth depend only on the
vertex contraction that the reader can be verify using the symmetry
properties in the eq. (\ref{AV4D}), and thus will not be exhibited here.

Before going to the calculations we must remind the reader that if the RAGF
were automatically valid, as one would be led to think because they are
linear relations coming from integration, then the choice of to get rid of
the surface terms could easily determine valid all the WI, the why\ of such
a thing do not happen is all the specialty of these amplitudes.

As the contractions with the external momenta in the explicitly computed
three rank amplitudes are concerned, it is necessary to state that the three
non-vanishing sign tensors have a set of contraction properties, first to
the momenta of each vertex one of them is null under contraction.

The finite part of the basic sign tensors on the eq.'s (\ref{T+-}) (\ref{T--}%
) and (\ref{T++}) have a set of contractions with the external momenta that
can are obtained using the appendix (\ref{AppInt4D}). The finite part of the
tensors $C_{i}$ can be shown to obey%
\begin{eqnarray*}
p_{21}^{\mu _{2}}C_{1\mu _{123}} &=&+\varepsilon _{\mu _{13}\nu _{1}\nu
_{2}}p_{32}^{\nu _{2}}p_{21}^{2}\left( J_{3}^{\nu _{1}}+p_{21}^{\nu
_{1}}J_{3}\right) -\frac{1}{2}\varepsilon _{\mu _{13}\nu _{1}\nu
_{2}}p_{21}^{\nu _{1}}p_{31}^{\nu _{2}}J_{2}\left( p_{31}\right) \\
p_{32}^{\mu _{3}}C_{1\mu _{123}} &=&-\varepsilon _{\mu _{12}\nu _{1}\nu
_{2}}p_{21}^{\nu _{2}}p_{32}^{2}J_{3}^{\nu _{1}}+\frac{1}{2}\varepsilon
_{\mu _{12}\nu _{1}\nu _{2}}p_{21}^{\nu _{1}}p_{31}^{\nu _{2}}J_{2}\left(
p_{31}\right)
\end{eqnarray*}%
\begin{eqnarray*}
p_{31}^{\mu _{1}}C_{2\mu _{123}} &=&\varepsilon _{\mu _{23}\nu _{1}\nu
_{2}}p_{32}^{\nu _{2}}p_{31}^{2}\left( J_{3}^{\nu _{1}}+p_{21}^{\nu
_{1}}J_{3}\right) -\frac{1}{2}\varepsilon _{\mu _{23}\nu _{1}\nu
_{2}}p_{21}^{\nu _{1}}p_{31}^{\nu _{2}}J_{2}\left( p_{21}\right) \\
p_{32}^{\mu _{3}}C_{2\mu _{123}} &=&\varepsilon _{\mu _{12}\nu _{1}\nu
_{2}}p_{31}^{\nu _{2}}p_{32}^{2}J_{3}^{\nu _{1}}+\frac{1}{2}\varepsilon
_{\mu _{12}\nu _{1}\nu _{2}}p_{21}^{\nu _{1}}p_{31}^{\nu _{2}}J_{2}\left(
p_{21}\right)
\end{eqnarray*}%
\begin{eqnarray*}
p_{31}^{\mu _{1}}C_{3\mu _{123}} &=&+\varepsilon _{\mu _{23}\nu _{1}\nu
_{2}}p_{21}^{\nu _{2}}p_{31}^{2}J_{3}^{\nu _{1}}-\frac{1}{2}\varepsilon
_{\mu _{23}\nu _{1}\nu _{2}}p_{21}^{\nu _{1}}p_{31}^{\nu _{2}}J_{2}\left(
p_{32}\right) \\
p_{21}^{\mu _{2}}C_{3\mu _{123}} &=&-\varepsilon _{\mu _{13}\nu _{1}\nu
_{2}}p_{31}^{\nu _{2}}p_{21}^{2}J_{3}^{\nu _{1}}-\frac{1}{2}\varepsilon
_{\mu _{13}\nu _{1}\nu _{2}}p_{21}^{\nu _{1}}p_{31}^{\nu _{2}}J_{2}\left(
p_{32}\right)
\end{eqnarray*}%
and the contraction when the momenta being contracted matches the version 
\begin{eqnarray*}
p_{31}^{\mu _{1}}C_{1\mu _{123}} &=&\varepsilon _{\mu _{2}\mu _{3}\nu
_{1}\nu _{2}}\left[ \left( p_{31}^{\nu _{2}}p_{21}^{2}-p_{21}^{\nu
_{2}}p_{31}^{2}\right) J_{3}^{\nu _{1}}+p_{21}^{\nu _{1}}p_{32}^{\nu
_{2}}J_{2}\left( p_{32}\right) \right] \\
&&+\varepsilon _{\mu _{2}\mu _{3}\nu _{1}\nu _{2}}p_{21}^{\nu
_{1}}p_{31}^{\nu _{2}}\left[ 2m^{2}J_{3}+i\left( 4\pi \right) ^{-2}\right]
\end{eqnarray*}%
\begin{eqnarray*}
p_{21}^{\mu _{2}}C_{2\mu _{123}} &=&\varepsilon _{\mu _{1}\mu _{3}\nu
_{1}\nu _{2}}\left[ \left( p_{31}^{\nu _{2}}p_{21}^{2}-p_{21}^{\nu
_{2}}p_{31}^{2}\right) J_{3}^{\nu _{1}}+p_{21}^{\nu _{1}}p_{32}^{\nu
_{2}}J_{2}\left( p_{32}\right) \right] \\
&&+\varepsilon _{\mu _{1}\mu _{3}\nu _{1}\nu _{2}}p_{21}^{\nu
_{1}}p_{31}^{\nu _{2}}\left[ 2m^{2}J_{3}+i\left( 4\pi \right) ^{-2}\right]
\end{eqnarray*}%
\begin{eqnarray*}
p_{32}^{\mu _{3}}C_{3\mu _{123}} &=&-\varepsilon _{\mu _{1}\mu _{2}\nu
_{1}\nu _{2}}\left[ \left( p_{31}^{\nu _{2}}p_{21}^{2}-p_{21}^{\nu
_{2}}p_{31}^{2}\right) J_{3}^{\nu _{1}}+p_{21}^{\nu _{1}}p_{32}^{\nu
_{2}}J_{2}\left( p_{32}\right) \right] \\
&&-\varepsilon _{\mu _{1}\mu _{2}\nu _{1}\nu _{2}}p_{21}^{\nu
_{1}}p_{31}^{\nu _{2}}\left[ 2m^{2}J_{3}+i\left( 4\pi \right) ^{-2}\right]
\end{eqnarray*}

The equations were grouped in this way to emphasize that in each $C_{i\mu
_{1}\mu _{2}\mu _{3}}$ the contraction with the momentum corresponding to $%
\mu _{i}$ needs the trace of $J_{3\mu \nu }$ see eq. (\ref{trJ3}), or
equivalently the reduction of $Z_{00}^{\left( 0\right) }$ in terms of $%
Z_{rs}^{\left( -1\right) }$ and $Z_{r}^{\left( 0\right) }$, eq. (\ref{Z00^0}%
), where appear a momentum-independent constant. It is essential to realize
that all the previous results refer to definite relations among finite
tensors and functions obtained free from any regularization procedure.

It was basically used%
\begin{equation}
\varepsilon _{\lbrack \mu _{1}\mu _{3}\nu _{1}\nu _{2}}p_{ij\nu
_{3}]}J_{3}^{\nu _{1}}=0;\quad \varepsilon _{\lbrack \mu _{a}\nu _{1}\nu
_{2}\nu _{3}}J_{3\mu _{c}]}^{\nu _{1}}=0,
\end{equation}%
in order to search contractions of the $J$ integrals with the momenta when
they are not directly available, which enable us to reduce the functions by
the formulas in the appendix. As a by-product, we get the trace of the
tensor integrals, and when there are three contractions involving the
momenta and the vector integrals with the $\varepsilon $, such terms vanish%
\footnote{%
A common situation is when appearing a term like, $\varepsilon _{\mu _{1}\mu
_{3}\nu _{1}\nu _{2}}\left[ \left( p_{21}\cdot p_{31}\right) p_{21}^{\nu
_{2}}-p_{31}^{\nu _{2}}p_{21}^{2}\right] J_{3}^{\nu _{1}}$, to find
contraction with the momentum one just use $\varepsilon _{\lbrack \mu
_{1}\mu _{3}\nu _{1}\nu _{2}}p_{21\nu _{3}]}=0\,,$ then multiplying by $%
p_{31}^{\nu _{2}}p_{21}^{\nu _{3}}J_{3}^{\nu _{1}}$, and get%
\begin{equation*}
\left[ \varepsilon _{\mu _{1}\mu _{3}\nu _{1}\nu _{2}}p_{21}^{2}p_{31}^{\nu
_{2}}J_{3}^{\nu _{1}}+\varepsilon _{\nu _{3}\mu _{1}\mu _{3}\nu _{1}}\left(
p_{21}\cdot p_{31}\right) p_{21}^{\nu _{3}}J_{3}^{\nu _{1}}\right]
=-\varepsilon _{\nu _{2}\nu _{3}\mu _{1}\mu _{3}}p_{31}^{\nu
_{2}}p_{21}^{\nu _{3}}\left( p_{21\nu _{1}}J_{3}^{\nu _{1}}\right)
\end{equation*}%
the RHS is the desired contraction, in this derivation terms like $%
\varepsilon _{\nu _{1}\nu _{2}\nu _{3}\mu _{1}}p_{31}^{\nu _{2}}p_{21}^{\nu
_{3}}J_{3}^{\nu _{1}}$ vanish because the number of contraction and that the
vector integral is proportional to the external momenta. Adjusting signs and
indexes, the equation is exactly the one sought.}.

For the contractions with the set of surface terms $S_{i\mu _{123}}$, see
eq.'s (\ref{ST1}), (\ref{ST2}) and (\ref{ST3}), must be noted that the index 
$\mu _{i}$ does not appear in the tensor $\Delta _{3\mu \nu }$, only in the $%
\varepsilon $-tensor, and thus the contraction with the other indexes can be
done directly, giving the term corresponding to the correct difference of
two point $AV$ functions, by example%
\begin{eqnarray}
p_{21}^{\mu _{2}}S_{1\mu _{1}\mu _{2}\mu _{3}} &=&-2ip_{21}^{\mu _{2}}\left(
\varepsilon _{\mu _{13}\nu _{12}}p_{32}^{\nu _{2}}\Delta _{3\mu _{2}}^{\nu
_{1}}+\varepsilon _{\mu _{12}\nu _{12}}p_{21}^{\nu _{2}}\Delta _{3\mu
_{3}}^{\nu _{1}}\right) +2i\varepsilon _{\mu _{123}\nu _{1}}p_{21}^{\mu
_{2}}P_{31}^{\nu _{2}}\Delta _{3\nu _{2}}^{\nu _{1}}  \notag \\
&=&2i\varepsilon _{\mu _{13}\nu _{12}}\left( -p_{21}^{\nu _{3}}p_{32}^{\nu
_{2}}+p_{21}^{\nu _{2}}P_{31}^{\nu _{3}}\right) \Delta _{3\nu _{3}}^{\nu
_{1}},
\end{eqnarray}%
organizing the momenta $\left( p_{ij},P_{ij}\right) $ by $%
p_{ij}=P_{ir}-P_{jr}$, see the defining eq.'s (\ref{pij}) and (\ref{Pij}),
follows%
\begin{equation}
p_{21}^{\mu _{2}}S_{1\mu _{123}}=-2i\varepsilon _{\mu _{13}\nu _{12}}\left(
p_{32}^{\nu _{2}}P_{32}^{\nu _{3}}-p_{31}^{\nu _{2}}P_{31}^{\nu _{3}}\right)
\Delta _{3\nu _{3}}^{\nu _{1}}.
\end{equation}

The important feature here is that the contraction with $p_{31}^{\mu _{1}}$
only hits the index $\mu _{1}$ in the $\varepsilon $-tensor letting the free
indexes inside the tensors $\Delta _{3\mu \nu }$, however the $AV$ functions
have the indexes of its surface terms both contracted with the momenta,
thereby to switch the indexes is just a matter of contract with that
momentum and of using%
\begin{equation}
\varepsilon _{\mu _{1}\mu _{3}\nu _{1}\nu _{2}}\Delta _{3\mu _{2}}^{\nu
_{1}}-\varepsilon _{\mu _{1}\mu _{2}\nu _{1}\nu _{2}}\Delta _{3\mu
_{3}}^{\nu _{1}}-\varepsilon _{\mu _{2}\mu _{3}\nu _{1}\nu _{2}}\Delta
_{3\mu _{1}}^{\nu _{1}}-\varepsilon _{\mu _{1}\mu _{2}\mu _{3}\nu
_{1}}\Delta _{3\nu _{2}}^{\nu _{1}}=-\varepsilon _{\mu _{1}\mu _{2}\mu
_{3}\nu _{2}}\Delta _{3\nu _{1}}^{\nu _{1}},
\end{equation}%
plus a simple reorganization of the momenta, to reach at%
\begin{equation}
p_{31}^{\mu _{1}}S_{1\mu _{123}}=-2i\varepsilon _{\mu _{23}\nu _{12}}\left(
p_{21}^{\nu _{2}}P_{12}^{\nu _{3}}+p_{32}^{\nu _{2}}P_{32}^{\nu _{3}}\right)
\Delta _{\nu _{3}}^{\nu _{1}}+2i\varepsilon _{\mu _{2}\mu _{3}\nu _{2}\nu
_{3}}p_{21}^{\nu _{2}}p_{31}^{\nu _{3}}\Delta _{3\nu _{1}}^{\nu _{1}},
\label{contS1}
\end{equation}%
the remaining contraction with $p_{32}^{\mu _{3}}$ follows the example of
the contraction with $p_{21}^{\mu _{2}}$, and is given by 
\begin{equation}
p_{32}^{\mu _{3}}S_{1\mu _{123}}=-2i\varepsilon _{\mu _{12}\nu _{12}}\left(
p_{31}^{\nu _{2}}P_{31}^{\nu _{3}}-p_{21}^{\nu _{2}}P_{21}^{\nu _{3}}\right)
\Delta _{3\nu _{3}}^{\nu _{1}}.
\end{equation}

The same phenomenon happens to the other two sets for which we quote the
results 
\begin{eqnarray}
p_{31}^{\mu _{1}}S_{2\mu _{123}} &=&-2i\varepsilon _{\mu _{23}\nu
_{12}}\left( p_{21}^{\nu _{2}}P_{12}^{\nu _{3}}+p_{32}^{\nu _{2}}P_{32}^{\nu
_{3}}\right) \Delta _{3\nu _{3}}^{\nu _{1}}  \notag \\
p_{21}^{\mu _{2}}S_{2\mu _{123}} &=&-2i\varepsilon _{\mu _{13}\nu
_{12}}\left( p_{32}^{\nu _{2}}P_{32}^{\nu _{3}}-p_{31}^{\nu _{2}}P_{31}^{\nu
_{3}}\right) \Delta _{3\nu _{3}}^{\nu _{1}}+2i\varepsilon _{\mu _{1}\mu
_{3}\nu _{2}\nu _{3}}p_{21}^{\nu _{2}}p_{31}^{\nu _{3}}\Delta _{3\nu
_{1}}^{\nu _{1}}  \label{contS2} \\
p_{32}^{\mu _{3}}S_{2\mu _{123}} &=&-2i\varepsilon _{\mu _{12}\nu
_{12}}\left( p_{31}^{\nu _{2}}P_{31}^{\nu _{3}}-p_{21}^{\nu _{2}}P_{21}^{\nu
_{3}}\right) \Delta _{3\nu _{3}}^{\nu _{1}}  \notag
\end{eqnarray}%
and for $S_{3\mu _{1}\mu _{2}\mu _{3}}$%
\begin{eqnarray*}
p_{31}^{\mu _{1}}S_{3\mu _{123}} &=&-2i\varepsilon _{\mu _{23}\nu
_{12}}\left( p_{21}^{\nu _{2}}P_{12}^{\nu _{3}}+p_{32}^{\nu _{2}}P_{32}^{\nu
_{3}}\right) \Delta _{3\nu _{3}}^{\nu _{1}} \\
p_{21}^{\mu _{2}}S_{3\mu _{123}} &=&-2i\varepsilon _{\mu _{13}\nu
_{12}}\left( p_{32}^{\nu _{2}}P_{32}^{\nu _{3}}-p_{31}^{\nu _{2}}P_{31}^{\nu
_{3}}\right) \Delta _{3\nu _{3}}^{\nu _{1}} \\
p_{32}^{\mu _{3}}S_{3\mu _{123}} &=&-2i\varepsilon _{\mu _{12}\nu
_{12}}\left( p_{31}^{\nu _{2}}P_{31}^{\nu _{3}}-p_{21}^{\nu _{2}}P_{21}^{\nu
_{3}}\right) \Delta _{3\nu _{3}}^{\nu _{1}}-2i\varepsilon _{\mu _{1}\mu
_{2}\nu _{2}\nu _{3}}p_{21}^{\nu _{2}}p_{31}^{\nu _{3}}\Delta _{3\nu
_{1}}^{\nu _{1}}
\end{eqnarray*}

Now that we have laid down the tools to analyze the integrated expression,
it is an issue of applying the devices presented until this moment to
establish the contraction with the first version of the calculated
three-point diagrams, starting with the $AVV$%
\begin{eqnarray*}
p_{31}^{\mu _{1}}\left( T_{\mu _{123}}^{AVV}\right) _{1} &=&T_{\mu
_{32}}^{AV}\left( 1,2\right) -T_{\mu _{23}}^{AV}\left( 2,3\right) -2mT_{\mu
_{23}}^{PVV}+\underline{2i\varepsilon _{\mu _{2}\mu _{3}\nu _{1}\nu
_{2}}p_{21}^{\nu _{1}}p_{31}^{\nu _{2}}\left[ \Delta _{3\alpha }^{\alpha
}+2i\left( 4\pi \right) ^{-2}\right] } \\
p_{21}^{\mu _{2}}\left( T_{\mu _{123}}^{AVV}\right) _{1} &=&T_{\mu
_{13}}^{AV}\left( 1,3\right) -T_{\mu _{13}}^{AV}\left( 2,3\right) \\
p_{32}^{\mu _{3}}\left( T_{\mu _{123}}^{AVV}\right) _{1} &=&T_{\mu
_{12}}^{AV}\left( 1,2\right) -T_{\mu _{12}}^{AV}\left( 1,3\right)
\end{eqnarray*}%
and noticing the same additional term will appear for the corresponding
version of $VAV$ in the contraction with the momentum $p_{31}^{\mu _{1}}$%
\begin{equation*}
p_{31}^{\mu _{1}}\left( T_{\mu _{123}}^{VAV}\right) _{1}=T_{\mu
_{23}}^{AV}\left( 2,1\right) -T_{\mu _{23}}^{AV}\left( 2,3\right) +%
\underline{2i\varepsilon _{\mu _{2}\mu _{3}\nu _{1}\nu _{2}}p_{21}^{\nu
_{1}}p_{31}^{\nu _{2}}\left( \Delta _{3\alpha }^{\alpha }+2i\left( 4\pi
\right) ^{-2}\right) },
\end{equation*}%
as the other RAGFs will be satisfied without conditions and exactly the same
thing for $VVA$ and $AAA$.

It can be of no surprise that the second and third versions of all
configurations of vertexes will show up the possibly violating factor $%
2i\varepsilon _{\mu _{i}\mu _{j}\nu _{1}\nu _{2}}p_{21}^{\nu
_{1}}p_{31}^{\nu _{2}}\left[ \Delta _{3\alpha }^{\alpha }+2i\left( 4\pi
\right) ^{-2}\right] $ in the contraction with the momenta $p_{21}^{\mu
_{2}} $ and $p_{32}^{\mu _{3}}$ entering in the second and third vertexes
respectively.

The explicitly computed equations below subsume all the results to for the
verification of the RAGFs 
\begin{eqnarray}
q_{1}^{\mu _{1}}\left[ T_{\mu _{123}}^{\Gamma _{123}}\right] _{1}^{\mathrm{%
viol}} &=&+2i\varepsilon _{\mu _{2}\mu _{3}\nu _{1}\nu _{2}}p_{21}^{\nu
_{1}}p_{32}^{\nu _{2}}\left[ \Delta _{3\alpha }^{\alpha }+2i\left( 4\pi
\right) ^{-2}\right]  \label{ragfViol4D} \\
q_{2}^{\mu _{2}}\left[ T_{\mu _{123}}^{\Gamma _{123}}\right] _{2}^{\mathrm{%
viol}} &=&+2i\varepsilon _{\mu _{1}\mu _{3}\nu _{1}\nu _{2}}p_{21}^{\nu
_{1}}p_{32}^{\nu _{2}}\left[ \Delta _{3\alpha }^{\alpha }+2i\left( 4\pi
\right) ^{-2}\right] \\
q_{3}^{\mu _{3}}\left[ T_{\mu _{123}}^{\Gamma _{123}}\right] _{3}^{\mathrm{%
viol}} &=&-2i\varepsilon _{\mu _{1}\mu _{2}\nu _{1}\nu _{2}}p_{21}^{\nu
_{1}}p_{32}^{\nu _{2}}\left[ \Delta _{3\alpha }^{\alpha }+2i\left( 4\pi
\right) ^{-2}\right] ,
\end{eqnarray}%
where we adopt the notation to the routing differences $q_{1}=p_{31}$, $%
q_{2}=p_{21}$, and $q_{3}=p_{32}$ to mark a convention for first, second,
and third vertexes respectively, this has already appeared in the Fig.(\ref%
{diag1}) for the general diagram. Whereas the symbol $\Gamma _{123}\equiv
\Gamma _{1}\Gamma _{2}\Gamma _{3}$ encode all the possibilities of
combination vertexes $\Gamma _{i}\in \left\{ A,V\right\} $ for which the
number of axial vertexes is odd. To visualize this violation pattern, we
offer the schematic graph

\begin{figure}[tbph]
\includegraphics[scale=0.8]{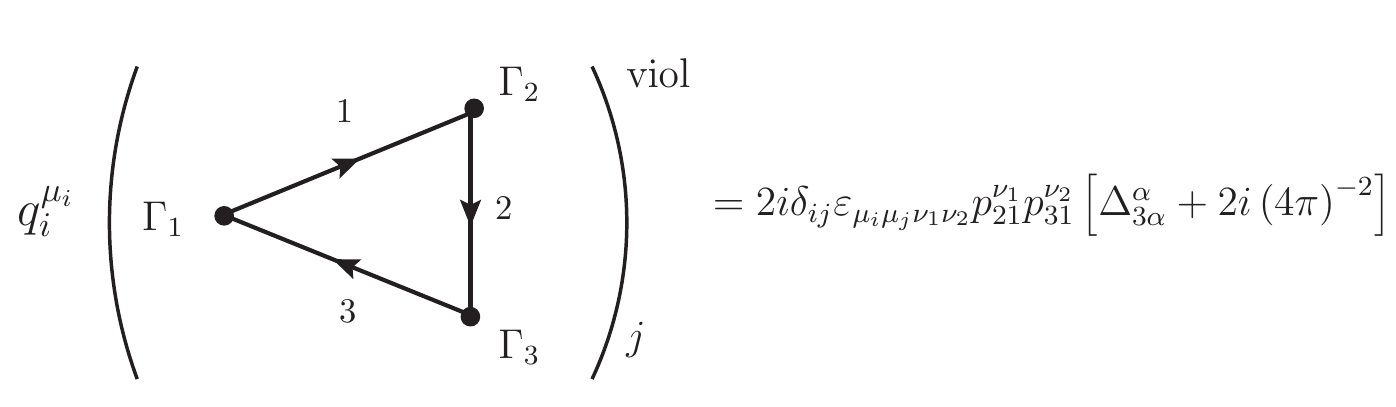}
\caption{The violation factor of the RAGF established for the contraction
with momenta $q_{i}^{\protect\mu _{1}}.$}
\label{diagviol}
\end{figure}

\textbf{Uniqueness: }The other subtle point is uniqueness. Any precise and
unambiguous discussion about this point needs a concrete definition of
uniqueness. Anyway, with the help of the explicit expressions we have
determined, it is established the equations%
\begin{eqnarray*}
\left[ T_{\mu _{123}}^{\Gamma _{123}}\right] _{1}-\left[ T_{\mu
_{123}}^{\Gamma _{123}}\right] _{2} &=&+2i\varepsilon _{\mu _{1}\mu _{2}\mu
_{3}\nu _{1}}p_{32}^{\nu _{1}}\left( \Delta _{3\alpha }^{\alpha }+2i\left(
4\pi \right) ^{-2}\right) \\
\left[ T_{\mu _{123}}^{\Gamma _{123}}\right] _{1}-\left[ T_{\mu
_{123}}^{\Gamma _{123}}\right] _{3} &=&-2i\varepsilon _{\mu _{1}\mu _{2}\mu
_{3}\nu _{1}}p_{21}^{\nu _{1}}\left( \Delta _{3\alpha }^{\alpha }+2i\left(
4\pi \right) ^{-2}\right) \\
\left[ T_{\mu _{123}}^{\Gamma _{123}}\right] _{2}-\left[ T_{\mu
_{123}}^{\Gamma _{123}}\right] _{3} &=&-2i\varepsilon _{\mu _{1}\mu _{2}\mu
_{3}\nu _{1}}p_{31}^{\nu _{1}}\left( \Delta _{3\alpha }^{\alpha }+2i\left(
4\pi \right) ^{-2}\right) ,
\end{eqnarray*}%
obtained by the same means as for the momenta contractions. Uniqueness then,
here, means to us, that any possible way to compute the same expression
returns the same result, in other words, the RHS is conditioned by this
definition to be zero.

Anyway, the integrands we have been treating so far are well-defined tensors
and they obey 
\begin{equation*}
\left( t_{\mu _{123}}^{\Gamma _{123}}\right) _{1}=\left( t_{\mu
_{123}}^{\Gamma _{123}}\right) _{2}=\left( t_{\mu _{123}}^{\Gamma
_{123}}\right) _{3},
\end{equation*}%
well, after separating two sectors, one in which physical parameters are
factored out of the integrals and another sector where it is not possible,
but every integral is finite. It could seem that the above equation should
be true and independent of interpretation given to the divergent aspects,
insofar as the same values are given to the undetermined parts. Nonetheless,
the sorting of indexes makes the results to the finite parts unequal at the
same time making the surface terms carry configuration of indexes that
condition the satisfaction of distinct sets of RAGF. What these features end
up showing is that if the computations furnish unique answers as we have
defined, then all RAGFs are satisfied and vice-versa.

Other identities of the Clifford algebra can lead to traces different from
the one we have started this argumentation, but it is provable they end up
in linear combinations of the ones we presented. Furthermore, in the section
(\ref{LED4DSTS}), it is going to become clear why the ones we chose are
enough to disentangle any feature of the anomalous odd tensor in even
space-time dimensions. In particular as the traces we employed can be
linearly combined it is possible to reach the following expression as a
result to the amplitude%
\begin{equation*}
\left[ T_{\mu _{123}}^{\Gamma _{123}}\right] _{\left\{ 12\right\} }=\frac{1}{%
2}\left[ \left( T_{\mu _{123}}^{\Gamma _{123}}\right) _{1}+\left( T_{\mu
_{123}}^{\Gamma _{123}}\right) _{2}\right] ,
\end{equation*}%
and the combination $13$ and $23$, which violates the RAGF to the vertexes $%
12$, $13$, and $23$ respectively if the surface term is any other than the
one producing unique amplitudes. These three combinations in reality are
enough to reproduce the computations using any substitution coded in the
identities among the antisymmetrized products, see eq.(\ref{Chiral-Id}), by
example, using a common substitution like 
\begin{equation*}
\gamma _{\ast }\gamma _{\mu _{i}\nu _{i}\mu _{i+1}}=i\varepsilon _{\mu
_{i}\nu _{i}\mu _{i+1}\nu }\gamma ^{\nu }+\gamma _{\ast }\left( g_{\nu
_{i}\mu _{i+1}}\gamma _{\mu _{i}}-g_{\mu _{i}\mu _{i+1}}\gamma _{\nu
_{i}}+g_{\mu _{i}\nu _{i}}\gamma _{\mu _{i+1}}\right) ,
\end{equation*}%
the difference between the integrated expression and the combination $\left[
T_{\mu _{123}}^{\Gamma _{123}}\right] _{\left\{ i,i+1\right\} }$ are finite
and identically vanishing integrals, in other words, the application of the
identity above or the linear combination have equal results when integrated,
without touching their surface terms. This happens to any manipulations
anywhere in the string of six gamma matrices and $\gamma _{\ast }$. This
proposition is outlined in the appendix (\ref{Tr6G4D}).

Another example, that illustrates the use of the definition of the chiral
matrix as nothing special, is the use of%
\begin{equation*}
\gamma _{\ast }\gamma _{\mu _{i}}=\frac{1}{3!}\varepsilon _{\mu _{i}\nu
_{123}}\gamma ^{\nu _{123}},
\end{equation*}%
in the level of the trace, of six matrices, it produces ten terms, and the
five terms different from the use of the version $i$ are finite integrals
that happens to vanish, precisely the $T_{\mu _{i}}^{\left( -+\right) }$ in
eq. (\ref{T-+}), and $T_{\mu }^{ASS}$ amplitude, eq. (\ref{ASS}). The
importance of defining the version of some amplitude in the way we did,
stems from the fact they satisfy automatically the maximum possible number
of RAGFs, and we can build arbitrary linear combinations of the basic
building-block versions, that reproduce all the ones deduced in the appendix
(\ref{Tr6G4D}), exploiting the substitutions encompassed in the formulas (%
\ref{Chiral-Id}), and any application of Schouten identities, precisely%
\begin{equation}
\left[ T_{\mu _{123}}^{\Gamma _{123}}\right] _{\left\{
r_{1}r_{2}r_{3}\right\} }=\frac{1}{r_{1}+r_{2}+r_{3}}\sum_{i=1}^{3}r_{i}%
\left( T_{\mu _{123}}^{\Gamma _{123}}\right) _{i},  \label{r123}
\end{equation}%
with their $r_{1}+r_{2}+r_{3}\not=0$, its significance is they correspond to
identical integrands, but if the surface term is forced to be zero, then all
of them become distinct and violate the RAGFs by arbitrary amounts, and as a
function of the external momenta same for the WIs. It helps to explain
certain violation amounts found in the literature as in \cite{Wu2006}.

\subsection{A Low Energy Theorem to the $T^{\Gamma _{1}\Gamma _{2}\Gamma
_{3}}$-Triangles and its relation with the Ward Identities}

\label{LE4D}

Hitherto, we have shown how the dynamic of the traces and the surface terms
interfere with the linearity of the contraction of the amplitudes with the
external momenta and the uniqueness of the perturbative expressions of the
nontrivial odd tensors in this dimension. Nonetheless, a question remains to
be answered. Could one anticipate such verified properties? And are they
unavoidable? And they are, then why? To spot the two main reasons why this
happens, we present a two-part analysis of the most general tensor that can
represent these amplitudes.

The cause of these phenomena start with a relation that a general tensor of
rank three, that is function of two variables, and odd parity is required to
have. Adopting the momenta such as $q_{1}=q_{2}+q_{3}$, where the $q_{2}$
and $q_{3}$ are incoming in the vertexes $\Gamma _{2}$ and $\Gamma _{3}$ a $%
q_{1}$ outgoing the vertex $\Gamma _{1}$, the general tensor will be
expressed as%
\begin{eqnarray}
F_{\mu _{1}\mu _{2}\mu _{3}} &=&\varepsilon _{\mu _{1}\mu _{2}\mu _{3}\nu
}\left( q_{2}^{\nu }F_{1}+q_{3}^{\nu }F_{2}\right) +\varepsilon _{\mu
_{1}\mu _{2}\nu _{1}\nu _{2}}q_{2}^{\nu _{1}}q_{3}^{\nu _{2}}\left( q_{2\mu
_{3}}G_{1}+q_{3\mu _{3}}G_{2}\right)  \label{GForm} \\
&&+\varepsilon _{\mu _{1}\mu _{3}\nu _{1}\nu _{2}}q_{2}^{\nu _{1}}q_{3}^{\nu
_{2}}\left( q_{2\mu _{2}}G_{3}+q_{3\mu _{2}}G_{4}\right) +\varepsilon _{\mu
_{2}\mu _{3}\nu _{1}\nu _{2}}q_{2}^{\nu _{1}}q_{3}^{\nu _{2}}\left( q_{2\mu
_{1}}G_{5}+q_{3\mu _{1}}G_{6}\right) ,  \notag
\end{eqnarray}%
that in the three contractions%
\begin{eqnarray*}
q_{1}^{\mu _{1}}F_{\mu _{1}\mu _{2}\mu _{3}} &=&\varepsilon _{\mu _{2}\mu
_{3}\nu _{1}\nu _{2}}q_{2}^{\nu _{1}}q_{3}^{\nu _{2}}V_{1} \\
q_{2}^{\mu _{2}}F_{\mu _{1}\mu _{2}\mu _{3}} &=&\varepsilon _{\mu _{1}\mu
_{3}\nu _{1}\nu _{2}}q_{2}^{\nu _{1}}q_{3}^{\nu _{2}}V_{2} \\
q_{3}^{\mu _{3}}F_{\mu _{1}\mu _{2}\mu _{3}} &=&\varepsilon _{\mu _{1}\mu
_{2}\nu _{1}\nu _{2}}q_{2}^{\nu _{1}}q_{3}^{\nu _{2}}V_{3},
\end{eqnarray*}%
can be identified three functions written exclusively as functions of the
ingredients if the general tensor, they are%
\begin{eqnarray*}
V_{1} &=&-F_{1}+F_{2}+\left( q_{1}\cdot q_{2}\right) G_{5}+\left( q_{1}\cdot
q_{3}\right) G_{6} \\
V_{2} &=&-F_{2}+q_{2}^{2}G_{3}+\left( q_{2}\cdot q_{3}\right) G_{4} \\
V_{3} &=&-F_{1}+\left( q_{2}\cdot q_{3}\right) G_{1}+q_{3}^{2}G_{2},
\end{eqnarray*}%
and thus, at the point $q_{i}\cdot q_{j}=0$, if the form factors are regular
at most discontinuous, we obtain%
\begin{equation*}
V_{1}\left( 0\right) =F_{2}-F_{1};\quad V_{2}\left( 0\right) =-F_{2};\quad
V_{3}\left( 0\right) =-F_{1},
\end{equation*}%
where the zero in our notation will always mean the point where all the
bilinears are zero $q_{i}\cdot q_{j}=0$, therefore from the equations above
follows%
\begin{equation}
V_{1}\left( 0\right) =V_{3}\left( 0\right) -V_{2}\left( 0\right) .
\label{VertsZero}
\end{equation}

These relations among the form factors have the information about the
symmetry or violation thereof, at one kinematical point, even if no
particular symmetry was need to such relation. Its satisfaction does not
guarantee symmetries, that this tensor may have, in all the points, but its
violation in this point, imply the violation of symmetry anyway. The crucial
feature is that if the divergence of the axial current is connected to the
pseudo-scalar density in a correlator with two other vector currents, as by
example in $AVV$, we have%
\begin{equation*}
\varepsilon _{\mu _{2}\mu _{3}\nu _{1}\nu _{2}}q_{2}^{\nu _{1}}q_{3}^{\nu
_{2}}V_{1}\left( 0\right) =-2mT_{\mu _{2}\mu _{3}}^{PVV}\left( 0\right)
=\varepsilon _{\mu _{2}\mu _{3}\nu _{1}\nu _{2}}q_{2}^{\nu _{1}}q_{3}^{\nu
_{2}}\Omega _{1}^{PVV}\left( 0\right)
\end{equation*}%
simply reworded as $V_{1}\left( 0\right) =1/\left( 2\pi \right) ^{2}$ the
result presented in eq. (\ref{LEPVV}), thereby at least one of the vector
currents will be violated as the eq.(\ref{VertsZero}) above does not allow
for $V_{2}=0$ and $V_{3}=0$ simultaneously in this case. Nonetheless, if the
vector currents are divergenceless $V_{2}=V_{3}=0$, then the parameters
defining the tensor under consideration, that means, rank, parity, number of
variables, and regularity, will imply the existence of an additional term in 
$V_{1}=1/\left( 2\pi \right) ^{2}+\mathcal{A}$, the anomaly, present simply
by the reason that the eq.(\ref{VertsZero}) relating the $V$'s requires it.
In which case $\mathcal{A}=-\Omega _{1}^{PVV}\left( 0\right) $, furnishing a
connection between an unavoidable property of a finite amplitude and the
symmetry content of another tensor, emerges.

This is the starting point of the violation dynamics of anomalous
amplitudes. If the AWI is preserved, that means, $V_{1}\left( 0\right)
=\Omega _{1}^{PVV}\left( 0\right) =\left( 2\pi \right) ^{-2}$, and if at the
same time VWIs are preserved $V_{2}\left( 0\right) =V_{3}\left( 0\right) =0$%
, this implies an obvious violation the linear-algebra type solution (\ref%
{VertsZero}), and the immediate conclusion is that no tensor, whatever its
origin, can be at the same time connected to the finite $PVV$ (take this as
a boundary condition) and have vanishing contraction with the momenta $%
q_{2}^{\mu _{2}}$ and $q_{3}^{\mu _{3}}$.

On the other hand, constructing $AVV$ such that $V_{1}=\Omega _{1}^{PVV}+%
\mathrm{Anomaly}$, for the function $V_{1}$ to has the necessary zero to be
compatible with the VWIs. This shows that whenever the function of the
axial-vertex contraction of $AVV$ is connected, anomalously or not, to the $%
PVV$ there will be an anomaly in at least one of the currents.

The particular ingredients of its perturbative expression fully corroborate
with these assertions, the computations cast in this contribution is one
more expression of these facts. However, we will show more, the RAGFs
embodying linearity of integration furnish an exact connection among the
ultraviolet and infrared features of the amplitudes, namely%
\begin{equation}
\Omega _{1}^{PVV}\left( 0\right) =2i\Delta _{3\alpha }^{\alpha }
\label{UV-IR}
\end{equation}%
it is the condition required by linearity and equivalently by a unique
expression as functions of the internal momenta, a condition that will be
detailed in the next section.

As will be demonstrated later in section (\ref{LED4DSTS}) assuming the form $%
V_{i}=\Omega _{i}+\mathcal{A}_{i}$, where the $V$'s are the result of the
contraction with the momenta corresponding to the $i$-th vertex and the $%
\Omega $'s are the odd rank-two finite amplitudes form factors. Even to
classically non-conserved vector currents and for an amplitude with three
arbitrary masses running in the loop, the $\mathcal{A}$'s obey%
\begin{equation}
\mathcal{A}_{3}-\mathcal{A}_{2}-\mathcal{A}_{1}=\left( 2\pi \right) ^{-2}
\label{Ans}
\end{equation}%
due to 
\begin{equation}
\Omega _{3}\left( 0\right) -\Omega _{2}\left( 0\right) -\Omega _{1}\left(
0\right) =-\left( 2\pi \right) ^{-2},  \label{GamatZero}
\end{equation}%
always and independent of the number of axial vertexes (odd ones for sure),
permutations thereof, and the masses and if the vector currents are
conserved or not. Independent of the masses is not the claim the anomaly is
independent of masses because that is known, for us, it means that even when
the finite amplitudes exhibit intricate dependency on the masses the
relation they have in the point zero is all the time incompatible with the
ones required to a tensor with the characteristics that the rank-three
triangles must have, eq. (\ref{VertsZero}), even if no divergence arises in
perturbation theory.

The last claim, implicit in the equation (\ref{Ans}), only explicit the
common result that when the vector currents are preserved the value of the
anomaly is unique, but even more, whatever explicit tensor obtained via
regularization, or not, violating by any quantity all the currents, it will
always obey the equation to the $\mathcal{A}$'s above, that in the very end
are determined by the combinations at zero of finite functions representing
the rank-two odd amplitudes.

Another striking feature of the way we pose the matter of divergences is the
fact that some restriction over the undetermined surface terms can be
anticipated only based on the characteristics of the general form this
tensor can have and its connection to the $AV$ two-point functions via
linearity of integration. They are the theme of the next section where, in
the end, the reasoning leading the the equation (\ref{UV-IR}) is unfolded.
After that we end up showing the simple, but non-obvious path leading to the
general equation (\ref{GamatZero})

\subsection{The RAGFs and the Kinematical Behavior of Arbitrary Amplitudes 
\label{LED4DSTS}}

In the section where the explicit result to the versions of the amplitudes
was shown to not automatically satisfy all their RAGFs, that means, a
condition connecting the surface terms and the finite part must be stated at
least in one of the contractions with the external momenta. This result can
be established without resorting to any explicit computation, only assuming
the most general form that the undetermined part of the amplitude can
assume, and studying the constraints given by the contractions relating it
to the differences of two-point $AV$ functions.

To demonstrate such a proposition, let us lay down the most general tensor
of mass dimension one, built out of the kinematical data, the arbitrary
momenta, and surface terms. First of all the $q_{i}$ vectors---the
kinematical data---are written as differences of the routings $k_{i}$, 
\textbf{but not the opposite}. Thus we replace the former by the latter and
use, in equivalent form, the combinations we have defined and used in this
paper, $P_{ij}=k_{i}+k_{j}$ in place of $k_{i}$, due to the size of the
expressions. Hence, the most general set of surface terms for the odd
amplitudes is%
\begin{eqnarray*}
F_{\mu _{123}}^{\Delta } &=&+\varepsilon _{\mu _{2}\mu _{3}\nu _{1}\nu
_{2}}\left( A_{11}P_{21}^{\nu _{2}}+A_{12}P_{31}^{\nu
_{2}}+A_{13}P_{32}^{\nu _{2}}\right) \Delta _{3\mu _{1}}^{\nu _{1}} \\
&&+\varepsilon _{\mu _{1}\mu _{3}\nu _{1}\nu _{2}}\left( A_{21}P_{21}^{\nu
_{2}}+A_{22}P_{31}^{\nu _{2}}+A_{23}P_{32}^{\nu _{2}}\right) \Delta _{3\mu
_{2}}^{\nu _{1}} \\
&&+\varepsilon _{\mu _{1}\mu _{2}\nu _{1}\nu _{2}}\left( A_{31}P_{21}^{\nu
_{2}}+A_{32}P_{31}^{\nu _{2}}+A_{33}P_{32}^{\nu _{2}}\right) \Delta _{3\mu
_{3}}^{\nu _{1}} \\
&&+\varepsilon _{\mu _{1}\mu _{2}\mu _{3}\nu _{1}}\left( B_{1}P_{21}^{\nu
_{2}}+B_{2}P_{31}^{\nu _{2}}+B_{3}P_{32}^{\nu _{2}}\right) \Delta _{3\nu
_{2}}^{\nu _{1}} \\
&&+\varepsilon _{\mu _{1}\mu _{2}\mu _{3}\nu _{1}}\left( C_{1}P_{21}^{\nu
_{1}}+C_{2}P_{31}^{\nu _{1}}+C_{3}P_{32}^{\nu _{1}}\right) \Delta _{3\nu
_{2}}^{\nu _{2}}.
\end{eqnarray*}

After contraction, we expect this tensor to be related to the two-point $AV$
tensors, by example, $q_{2}^{\mu _{2}}T_{\mu _{123}}^{AVV}=T_{\mu
_{13}}^{AV}\left( 1,3\right) -T_{\mu _{13}}^{AV}\left( 2,3\right) $. Those
amplitudes have their indexes in the surface term both in contracted mode
and without a term like the last one, a trace of the surface term. Thus one
more property of this general tensor will be used, it has an identity that
reduce the number of linearly independent constants%
\begin{equation*}
\varepsilon _{\mu _{1}\mu _{2}\mu _{3}\nu _{1}}\Delta _{3\nu _{2}}^{\nu
_{2}}=\varepsilon _{\mu _{1}\mu _{2}\mu _{3}\nu _{2}}\Delta _{3\nu
_{1}}^{\nu _{2}}-\varepsilon _{\mu _{1}\mu _{2}\nu _{1}\nu _{2}}\Delta
_{3\mu _{3}}^{\nu _{2}}-\varepsilon _{\mu _{1}\mu _{3}\nu _{1}\nu
_{2}}\Delta _{3\mu _{2}}^{\nu _{2}}-\varepsilon _{\mu _{2}\mu _{3}\nu
_{1}\nu _{2}}\Delta _{3\mu _{1}}^{\nu _{2}}
\end{equation*}%
using it, the most general tensor of these variables under the conditions
stated become 
\begin{eqnarray*}
F_{\mu _{123}}^{\Delta } &=&+\varepsilon _{\mu _{2}\mu _{3}\nu _{1}\nu
_{2}}\left( a_{11}P_{21}+a_{12}P_{31}+a_{13}P_{32}\right) ^{\nu _{2}}\Delta
_{3\mu _{1}}^{\nu _{1}} \\
&&+\varepsilon _{\mu _{1}\mu _{3}\nu _{1}\nu _{2}}\left(
a_{21}P_{21}+a_{22}P_{31}+a_{23}P_{32}\right) ^{\nu _{2}}\Delta _{3\mu
_{2}}^{\nu _{1}} \\
&&+\varepsilon _{\mu _{1}\mu _{2}\nu _{1}\nu _{2}}\left(
a_{31}P_{21}+a_{32}P_{31}+a_{33}P_{32}\right) ^{\nu _{2}}\Delta _{3\mu
_{3}}^{\nu _{1}} \\
&&+\varepsilon _{\mu _{1}\mu _{2}\mu _{3}\nu _{1}}\left(
b_{1}P_{21}+b_{2}P_{31}+b_{3}P_{32}\right) ^{\nu _{2}}\Delta _{3\nu
_{2}}^{\nu _{1}}
\end{eqnarray*}%
where the new constants are now given by $a_{ij}=A_{ij}-C_{j}$ and $%
b_{i}=B_{i}+C_{i}$ and the range of $i,j\in \left\{ 1,2,3\right\} $.

The $a_{ij}$ and $b_{j}$ are twelve arbitrary constants that embody all the
freedom present by such tensor: Function of three variables, essentially the
routings of the diagram, the rank, and parity of the tensor, the power
counting and mass dimension equal one. The $j$ captures the $P$ momenta in
the order $\left( P_{21},P_{31},P_{32}\right) $, and the index $i$ is
clearly linked to the index $\mu _{i}$ that is turn is associated with the
vertex that will appear in the amplitudes $T_{\mu _{123}}^{\Gamma _{123}}$.

Now expressing for convenience the three independent differences of
two-point $AV$ functions (\ref{AV(-)1}),(\ref{AV(-)2}) and (\ref{AV(-)3}) as%
\begin{eqnarray}
T_{1\left( -\right) \mu _{23}}^{AV} &=&-2i\varepsilon _{\mu _{23}\nu _{12}} 
\left[ -P_{21}^{\nu _{2}}P_{32}^{\nu _{3}}+P_{31}^{\nu _{2}}\left(
P_{32}^{\nu _{3}}-P_{21}^{\nu _{3}}\right) +P_{32}^{\nu _{2}}P_{21}^{\nu
_{3}}\right] \Delta _{3\nu _{3}}^{\nu _{1}} \\
T_{2\left( -\right) \mu _{13}}^{AV} &=&-2i\varepsilon _{\mu _{13}\nu _{12}} 
\left[ +P_{21}^{\nu _{2}}\left( P_{31}^{\nu _{3}}-P_{32}^{\nu _{3}}\right)
+P_{31}^{\nu _{2}}P_{32}^{\nu _{3}}-P_{32}^{\nu _{2}}P_{31}^{\nu _{3}}\right]
\Delta _{3\nu _{3}}^{\nu _{1}} \\
T_{3\left( -\right) \mu _{12}}^{AV} &=&-2i\varepsilon _{\mu _{12}\nu _{12}} 
\left[ -P_{21}^{\nu _{2}}P_{31}^{\nu _{3}}+P_{31}^{\nu _{2}}P_{21}^{\nu
_{3}}+P_{32}^{\nu _{2}}\left( P_{31}^{\nu _{3}}-P_{21}^{\nu _{3}}\right) %
\right] \Delta _{3\nu _{3}}^{\nu _{1}},
\end{eqnarray}%
where the notation $T_{i\left( -\right) }^{AV}$ means they came from the $%
\mu _{i}$-th contraction with the corresponding momenta.

Let us lay down the impossibility of satisfying all these relations without
further conditions, beginning with the most general tensor of surface terms $%
F_{\mu _{1}\mu _{2}\mu _{3}}^{\Delta }$. Contracting $F_{\mu _{1}\mu _{2}\mu
_{3}}^{\Delta }$ with $q_{1}^{\mu _{1}}=p_{31}^{\mu _{1}}=P_{32}^{\mu
_{1}}-P_{21}^{\mu _{1}}$

\begin{eqnarray*}
p_{31}^{\mu _{1}}F_{\mu _{123}}^{\Delta } &=&+\varepsilon _{\mu _{3}\nu
_{1}\nu _{2}\nu _{3}}\left[ -\left( a_{21}+a_{23}\right) P_{21}^{\nu
_{2}}P_{32}^{\nu _{3}}+a_{22}P_{31}^{\nu _{2}}\left( P_{21}^{\nu
_{3}}-P_{32}^{\nu _{3}}\right) \right] \Delta _{3\mu _{2}}^{\nu _{1}} \\
&&+\varepsilon _{\mu _{2}\nu _{1}\nu _{2}\nu _{3}}\left[ -\left(
a_{31}+a_{33}\right) P_{21}^{\nu _{2}}P_{32}^{\nu _{3}}+a_{32}P_{31}^{\nu
_{2}}\left( P_{21}^{\nu _{3}}-P_{32}^{\nu _{3}}\right) \right] \Delta _{3\mu
_{3}}^{\nu _{1}} \\
&&+\varepsilon _{\mu _{2}\mu _{3}\nu _{1}\nu _{2}}\left[ -\left(
a_{11}-b_{1}\right) P_{21}^{\nu _{2}}P_{21}^{\nu _{3}}+\left(
a_{13}-b_{3}\right) P_{32}^{\nu _{2}}P_{32}^{\nu _{3}}\right] \Delta _{3\nu
_{3}}^{\nu _{1}} \\
&&+\varepsilon _{\mu _{2}\mu _{3}\nu _{1}\nu _{2}}\left[ +\left(
a_{11}+b_{3}\right) P_{21}^{\nu _{2}}P_{32}^{\nu _{3}}-a_{12}P_{31}^{\nu
_{2}}\left( P_{21}^{\nu _{3}}-P_{32}^{\nu _{3}}\right) \right] \Delta _{3\nu
_{3}}^{\nu _{1}} \\
&&+\varepsilon _{\mu _{2}\mu _{3}\nu _{1}\nu _{2}}\left[ -\left(
a_{13}+b_{1}\right) P_{32}^{\nu _{2}}P_{21}^{\nu _{3}}+b_{2}\left(
P_{21}^{\nu _{2}}-P_{32}^{\nu _{2}}\right) P_{31}^{\nu _{3}}\right] \Delta
_{3\nu _{3}}^{\nu _{1}}
\end{eqnarray*}%
and from the first two rows we obtain $\boldsymbol{a}_{2}=\left(
-a_{23},0,a_{23}\right) $ and $\boldsymbol{a}_{3}=\left(
-a_{33},0,a_{33}\right) $, the remaining must be compared with 
\begin{equation*}
T_{1\left( -\right) \mu _{32}}^{AV}=2i\varepsilon _{\mu _{23}\nu
_{12}}\left( P_{21}^{\nu _{2}}P_{32}^{\nu _{3}}+P_{31}^{\nu _{2}}\left(
P_{21}^{\nu _{3}}-P_{32}^{\nu _{3}}\right) -P_{32}^{\nu _{2}}P_{21}^{\nu
_{3}}\right) \Delta _{3\nu _{3}}^{\nu _{1}}
\end{equation*}%
giving%
\begin{equation*}
a_{11}+b_{3}=2i,\quad a_{12}=-2i,\quad a_{13}+b_{1}=2i,\quad b_{2}=0\quad
b_{3}=2i-b_{1}
\end{equation*}%
which rephrased in vector notation the full solution is%
\begin{equation}
\left( 
\begin{array}{c}
\boldsymbol{b} \\ 
\boldsymbol{a}_{1} \\ 
\boldsymbol{a}_{2} \\ 
\boldsymbol{a}_{3}%
\end{array}%
\right) =%
\begin{pmatrix}
b_{1} & 0 & 2i-b_{1} \\ 
b_{1} & -2i & 2i-b_{1} \\ 
-a_{23} & 0 & a_{23} \\ 
-a_{33} & 0 & a_{33}%
\end{pmatrix}%
.  \label{Sol1}
\end{equation}

First thing to note is the reduction from twelve parameters to just three,
here the constants $\left\{ a_{23},a_{33},b_{1}\right\} $, by requiring just
one of the relations to be satisfied.

If it is asked for any other relation to be satisfied, the solution will be
unique and it will have consequences over the last one. Repeating the
analysis to the contraction of $F_{\mu _{1}\mu _{2}\mu _{3}}^{\Delta }$ with 
$q_{2}^{\mu _{2}}=p_{21}^{\mu _{2}}=P_{32}^{\mu _{2}}-P_{31}^{\mu _{2}}$,
forming the system of linear equation by comparing $q_{2}^{\mu _{2}}F_{\mu
_{123}}^{\Delta }$ with $T_{2\left( -\right) \mu _{13}}^{AV}$follows the
complete solution for the automatic satisfaction of the RAGF born out of the
contraction with $q_{2}^{\mu _{2}}$%
\begin{equation}
\left( 
\begin{array}{c}
\boldsymbol{b} \\ 
\boldsymbol{a}_{1} \\ 
\boldsymbol{a}_{2} \\ 
\boldsymbol{a}_{3}%
\end{array}%
\right) =%
\begin{pmatrix}
0 & b_{2} & 2i-b_{2} \\ 
0 & -a_{13} & a_{13} \\ 
2i & -b_{2} & b_{2}-2i \\ 
0 & -a_{33} & a_{33}%
\end{pmatrix}%
.  \label{Sol2}
\end{equation}

The contraction with $q_{3}^{\mu _{3}}=p_{32}^{\mu _{3}}=P_{31}^{\mu
_{3}}-P_{21}^{\mu _{3}}$, allows us to determine the conditions for $%
q_{3}^{\mu _{3}}F_{\mu _{123}}^{\Delta }=T_{3\left( -\right) \mu _{12}}^{AV}$%
, and then emerges the solution to the automatic satisfaction of the RAGF,
product of the contraction with the index and momenta of the third vertex%
\begin{equation}
\left( 
\begin{array}{c}
\boldsymbol{b} \\ 
\boldsymbol{a}_{1} \\ 
\boldsymbol{a}_{2} \\ 
\boldsymbol{a}_{3}%
\end{array}%
\right) =%
\begin{pmatrix}
b_{1} & 2i-b_{1} & 0 \\ 
a_{11} & -a_{11} & 0 \\ 
a_{21} & -a_{21} & 0 \\ 
b_{1} & 2i-b_{1} & -2i%
\end{pmatrix}%
.  \label{Sol3}
\end{equation}

The intersection of the solution (\ref{Sol1}) and (\ref{Sol2}), that means,
the ones that satisfies automatically their RAGFs coming from the
contraction with $q_{1}^{\mu _{1}}$ and $q_{2}^{\mu _{2}}$, leads to a
unique solution with $b_{1}=0$, $b_{2}=0$, $b_{3}=2i$ and all the other
coefficients determined. Putting all that values in the tensor, we get%
\begin{equation*}
\left( F_{\mu _{123}}^{\Delta }\right) _{12}=-2i\left[ \varepsilon _{\mu
_{2}\mu _{3}\nu _{1}\nu _{2}}\left( P_{32}^{\nu _{2}}-P_{31}^{\nu
_{2}}\right) \Delta _{3\mu _{1}}^{\nu _{1}}+\varepsilon _{\mu _{1}\mu
_{3}\nu _{1}\nu _{2}}\left( P_{21}^{\nu _{2}}-P_{32}^{\nu _{2}}\right)
\Delta _{3\mu _{2}}^{\nu _{1}}+\varepsilon _{\mu _{1}\mu _{2}\mu _{3}\nu
_{1}}P_{32}^{\nu _{2}}\Delta _{3\nu _{2}}^{\nu _{1}}\right] ,
\end{equation*}%
where sub-index $ij$ in $\left( F_{\mu _{123}}^{\Delta }\right) _{ij}$
stands for the vertexes where the RAGFs are satisfied without further
assumptions.

Each of the solutions presented depends on three parameters and are
compatible with one another. However, once the coefficients are determined
to the unique and unrestricted satisfaction of two RAGFs, the third solution
will always exhibit an additional term, in other words%
\begin{equation*}
\left( F_{\mu _{123}}^{\Delta }\right) _{23}\not\equiv \left( F_{\mu
_{123}}^{\Delta }\right) _{13}\not\equiv \left( F_{\mu _{123}}^{\Delta
}\right) _{12}.
\end{equation*}

It is an easy task to verify that the solution obtained requiring at most
two RAGF satisfied corresponds to versions of the traces computed
explicitly, see (\ref{ST3}), 
\begin{equation*}
\left( F_{\mu _{123}}^{\Delta }\right) _{23}=S_{1\mu _{123}};\quad \left(
F_{\mu _{123}}^{\Delta }\right) _{13}=S_{2\mu _{123}};\quad \left( F_{\mu
_{123}}^{\Delta }\right) _{12}=S_{3\mu _{123}}
\end{equation*}

Owing to this behavior, whose hypotheses can be verified in the explicit
computation, if desired, it can be drawn a parallel conclusion to the one
stated in the section (\ref{LE4D}) where the value at zero of $PVV$ has
consequences over the symmetries, by example. Here this finite amplitude
will establish a connection between the linearity embodied in the RAGFs and
the low energy behavior of the same finite $PVV$, amplitude taken as example.

For such end, we have got to read this result in the light of the general
form factors in the eq. (\ref{GForm}). At this point, we will take the form
factors as representing only the strictly finite part of any tensor under
consideration.

Choosing the solution satisfying the RAGFs in the vertexes two and three%
\begin{equation*}
T_{\mu _{123}}^{\Gamma _{123}}=F_{\mu _{123}}+S_{1\mu _{123}}
\end{equation*}%
to any combination that the vertexes might assume, once again due to the
fact the two-point functions depend only on the vertex where we effect the
contraction. Contracting with the momenta of the respective vertexes, we
obtain%
\begin{eqnarray*}
q_{1}^{\mu _{1}}T_{\mu _{123}}^{\Gamma _{123}} &=&T_{1\left( -\right) \mu
_{23}}^{AV}+\varepsilon _{\mu _{2}\mu _{3}\nu _{1}\nu _{2}}q_{2}^{\nu
_{1}}q_{3}^{\nu _{2}}\left( V_{1}+2i\Delta _{3\alpha }^{\alpha }\right) \\
q_{2}^{\mu _{2}}T_{\mu _{123}}^{\Gamma _{123}} &=&T_{2\left( -\right) \mu
_{13}}^{AV}+\varepsilon _{\mu _{1}\mu _{3}\nu _{1}\nu _{2}}q_{2}^{\nu
_{1}}q_{3}^{\nu _{2}}V_{2} \\
q_{3}^{\mu _{3}}T_{\mu _{123}}^{\Gamma _{123}} &=&T_{3\left( -\right) \mu
_{12}}^{AV}+\varepsilon _{\mu _{1}\mu _{2}\nu _{1}\nu _{2}}q_{2}^{\nu
_{1}}q_{3}^{\nu _{2}}V_{3},
\end{eqnarray*}%
remember here the $V_{i}$ contains only dependence on the finite parts. The
trace of the surface term must be put together with the finite part of the
first contraction due to the equation obtained in (\ref{contS1})%
\begin{equation*}
q_{1}^{\mu _{1}}S_{1\mu _{123}}=T_{1\left( -\right) \mu
_{23}}^{AV}+\varepsilon _{\mu _{2}\mu _{3}\nu _{2}\nu _{3}}q_{2}^{\nu
_{2}}q_{3}^{\nu _{3}}\left( 2i\Delta _{3\nu _{1}}^{\nu _{1}}\right) .
\end{equation*}

Take an arbitrary sorting of vertexes that yields an odd tensor and let the $%
\Omega _{i}$ represent the scalar invariant that corresponds to the rank-two
tensor obtained in the RAGF, which are finite functions, and write the
equations that signify the hypothesis of satisfaction of the RAGF%
\begin{eqnarray*}
q_{1}^{\mu _{1}}T_{\mu _{123}}^{\Gamma _{123}} &=&T_{1\left( -\right) \mu
_{23}}^{AV}+\varepsilon _{\mu _{2}\mu _{3}\nu _{1}\nu _{2}}q_{2}^{\nu
_{1}}q_{3}^{\nu _{2}}\Omega _{1} \\
q_{2}^{\mu _{2}}T_{\mu _{123}}^{\Gamma _{123}} &=&T_{2\left( -\right) \mu
_{13}}^{AV}+\varepsilon _{\mu _{1}\mu _{3}\nu _{1}\nu _{2}}q_{2}^{\nu
_{1}}q_{3}^{\nu _{2}}\Omega _{2} \\
q_{3}^{\mu _{3}}T_{\mu _{123}}^{\Gamma _{123}} &=&T_{3\left( -\right) \mu
_{12}}^{AV}+\varepsilon _{\mu _{1}\mu _{2}\nu _{1}\nu _{2}}q_{2}^{\nu
_{1}}q_{3}^{\nu _{2}}\Omega _{3},
\end{eqnarray*}%
then look at the deduced behavior above where it is admitted that relation
two and three are unrestricted satisfied, from there we get the final
condition to our relations, namely 
\begin{equation*}
V_{1}+2i\Delta _{3\alpha }^{\alpha }=\Omega _{1};\quad V_{2}=\Omega
_{2};\quad V_{3}=\Omega _{3},
\end{equation*}%
remain to observe the formulas%
\begin{eqnarray*}
V_{1} &=&-F_{1}+F_{2}+q_{2}^{2}G_{5}+q_{3}^{2}G_{6}+\left( q_{2}\cdot
q_{3}\right) \left( G_{5}+G_{6}\right) \\
V_{2} &=&-F_{2}+q_{2}^{2}G_{3}+\left( q_{2}\cdot q_{3}\right) G_{4} \\
V_{3} &=&-F_{1}+\left( q_{2}\cdot q_{3}\right) G_{1}+q_{3}^{2}G_{2},
\end{eqnarray*}%
and eliminate the $F_{i}$ form factor to reach at%
\begin{eqnarray*}
&&-q_{2}^{2}\left( G_{3}+G_{5}\right) +q_{3}^{2}\left( G_{2}-G_{6}\right)
+\left( q_{2}\cdot q_{3}\right) \left( G_{1}-G_{4}-G_{5}-G_{6}\right) \\
&=&2i\Delta _{3\alpha }^{\alpha }+\Omega _{3}-\Omega _{2}-\Omega _{1}
\end{eqnarray*}%
now under the condition, that the $G_{i}$ functions are regular at zero%
\footnote{%
The functions $Z_{nm}^{\left( 0\right) },Z_{nm}^{\left( -1\right)
},Z_{n}^{\left( 0\right) }$ that comprise the finite part of any of these
amplitudes do not have kinematical singularities at the point $q_{i}\cdot
q_{j}=0$.}, follows the master equation%
\begin{equation}
-2i\Delta _{3\alpha }^{\alpha }=\Omega _{3}\left( 0\right) -\Omega
_{2}\left( 0\right) -\Omega _{1}\left( 0\right) ,  \label{ME}
\end{equation}%
and this trivially will happen irrespective of what set of RAGF is chosen to
be satisfied without restriction, suppose one starts with a version with $%
S_{2\mu _{123}}$ that satisfies the RAGFs in the first and third vertex, to
this tensor the term $\Delta _{3\alpha }^{\alpha }$ will appear in 
\begin{equation*}
q_{2}^{\mu _{2}}S_{2\mu _{123}}=T_{2\left( -\right) \mu
_{13}}^{AV}+\varepsilon _{\mu _{1}\mu _{3}\nu _{1}\nu _{2}}q_{2}^{\nu
_{1}}q_{3}^{\nu _{2}}\left( 2i\Delta _{3\alpha }^{\alpha }\right) ,
\end{equation*}%
a relation established in eq. (\ref{contS2}), then from $V_{1}=\Omega _{1}$
and $V_{3}=\Omega _{3}$ because the RAGFs are unrestricted by hypothesis, we
exchange the $F_{1}$ and $F_{2}$ by $G_{i}$ multiplied by bilinears plus
finite functions and then again in the point $q_{i}\cdot q_{j}=0$ we
retrieve the previous result.

Here is important to be aware of the fact that only hypotheses were
employed, a tensor that has two RAGFs satisfied without restriction, that
means connected to known differences of $AV$ functions and $PVV$/$PAA$-like
amplitudes without further ado, and from that point on the third is
necessarily bounded by the zero value of the finite rank-two amplitude. In
any case, this requirements are always possible to attain in the explicit
computations we have presented and shown how.

When assessing the values of $\Omega _{i}\left( 0\right) $, see the results (%
\ref{LEPVV}), $\Omega ^{PVV}=\Omega ^{VPV}=-\Omega ^{VVP}=\left( 2\pi
\right) ^{-2}$, we find out 
\begin{equation*}
\Omega _{3}\left( 0\right) -\Omega _{2}\left( 0\right) -\Omega _{1}\left(
0\right) =-\left( 2\pi \right) ^{-2},
\end{equation*}%
notice that for the $AVV$, $VAV$, and $VVA$ where only one axial current
appears, two of the $\Omega _{i}$ are zero to each amplitude or more
precisely the result above represents in reality three situations. The same
happens to the $AAA$ triangle, on the other hand in the case the three
contractions of the \textit{same} amplitude are related to $T_{\mu
_{23}}^{PAA},\quad T_{\mu _{13}}^{APA},\quad T_{\mu _{12}}^{AAP}$, thereby
it only a matter of combining the constants cast in eq. (\ref{LEPAA}), $%
\Omega ^{PAA}\left( 0\right) =\Omega ^{APA}\left( 0\right) =-\Omega
^{AAP}\left( 0\right) =\frac{1}{3\left( 2\pi \right) ^{2}}$, in other words%
\begin{equation*}
\Omega ^{AAP}\left( 0\right) -\Omega ^{APA}\left( 0\right) -\Omega
^{PAA}\left( 0\right) =-\left( 2\pi \right) ^{-2},
\end{equation*}%
it shows that because the differences of $AV$ structures depend only on the
contraction with the momenta, eq.'s(\ref{AV(-)1}, \ref{AV(-)2} and \ref%
{AV(-)3}), they are common to all cases analyzed, but as the correlators
with the $P$ density are finite but distinct, it could be (very unlikely)
that distinct diagrams would require different numerical values to the
surface term, notwithstanding what we get is%
\begin{equation*}
\text{RAGF}\Leftrightarrow \Delta _{3\alpha }^{\alpha }=-\frac{i}{2\left(
2\pi \right) ^{2}},
\end{equation*}%
always, and to amplitudes where three distinct masses run in the internal
lines, in that case, the vector currents are not classically conserved, the
constrain remains the same.

To see this, let us consider, as an example, that three arbitrary masses run
in the inner lines of the amplitude $AVV$, where the indexes of the
propagators now account for the masses too, $S\left( a\right) =\left( %
\slashed{K}_{a}-m_{a}\right) ^{-1}$, and then the standard identity $\slashed%
{p}_{ij}=S^{-1}\left( i\right) -S^{-1}\left( j\right) +\left(
m_{i}-m_{j}\right) $ is applied to derive the relations%
\begin{eqnarray*}
p_{31}^{\mu _{1}}t_{\mu _{123}}^{AVV} &=&t_{\mu _{32}}^{AV}\left( 1,2\right)
-t_{\mu _{23}}^{AV}\left( 2,3\right) -\left( m_{1}+m_{3}\right) t_{\mu
_{23}}^{PVV} \\
p_{21}^{\mu _{2}}t_{\mu _{123}}^{AVV} &=&t_{\mu _{13}}^{AV}\left( 1,3\right)
-t_{\mu _{13}}^{AV}\left( 2,3\right) +\left( m_{2}-m_{1}\right) t_{\mu
_{13}}^{ASV} \\
p_{32}^{\mu _{3}}t_{\mu _{123}}^{AVV} &=&t_{\mu _{12}}^{AV}\left( 1,2\right)
-t_{\mu _{12}}^{AV}\left( 1,3\right) +\left( m_{3}-m_{2}\right) t_{\mu
_{12}}^{AVS},
\end{eqnarray*}%
now the vector currents aren't conserved, however the $ASV$, $AVS$, and $PVV$
of arbitrary masses will not comply with the eq. (\ref{VertsZero}), repeated
here for clarity of the arguments, $V_{3}\left( 0\right) -V_{2}\left(
0\right) -V_{1}\left( 0\right) =0$.

Taking the integrals of the three-point rank-two finite amplitudes, we get%
\begin{eqnarray*}
T_{\mu _{23}}^{PVV} &=&\varepsilon _{\mu _{23}\nu _{12}}p_{21}^{\nu
_{1}}p_{32}^{\nu _{2}}\left[ \left( m_{1}-m_{2}\right) Z_{10}^{\left(
-1\right) }+\left( m_{1}-m_{3}\right) Z_{01}^{\left( -1\right)
}-m_{1}Z_{00}^{\left( -1\right) }\right] \\
T_{\mu _{13}}^{ASV} &=&\varepsilon _{\mu _{13}\nu _{12}}p_{21}^{\nu
_{1}}p_{32}^{\nu _{2}}\left[ \left( m_{1}+m_{2}\right) Z_{10}^{\left(
-1\right) }+\left( m_{1}+m_{3}\right) Z_{01}^{\left( -1\right)
}-m_{1}Z_{00}^{\left( -1\right) }\right] \\
T_{\mu _{12}}^{AVS} &=&\varepsilon _{\mu _{12}\nu _{12}}p_{21}^{\nu
_{1}}p_{32}^{\nu _{2}}\left[ \left( m_{2}-m_{1}\right) Z_{10}^{\left(
-1\right) }-\left( m_{3}+m_{1}\right) Z_{01}^{\left( -1\right)
}+m_{1}Z_{00}^{\left( -1\right) }\right] ,
\end{eqnarray*}%
identifying the form factor through%
\begin{eqnarray*}
\varepsilon _{\mu _{23}\nu _{12}}p_{21}^{\nu _{1}}p_{32}^{\nu _{2}}\Omega
_{1}^{PVV} &=&-\left( m_{1}+m_{3}\right) T_{\mu _{23}}^{PVV} \\
\varepsilon _{\mu _{13}\nu _{12}}p_{21}^{\nu _{1}}p_{32}^{\nu _{2}}\Omega
_{2}^{ASV} &=&+\left( m_{2}-m_{1}\right) T_{\mu _{13}}^{ASV} \\
\varepsilon _{\mu _{12}\nu _{12}}p_{21}^{\nu _{1}}p_{32}^{\nu _{2}}\Omega
_{3}^{AVS} &=&+\left( m_{3}-m_{2}\right) T_{\mu _{12}}^{AVS},
\end{eqnarray*}%
and combining them as dictated by eq. (\ref{VertsZero}), we have%
\begin{equation}
-\left( \Omega _{3}^{AVS}-\Omega _{2}^{ASV}-\Omega _{1}^{PVV}\right) =\frac{1%
}{2\pi ^{2}}\left[ \left( m_{1}^{2}-m_{2}^{2}\right) Z_{10}^{\left(
-1\right) }+\left( m_{1}^{2}-m_{3}^{2}\right) Z_{01}^{\left( -1\right)
}-m_{1}^{2}Z_{00}^{\left( -1\right) }\right]
\end{equation}%
in the kinematical limit $q_{i}\cdot q_{j}=0$, follows%
\begin{equation}
\left( \Omega _{3}^{AVS}-\Omega _{2}^{ASV}-\Omega _{1}^{PVV}\right) \left(
0\right) =-\left( 2\pi \right) ^{-2},
\end{equation}%
since in the definition for distinct masses we have\footnote{%
To arbitrary masses, the Feynman polynomial for the function involved in
this derivation, reads 
\begin{equation*}
Q=q_{1}^{2}x_{1}\left( 1-x_{1}\right) +q_{2}^{2}x_{2}\left( 1-x_{2}\right)
-2q_{1}\cdot q_{2}x_{1}x_{2}+\left( m_{1}^{2}-m_{2}^{2}\right) x_{1}+\left(
m_{1}^{2}-m_{3}^{2}\right) x_{2}-m_{1}^{2}
\end{equation*}%
and 
\begin{equation*}
Z_{rs}^{\left( -1\right) }=\int_{0}^{1}\mathrm{d}x_{1}\int_{0}^{1-x_{1}}%
\mathrm{d}x_{2}\frac{x_{1}^{r}x_{2}^{s}}{Q\left(
q_{i}^{2},m_{1}^{2},m_{2}^{2},m_{3}^{2}\right) }
\end{equation*}%
} $Q\left( 0\right) =\left( m_{1}^{2}-m_{2}^{2}\right) x_{1}+\left(
m_{1}^{2}-m_{3}^{2}\right) x_{2}-m_{1}^{2}$, hence%
\begin{equation*}
\left[ \left( m_{1}^{2}-m_{2}^{2}\right) Z_{10}^{\left( -1\right) }+\left(
m_{1}^{2}-m_{3}^{2}\right) Z_{01}^{\left( -1\right)
}-m_{1}^{2}Z_{00}^{\left( -1\right) }\right] _{q_{i}\cdot q_{j}=0}=1/2,
\end{equation*}%
the integrals with distinct masses are more laborious, but this result can
be obtained by integrating until the end all these functions in the limit
under consideration.

What this says is that the kinematical limits of all rank-two odd and finite
amplitudes are not compatible with the satisfaction of all Ward identities
as already noted in the $2D$-$AV$ amplitude. This is independent of
divergences or the particular aspects of perturbative solutions\footnote{%
With the exception these finite functions are perturbative solutions}, even
if a solution coming from a divergent-free formalism were provided, but
their contractions would have to be related to the finite tensors we have
described, there should be an anomaly.

\subsection{Parameter Landscape of the Violations\protect\footnote{%
Throughout this section we factored out the three-point rank-two finite
amplitudes from the discussion.} and Commentary}

From the consequences derived in the last section, we see that if the value
attributed to our tensors is done in a form that obeys the linearity of
integration, then the surface term present in their perturbative expressions
must be non-zero. Since its coefficients are not combinations of only
physical momenta, therefore choices must be made. As the choice dictates the
symmetry and vice-versa, we could be guided by them.

Talking about choices, by example, the eventual role of a regularization
argument in our strategy could be summarized schematically as: Consider a
regularization obeying $\lim_{\Lambda \rightarrow \infty }G\left( \Lambda
_{i},k^{2}\right) =1$, thus it yields tensors that are immune to the Dirac
traces, 
\begin{equation*}
\lim_{\Lambda \rightarrow \infty }\int \frac{\mathrm{d}^{4}k}{\left( 2\pi
\right) ^{4}}G\left( \Lambda _{i},k^{2}\right) \left( t_{\mu _{123}}^{\Gamma
_{123}}\right) _{r}=\lim_{\Lambda \rightarrow \infty }\int \frac{\mathrm{d}%
^{4}k}{\left( 2\pi \right) ^{4}}G\left( \Lambda _{i},k^{2}\right) \left(
t_{\mu _{123}}^{\Gamma _{123}}\right) _{s},
\end{equation*}%
if and only if $\Delta _{3\mu \nu }=-ig_{\mu \nu }/2\left( 4\pi \right) ^{2}$%
.

But, if in the process we get $\Delta _{3\mu \nu }=0$, then the linearity is
violated by $q_{i}^{\mu _{i}}\left[ T_{\mu _{123}}^{\Gamma _{123}}\right]
_{i}^{\mathrm{viol}}=\pm \left( 2\pi \right) ^{-2}\varepsilon _{\mu _{k}\mu
_{l}\nu _{1}\nu _{2}}q_{2}^{\nu _{1}}q_{3}^{\nu _{2}}$, see these results in
the eq.'s (\ref{ragfViol4D}), the sign is negative for $i=1,2$ and positive
for $i=3$, plus we arrange the terms with $k<l$. What comes next is
independent of this type of arguments, one example of rules to achieve this
is saw in \cite{Cynolter2011}.

Therefore, combining arbitrarily, and before integration, the versions that
saves the maximum number of RAGFs, observe that is clear in this point this
claim can be stated independent of Dirac traces computation, we have%
\begin{eqnarray}
\left[ T_{\mu _{123}}^{\Gamma _{123}}\right] _{\left\{
r_{1}r_{2}r_{3}\right\} } &=&\frac{1}{R}\left\{ r_{1}\left[ T_{\mu
_{123}}^{\Gamma _{123}}\right] _{1}+r_{2}\left[ T_{\mu _{123}}^{\Gamma
_{123}}\right] _{2}+r_{3}\left[ T_{\mu _{123}}^{\Gamma _{123}}\right]
_{3}\right\} \\
R &=&r_{1}+r_{2}+r_{3}\not=0,  \notag
\end{eqnarray}%
and as, the derivations about uniqueness in the end of section (\ref{unique}%
) showed, they are identically equal before integration, but when $\Delta
_{3\mu \nu }=0$, they become an infinity set of different tensors, obtained
by the same rules from the same integrand. For zero surface term, their
symmetry violations are 
\begin{equation*}
q_{1}^{\mu _{1}}\left[ T_{\mu _{123}}^{\Gamma _{123}}\right] _{\left\{
r_{1}r_{2}r_{3}\right\} }\sim -\frac{r_{1}}{R}\frac{1}{\left( 2\pi \right)
^{2}};\quad q_{2}^{\mu _{2}}\left[ T_{\mu _{123}}^{\Gamma _{123}}\right]
_{\left\{ r_{1}r_{2}r_{3}\right\} }\sim -\frac{r_{2}}{R}\frac{1}{\left( 2\pi
\right) ^{2}};\quad q_{3}^{\mu _{3}}\left[ T_{\mu _{123}}^{\Gamma _{123}}%
\right] _{\left\{ r_{1}r_{2}r_{3}\right\} }\sim \frac{r_{3}}{R}\frac{1}{%
\left( 2\pi \right) ^{2}}
\end{equation*}%
and satisfies the equation determined to its anomalies, $\mathcal{A}_{3}-%
\mathcal{A}_{2}-\mathcal{A}_{1}=\left( 2\pi \right) ^{-2}$, since the basic
versions satisfies it.

If the surface term is considered an arbitrary parameter, any result and why
they are such ones derives systematically from our strategy. Take the
surface term determined by an arbitrary constant $c_{1}$, see below, that is
equal one for the satisfaction of RAGFs, or zero for the momentum-space
translational invariance. Parametrize the internal lines chosing any of the
sums $P_{ij}=k_{i}+k_{j}$, since any other is determined by the differences
equal to the external momenta,%
\begin{eqnarray}
P_{31} &=&c_{2}q_{2}+c_{3}q_{3}\rightarrow P_{21}=c_{2}q_{2}+\left(
c_{3}-1\right) q_{3};\quad P_{32}=\left( c_{2}+1\right) q_{2}+c_{3}q_{3}, 
\notag \\
\text{and }\Delta _{3\mu _{1}\mu _{2}} &=&c_{1}\left[ -i/2\left( 4\pi
\right) ^{2}\right] g_{\mu _{1}\mu _{2}}.
\end{eqnarray}%
the $AV$ structures, see the derivations of section (\ref{unique}) and eq'.s
(\ref{AV(-)1}, \ref{AV(-)2}, and \ref{AV(-)3}), assume the value%
\begin{equation}
T_{1\left( -\right) \mu _{23}}^{AV}=\frac{c_{1}\left( c_{3}-c_{2}-2\right) }{%
4\left( 2\pi \right) ^{2}};\quad T_{2\left( -\right) \mu _{13}}^{AV}=-\frac{%
c_{1}\left( c_{3}+1\right) }{4\left( 2\pi \right) ^{2}};\quad T_{3\left(
-\right) \mu _{12}}^{AV}=-\frac{c_{1}\left( c_{2}-1\right) }{4\left( 2\pi
\right) ^{2}},
\end{equation}%
and the basic violations of the RAGFs, eq'.s (\ref{ragfViol4D}), turn into%
\begin{equation}
q_{1}^{\mu _{1}}\left[ T_{\mu _{123}}^{\Gamma _{123}}\right] _{1}^{\mathrm{%
ragf}}=\frac{\left( c_{1}-1\right) }{\left( 2\pi \right) ^{2}};\quad
q_{2}^{\mu _{2}}\left[ T_{\mu _{123}}^{\Gamma _{123}}\right] _{2}^{\mathrm{%
ragf}}=\frac{\left( c_{1}-1\right) }{\left( 2\pi \right) ^{2}};\quad
q_{3}^{\mu _{3}}\left[ T_{\mu _{123}}^{\Gamma _{123}}\right] _{3}^{\mathrm{%
ragf}}=-\frac{\left( c_{1}-1\right) }{\left( 2\pi \right) ^{2}},
\end{equation}%
all the terms are multiplied by the adequate tensor, $\varepsilon _{\mu
_{i}\mu _{j}\nu _{1}\nu _{2}}q_{2}^{\nu _{1}}q_{3}^{\nu _{2}}$ with $i<j$.

Therefore, this is enough to get the total set of possible values to the
contractions of the expression of the basic versions, with the caveat that
only the contraction where both the $AV$'s and the term that break of
linearity appear are the $i$-th version with $q_{i}^{\mu _{i}}$, to the $i$%
-th version a contraction with $q_{j}$, $j\not=i$, only the $AV$'s
contributes. The arbitrary linear combination of the versions will assume
the form%
\begin{eqnarray}
q_{1}^{\mu _{1}}\left[ T_{\mu _{123}}^{\Gamma _{123}}\right] _{\left\{
r_{1}r_{2}r_{3}\right\} } &=&\frac{1}{\left( 2\pi \right) ^{2}}\left\{ \frac{%
r_{1}\left( c_{1}-1\right) }{R}+\frac{1}{4}c_{1}\left( c_{3}-c_{2}-2\right)
\right\}  \label{genViol} \\
q_{2}^{\mu _{2}}\left[ T_{\mu _{123}}^{\Gamma _{123}}\right] _{\left\{
r_{1}r_{2}r_{3}\right\} } &=&\frac{1}{\left( 2\pi \right) ^{2}}\left\{ \frac{%
r_{2}\left( c_{1}-1\right) }{R}-\frac{1}{4}c_{1}\left( c_{3}+1\right)
\right\}  \notag \\
q_{3}^{\mu _{3}}\left[ T_{\mu _{123}}^{\Gamma _{123}}\right] _{\left\{
r_{1}r_{2}r_{3}\right\} } &=&\frac{1}{\left( 2\pi \right) ^{2}}\left\{ -%
\frac{r_{3}\left( c_{1}-1\right) }{R}-\frac{1}{4}c_{1}\left( c_{2}-1\right)
\right\} ,  \notag
\end{eqnarray}%
where in the first terms we have the contributions associated to the unique
perspective, as we posed, as the second ones representing routing choice.
Showing that the number of independent parameter is two%
\begin{eqnarray}
A &=&-\frac{1}{4}c_{1}\left( c_{3}+1\right) +\frac{r_{2}}{R}\left(
c_{1}-1\right) \\
B &=&-\frac{1}{4}c_{1}\left( c_{2}-1\right) -\frac{r_{3}}{R}\left(
c_{1}-1\right)
\end{eqnarray}%
because the first line is simply%
\begin{equation}
B-A-4=\frac{1}{4}c_{1}\left( c_{3}-c_{2}-2\right) +\frac{r_{1}}{R}\left(
c_{1}-1\right) ,
\end{equation}%
since, as we deduced in the former sections, when two violations are given,
no matter the path leading to them, a third is determined without ambiguity
of any nature.

If $c_{1}=1$, there is no dependence in $r_{i}$, we have the unique solution
that satisfies linearity but is momenta ambiguous. However, if $c_{1}=0$
will be no dependence in $c_{2}$ and $c_{3}$, that account for the internal
momenta ambiguities. This is the full range of possibilities and they comply
with the constraint determined only by utilizing finite integrals, eq. (\ref%
{Ans}) for the eventual arbitray anomalies. The crossed diagrams work in the
same form, and just add more parameters to the discussion, the break of
linearity, ambiguities, and unavoidable violation of symmetries in the way
we have described is independent of this feature.

The divergent character of the perturbative amplitudes is not the
determining factor of the anomaly phenomena, but yes, it plays a role. The
crucial factor is the kinematical behavior of finite, and as such
independent of interpretation, functions that code the amplitudes for the
pseudo-scalar density. Not demonstrated here is that this aspect gets
mutated, in the massless limit, in the values to the residue of poles of the
form factors, form factors that are regular in the massive case.

Breaking linearity has a function in the divergent amplitudes that
corroborates, and should be, with the finite $PV^{n}$ amplitude in dimension 
$d=2n$. As implicitly said, it rises from the situation where integrating
zero we obtain a non-zero result. To begin with, the unique surface-term
value will guarantee the following identity

\begin{eqnarray}
&&\left( \varepsilon _{\mu _{5123}}K_{1}^{\mu _{5}}K_{1\mu _{4}}+\varepsilon
_{\mu _{4512}}K_{1}^{\mu _{5}}K_{1\mu _{3}}+\varepsilon _{\mu
_{3451}}K_{1}^{\mu _{5}}K_{1\mu _{2}}+\varepsilon _{\mu _{2345}}K_{1}^{\mu
_{5}}K_{1\mu _{1}}\right) \frac{1}{D_{123}}+  \notag \\
+ &&\varepsilon _{\mu _{1234}}\frac{1}{D_{23}}+\varepsilon _{\mu _{1234}}%
\frac{m^{2}}{D_{123}}=0,
\end{eqnarray}%
in four dimensions. That formula comes from the fact that $\varepsilon
_{\lbrack \mu _{1234}}K_{1\mu _{5}]}=0$, that was multiplied by $K_{1}^{\mu
_{5}}/D_{123}$, and used $K_{1}^{2}=D_{1}+m^{2}$, it is a Schouten identity
to the integrand of bare Feynman integral $\bar{J}_{3\mu \nu }$. The
critical step arises when we separate, without commitment with a particular
interpretation to the divergences, the finite and divergent parts, 
\begin{eqnarray*}
\bar{J}_{2}\left( 2,3\right) &=&J_{2}\left( p_{32}\right) +I_{\log } \\
J_{3}\left( 1,2,3\right) &=&J_{3}\left( p_{21},p_{31}\right) \\
\bar{J}_{3\mu \nu }\left( 1,2,3\right) &=&J_{3\mu \nu }\left(
p_{21},p_{31}\right) +\left( \Delta _{3\mu \nu }+g_{\mu \nu }I_{\log
}\right) /4
\end{eqnarray*}%
and recombine them by means of $\varepsilon _{\lbrack \mu _{1235}}\Delta
_{3\mu _{5}]}^{\mu _{5}}=0$ and $\varepsilon _{\lbrack \mu _{1235}}J_{3\mu
_{5}]}^{\mu _{5}}=0$. Thus, although the identity for the surface term is
consistent to any value, constrained only by $\Delta _{3\mu \nu }=\left[
g^{\alpha \beta }\Delta _{3\alpha \beta }\right] /4$, the same is not true
to the bare integral $\bar{J}_{3\mu \nu }$, after followed the steps
indicated, the linearity-breaking phenomena arrive, 
\begin{equation}
\text{Eq. for }\bar{J}_{3\mu \nu }\text{ above}\rightarrow -\frac{1}{4}%
\varepsilon _{\mu _{1234}}\left[ \Delta _{3\mu _{5}}^{\mu _{5}}+2i/\left(
4\pi \right) ^{2}\right] =0,
\end{equation}%
in other words, as a part of the Feynman integrals the mere satisfaction of
the Schouten identity, which implies $\Delta _{3\mu \nu }=\left[ g^{\alpha
\beta }\Delta _{3\alpha \beta }\right] /4$, is not enough to make it valid
when the full integrals are brought into light. We have used, see eq. (\ref%
{trJ3}) in the appendix easily obtained with the formulas present there, the
result%
\begin{equation}
g^{\alpha \beta }J_{3\alpha \beta }=m^{2}J_{3}+J_{2}\left( p_{32}\right) +%
\frac{i}{2\left( 4\pi \right) ^{2}},
\end{equation}%
hence, the identity is respected only and only if 
\begin{equation}
\Delta _{3\mu _{5}}^{\mu _{5}}=-2i/\left( 4\pi \right) ^{2},
\end{equation}%
derived without ever manipulating a divergent integral.

We want to mention here the mere violation by an evanescent term as occur in
dimensional methods\footnote{%
See \cite{EliasMckeon1983}\cite{Chowdhury1986} for this type of view in
traditional approaches.}, does not spoils the paradigm of linearity
breaking, the finite value we are demonstrating to be necessary is not a
function of dimension and in truth it correponds to the low-energy limit of
the integral $J_{3}$, it can not be nullifyed in any limit and if not
adopted violates the linearity and uniqueness of the perturbative amplitudes.

There is no more space to show the clear effect of what we claimed here, but
as a title of comment, when one tries to establish the RAGF in the explicit
traces, take by example the first version, before integration, for the
contractions $p_{21}$ and $p_{32}$ no question arises, it is as if no trace
need to be explicited at all, the startling thing is that in the contraction
with $p_{31}$ it is possible to organize the integrand in such a way the
RAGF follows from the identity that we started this final discussion.
Thereby, the condition to the validity of RAGFs is the validity, in general,
of the vanishing of a total antisymmetric tensor of rank five or bigger,
independent of its nature.

\section{Six Dimensional Box Amplitude $AVVV$}

\label{6Dim4Pt}To complete our line of reasoning, let us make an exercise of
our notation in two versions of the $AVVV$ box. The sub-amplitudes for this
tensor in six dimensions require vertexes not present in the model, the
tensor and pseudo-tensor vertexes 
\begin{equation*}
T=\gamma _{\lbrack \alpha _{12}]}=\frac{1}{2}\left[ \gamma _{\alpha
_{1}},\gamma _{\alpha _{2}}\right] ;\text{ and}\quad \tilde{T}=\gamma _{\ast
}\gamma _{\lbrack \alpha _{12}]}.
\end{equation*}

Making the choices $\Gamma _{1}=\gamma _{\ast }\gamma _{\mu _{1}}$ and $%
\Gamma _{i}=\gamma _{\mu _{i}}$ for $i=2,3,4$, and keeping the non-zero
terms we have%
\begin{eqnarray*}
t_{\mu _{1234}}^{AVVV} &=&K_{1234}^{\nu _{1234}}\mathrm{tr}(\gamma _{\ast
\mu _{1}\nu _{1}\mu _{2}\nu _{2}\mu _{3}\nu _{3}\mu _{4}\nu _{4}})\frac{1}{%
D_{1234}} \\
&&+m^{2}\mathrm{tr}(\gamma _{\ast \mu _{1}\mu _{2}\mu _{3}\mu _{4}\nu
_{1}\nu _{2}})\left( -K_{12}^{\nu _{12}}+K_{13}^{\nu _{12}}-K_{14}^{\nu
_{12}}-K_{23}^{\nu _{12}}+K_{24}^{\nu _{12}}-K_{34}^{\nu _{12}}\right) \frac{%
1}{D_{1234}}.
\end{eqnarray*}%
This tensor satisfies a class of relations, its RAGF, through the standard
procedure outline in section (\ref{ModlDef}) 
\begin{eqnarray*}
p_{21}^{\mu _{2}}t_{\mu _{1234}}^{AVVV} &=&t_{\mu _{134}}^{AVV}\left(
1,3,4\right) -t_{\mu _{134}}^{AVV}\left( 2,3,4\right) \\
p_{32}^{\mu _{3}}t_{\mu _{1234}}^{AVVV} &=&t_{\mu _{124}}^{AVV}\left(
1,2,4\right) -t_{\mu _{124}}^{AVV}\left( 1,3,4\right) \\
p_{43}^{\mu _{4}}t_{\mu _{1234}}^{AVVV} &=&t_{\mu _{123}}^{AVV}\left(
1,2,3\right) -t_{\mu _{123}}^{AVV}\left( 1,2,4\right) \\
p_{41}^{\mu _{1}}t_{\mu _{1234}}^{AVVV} &=&t_{\mu _{423}}^{AVV}\left(
1,2,3\right) -t_{\mu _{234}}^{AVV}\left( 2,3,4\right) -2mt_{\mu
_{234}}^{PVVV}.
\end{eqnarray*}%
The integrated three point functions assume the setting%
\begin{equation}
T_{\mu _{1}\mu _{2}\mu _{3}}^{AVV}\left( i,j,l\right) =\frac{4}{3}%
\varepsilon _{\mu _{123}\nu _{123}}p_{ji}^{\nu _{2}}p_{li}^{\nu _{3}}\left(
P_{ij}+P_{il}+P_{jl}\right) ^{\nu _{4}}\Delta _{\nu _{4}}^{\nu _{1}}
\end{equation}%
and the finite box arising from the contraction with axial-vertex momentum
is 
\begin{equation}
T_{\mu _{234}}^{PVVV}=-8m\varepsilon _{\mu _{234}\nu _{123}}p_{21}^{\nu
_{1}}p_{32}^{\nu _{2}}p_{42}^{\nu _{3}}J_{4},
\end{equation}%
where $J_{4}$ has the property%
\begin{equation*}
J_{4}\left( p_{ij}^{2}=0\right) =-\frac{i}{3!\left( 4\pi \right) ^{3}m^{2}}.
\end{equation*}

To start with, one can use the general formula in the dimension $d=2n$, to
the string of $2n+2$ gamma matrices plus $\gamma _{\ast }$ using the
definition, as follows%
\begin{equation*}
\mathrm{tr}(\gamma _{\ast }\gamma _{a_{1}a_{2}\cdots a_{2n+1}a_{2n+2}})=%
\mathrm{tr}\left( \mathbf{1}_{2^{n}\times 2^{n}}\right)
i^{3n-1}\sum_{k=1}^{2n+1}\sum_{j=k+1}^{2n+2}\left( -1\right)
^{j+k+1}g_{a_{k}a_{j}}\varepsilon _{a_{1}\cdots \hat{a}_{k}\cdots \hat{a}%
_{j}\cdots (2n+2)},
\end{equation*}%
where the hat means omission of that indexes, the formula was indexed in
such a way to facilitate its use by means of substitutions, effectively
encompassing all the results present in this paper. The first version, that
comes from the substitution adjacent to the matrix $\gamma _{\mu _{1}}$, is
obtained through the choices of $n=3$ and the sequence $a_{1}a_{2}\cdots
a_{8}$ as $\mu _{1}\nu _{1}\mu _{2}\nu _{2}\mu _{3}\nu _{3}\mu _{4}\nu _{4}$.

\textbf{First Version: }The sign tensors came from the first four rows of
the trace of eight matrices below%
\begin{eqnarray*}
&&\frac{1}{8}K_{1234}^{\nu _{1234}}\mathrm{tr}(\gamma _{\ast }\gamma _{\mu
_{1}\nu _{1}\mu _{2}\nu _{2}\mu _{3}\nu _{3}\mu _{4}\nu _{4}}) \\
&=&-\varepsilon _{\mu _{2}\mu _{3}\mu _{4}\nu _{123}}\left[ K_{234}^{\nu
_{123}}K_{1\mu _{1}}-K_{134}^{\nu _{123}}K_{2\mu _{1}}+K_{124}^{\nu
_{123}}K_{3\mu _{1}}-K_{123}^{\nu _{123}}K_{4\mu _{1}}\right] \\
&&-\varepsilon _{\mu _{1}\mu _{3}\mu _{4}\nu _{123}}\left[ K_{234}^{\nu
_{123}}K_{1\mu _{2}}+K_{134}^{\nu _{123}}K_{2\mu _{2}}-K_{124}^{\nu
_{123}}K_{3\mu _{2}}+K_{123}^{\nu _{123}}K_{4\mu _{2}}\right] \\
&&+\varepsilon _{\mu _{1}\mu _{2}\mu _{4}\nu _{123}}\left[ K_{234}^{\nu
_{123}}K_{1\mu _{3}}-K_{134}^{\nu _{123}}K_{2\mu _{3}}-K_{124}^{\nu
_{123}}K_{3\mu _{3}}+K_{123}^{\nu _{123}}K_{4\mu _{3}}\right] \\
&&-\varepsilon _{\mu _{1}\mu _{2}\mu _{3}\nu _{123}}\left[ K_{234}^{\nu
_{123}}K_{1\mu _{4}}-K_{134}^{\nu _{123}}K_{2\mu _{4}}+K_{124}^{\nu
_{123}}K_{3\mu _{4}}+K_{123}^{\nu _{123}}K_{4\mu _{4}}\right] \\
&&+\varepsilon _{\mu _{1}\mu _{2}\mu _{3}\mu _{4}\nu _{1}\nu _{2}}\left[
K_{34}^{\nu _{12}}\left( K_{1}\cdot K_{2}\right) -K_{24}^{\nu _{12}}\left(
K_{1}\cdot K_{3}\right) +K_{23}^{\nu _{12}}\left( K_{1}\cdot K_{4}\right) %
\right] \\
&&+\varepsilon _{\mu _{1}\mu _{2}\mu _{3}\mu _{4}\nu _{1}\nu _{2}}\left[
K_{14}^{\nu _{12}}\left( K_{2}\cdot K_{3}\right) -K_{13}^{\nu _{12}}\left(
K_{2}\cdot K_{4}\right) +K_{12}^{\nu _{12}}\left( K_{3}\cdot K_{4}\right) %
\right] \\
&&+\left[ g_{\mu _{1}\mu _{2}}\varepsilon _{\mu _{3}\mu _{4}\nu _{1}\nu
_{2}\nu _{3}\nu _{4}}-g_{\mu _{1}\mu _{3}}\varepsilon _{\mu _{2}\mu _{4}\nu
_{1}\nu _{2}\nu _{3}\nu _{4}}+g_{\mu _{1}\mu _{4}}\varepsilon _{\mu _{2}\mu
_{3}\nu _{1}\nu _{2}\nu _{3}\nu _{4}}\right] K_{1234}^{\nu _{1234}} \\
&&+\left[ g_{\mu _{2}\mu _{3}}\varepsilon _{\mu _{1}\mu _{4}\nu _{1}\nu
_{2}\nu _{3}\nu _{4}}-g_{\mu _{2}\mu _{4}}\varepsilon _{\mu _{1}\mu _{3}\nu
_{1}\nu _{2}\nu _{3}\nu _{4}}+g_{\mu _{3}\mu _{4}}\varepsilon _{\mu _{1}\mu
_{2}\nu _{1}\nu _{2}\nu _{3}\nu _{4}}\right] K_{1234}^{\nu _{1234}},
\end{eqnarray*}%
the general form of that tensors is given by%
\begin{equation*}
\varepsilon _{\mu _{abc}\nu _{123}}t_{\mu _{d}}^{\left(
s_{1}s_{2}s_{3}\right) \nu _{123}}=\varepsilon _{\mu _{abc}\nu _{123}}\left[
K_{1\mu _{d}}K_{234}^{\nu _{123}}+s_{1}K_{2\mu _{d}}K_{134}^{\nu
_{123}}+s_{2}K_{3\mu _{d}}K_{124}^{\nu _{123}}+s_{3}K_{4\mu
_{d}}K_{123}^{\nu _{123}}\right] \frac{1}{D_{1234}},
\end{equation*}%
and their integrals, following our procedure, are 
\begin{eqnarray*}
\varepsilon _{\mu _{abc}\nu _{123}}T_{\mu _{d}}^{\left(
s_{1}s_{2}s_{3}\right) \nu _{123}} &=&\left( 1+s_{1}\right) \varepsilon
_{\mu _{abc}\nu _{123}}p_{31}^{\nu _{2}}p_{41}^{\nu _{3}}\left( J_{4\mu
_{d}}^{\nu _{1}}+p_{21\mu _{d}}J_{4}^{\nu _{1}}\right) \\
&&-\left( 1-s_{2}\right) \varepsilon _{\mu _{abc}\nu _{123}}p_{21}^{\nu
_{2}}p_{41}^{\nu _{3}}\left( J_{4\mu _{d}}^{\nu _{1}}+p_{31\mu
_{d}}J_{4}^{\nu _{1}}\right) \\
&&+\left( 1+s_{3}\right) \varepsilon _{\mu _{abc}\nu _{123}}p_{21}^{\nu
_{2}}p_{31}^{\nu _{3}}\left( J_{4\mu _{d}}^{\nu _{1}}+p_{41\mu
_{d}}J_{4}^{\nu _{1}}\right) \\
&&+\frac{1}{6}\varepsilon _{\mu _{abc}\nu _{123}}\left[ \left(
1+s_{1}\right) p_{31}^{\nu _{2}}p_{41}^{\nu _{3}}-\left( 1-s_{2}\right)
p_{21}^{\nu _{2}}p_{41}^{\nu _{3}}+\left( 1+s_{3}\right) p_{21}^{\nu
_{2}}p_{31}^{\nu _{3}}\right] \Delta _{4\mu _{d}}^{\nu _{1}} \\
&&+\frac{1}{6}\varepsilon _{\mu _{abcd}\nu _{12}}\left[ \left(
1+s_{1}\right) p_{31}^{\nu _{1}}p_{41}^{\nu _{2}}-\left( 1-s_{2}\right)
p_{21}^{\nu _{1}}p_{41}^{\nu _{2}}+\left( 1+s_{3}\right) p_{21}^{\nu
_{1}}p_{31}^{\nu _{2}}\right] I_{\log },
\end{eqnarray*}%
a general property to note is that one of the tensors is finite and zero%
\begin{equation*}
\varepsilon _{\mu _{abc}\nu _{123}}T_{\mu _{d}}^{\left( -+-\right) \nu
_{123}}=0,
\end{equation*}%
and as always some odd and rank-two sub-amplitudes are finite and vanishing,
all equal to\textit{\ }$\pm T_{\mu \nu }^{AVPP}$, as in $4D$ is the case of
the $T_{\mu }^{ASS}$ like terms.

To the first version, we obtain the result%
\begin{equation*}
\left( T_{\mu _{1234}}^{AVVV}\right) _{1}=-8\varepsilon _{\mu _{134}\nu
_{123}}T_{\mu _{2}}^{\left( +-+\right) \nu _{123}}+8\varepsilon _{\mu
_{124}\nu _{123}}T_{\mu _{3}}^{\left( --+\right) \nu _{123}}-8\varepsilon
_{\mu _{123}\nu _{123}}T_{\mu _{4}}^{\left( -++\right) \nu _{123}}-\frac{1}{2%
}\varepsilon _{\mu _{1234}}^{\qquad \nu _{12}}T_{\nu _{12}}^{\tilde{T}PPP},
\end{equation*}%
where the sub-amplitude, resulting of sum of the bilinear terms above with
the mass terms, coming from the trace $\mathrm{tr}(\gamma _{\ast }\gamma
_{\mu _{1234}\nu _{12}})=8\varepsilon _{\mu _{1234}\nu _{12}}$, has the
integrand 
\begin{eqnarray*}
\frac{1}{2}\varepsilon _{\mu _{1234}}^{\qquad \nu _{12}}t_{\nu _{12}}^{%
\tilde{T}PPP} &=&+8\varepsilon _{\mu _{1234}\nu _{12}}\left[ -K_{12}^{\nu
_{12}}S_{34}+K_{13}^{\nu _{12}}S_{24}-K_{14}^{\nu _{12}}S_{23}\right] \frac{1%
}{D_{1234}} \\
&&+8\varepsilon _{\mu _{1234}\nu _{12}}\left[ -K_{23}^{\nu
_{12}}S_{14}+K_{24}^{\nu _{12}}S_{13}-K_{34}^{\nu _{12}}S_{12}\right] \frac{1%
}{D_{1234}},
\end{eqnarray*}
its finite part is given by 
\begin{eqnarray*}
-\frac{1}{2}\varepsilon _{\mu _{1234}}^{\qquad \nu _{12}}T_{\nu _{12}}^{%
\tilde{T}PPP} &=&8\varepsilon _{\mu _{1234}\nu _{12}}\left[ \left(
p_{31}\cdot p_{43}\right) p_{21}^{\nu _{2}}-\left( p_{21}\cdot p_{42}\right)
p_{31}^{\nu _{2}}+\left( p_{21}\cdot p_{32}\right) p_{41}^{\nu _{2}}\right]
J_{4}^{\nu _{1}} \\
&&+4\varepsilon _{\mu _{1234}\nu _{12}}\left( p_{21}^{\nu _{1}}p_{41}^{\nu
_{2}}p_{31}^{2}-p_{31}^{\nu _{1}}p_{41}^{\nu _{2}}p_{21}^{2}-p_{21}^{\nu
_{1}}p_{31}^{\nu _{2}}p_{41}^{2}\right) J_{4} \\
&&+4\varepsilon _{\mu _{1234}\nu _{12}}\left[ 2p_{43}^{\nu _{2}}J_{3}^{\nu
_{1}}\left( p_{31},p_{41}\right) +p_{31}^{\nu _{1}}p_{41}^{\nu
_{2}}J_{3}\left( p_{31},p_{41}\right) \right] \\
&&+4\varepsilon _{\mu _{1234}\nu _{12}}\left[ 2p_{21}^{\nu _{2}}J_{3}^{\nu
_{1}}\left( p_{21},p_{41}\right) -p_{21}^{\nu _{1}}p_{41}^{\nu
_{2}}J_{3}\left( p_{21},p_{41}\right) \right] \\
&&+4\varepsilon _{\mu _{1234}\nu _{12}}\left[ p_{32}^{\nu _{1}}p_{43}^{\nu
_{2}}J_{3}\left( p_{32},p_{42}\right) +p_{21}^{\nu _{1}}p_{31}^{\nu
_{2}}J_{3}\left( p_{21},p_{31}\right) \right] ,
\end{eqnarray*}%
note the fact that this vertex is a pseudo-tensor one, as we anticipated.
Gathering all the divergent objects that appear in the sub-structure and the
combination of sign tensors, we get the set of surface terms 
\begin{eqnarray*}
S_{1\mu _{1234}} &=&-\frac{8}{3}\left[ \varepsilon _{\mu _{134}\nu
_{123}}p_{32}^{\nu _{2}}p_{42}^{\nu _{3}}\Delta _{4\mu _{2}}^{\nu
_{1}}+\varepsilon _{\mu _{124}\nu _{123}}p_{21}^{\nu _{2}}p_{43}^{\nu
_{3}}\Delta _{4\mu _{3}}^{\nu _{1}}+\varepsilon _{\mu _{123}\nu
_{123}}p_{21}^{\nu _{2}}p_{31}^{\nu _{3}}\Delta _{4\mu _{4}}^{\nu _{1}}%
\right] \\
&&-\frac{8}{6}\varepsilon _{\mu _{1234}\nu _{12}}\left[ p_{43}^{\nu
_{2}}\left( P_{13}+P_{34}+P_{14}\right) ^{\nu _{4}}+p_{21}^{\nu _{2}}\left(
P_{12}+P_{24}+P_{14}\right) ^{\nu _{4}}\right] \Delta _{4\nu _{4}}^{\nu _{1}}
\end{eqnarray*}%
meanwhile the irreducible object cancels exactly. With the tools, we have
been using up to now, these results are a quite direct, but a long, task.

\textbf{Second Version:} Without any new protocol, only computing the trace
with the definition of $\gamma _{\ast }$ in the right or left of $\gamma
_{\mu _{2}}$, and following exactly the same steps for two, four, and that
first case in six dimensions, we have the formula%
\begin{equation*}
\left( T_{\mu _{1234}}^{AVVV}\right) _{2}=-8\varepsilon _{\mu _{234}\nu
_{123}}T_{\mu _{1}}^{\left( +-+\right) \nu _{123}}-8\varepsilon _{\mu
_{124}\nu _{123}}T_{\mu _{3}}^{\left( ++-\right) \nu _{123}}+8\varepsilon
_{\mu _{123}\nu _{123}}T_{\mu _{4}}^{\left( +--\right) \nu _{123}}-\frac{1}{2%
}\varepsilon _{\mu _{1234}}^{\qquad \nu _{12}}T_{\nu _{1}\nu _{2}}^{STPP},
\end{equation*}%
here the sub-amplitude is $STPP$, where we see another vertex a two-rank
tensor one, its finite part is 
\begin{eqnarray*}
&&-\frac{1}{2}\varepsilon _{\mu _{1234}}^{\qquad \nu _{12}}T_{\nu _{1}\nu
_{2}}^{STPP} \\
&=&8\varepsilon _{\mu _{1234}\nu _{12}}\left[ -\left( p_{41}\cdot
p_{43}\right) p_{21}^{\nu _{2}}+\left( p_{41}\cdot p_{42}\right) p_{31}^{\nu
_{2}}-\left( p_{31}\cdot p_{32}\right) p_{41}^{\nu _{2}}\right] J_{4}^{\nu
_{1}} \\
&&+4\varepsilon _{\mu _{1234}\nu _{12}}\left[ p_{31}^{\nu _{1}}p_{43}^{\nu
_{2}}p_{21}^{2}-p_{21}^{\nu _{1}}p_{42}^{\nu _{2}}p_{31}^{2}+p_{21}^{\nu
_{1}}p_{32}^{\nu _{2}}p_{41}^{2}-4m^{2}p_{32}^{\nu _{1}}p_{42}^{\nu _{2}}%
\right] J_{4} \\
&&+8\varepsilon _{\mu _{1234}\nu _{12}}\left[ p_{41}^{\nu _{2}}J_{3}^{\nu
_{1}}\left( p_{21},p_{41}\right) -p_{32}^{\nu _{2}}J_{3}^{\nu _{1}}\left(
p_{21},p_{31}\right) \right] -4\varepsilon _{\mu _{1234}\nu
_{12}}p_{32}^{\nu _{1}}p_{43}^{\nu _{2}}J_{3}\left( p_{42},p_{43}\right) \\
&&+4\varepsilon _{\mu _{1234}\nu _{12}}\left[ -p_{31}^{\nu _{1}}p_{43}^{\nu
_{2}}J_{3}\left( p_{31},p_{41}\right) +p_{21}^{\nu _{1}}p_{42}^{\nu
_{2}}J_{3}\left( p_{21},p_{41}\right) -p_{21}^{\nu _{1}}p_{32}^{\nu
_{2}}J_{3}\left( p_{21},p_{31}\right) \right] .
\end{eqnarray*}%
The divergent terms are summed between the two sectors, and provide us with%
\begin{eqnarray*}
S_{2\mu _{1234}} &=&+\frac{8}{3}\left[ -\varepsilon _{\mu _{234}\nu
_{123}}p_{32}^{\nu _{2}}p_{43}^{\nu _{3}}\Delta _{4\mu _{1}}^{\nu
_{1}}-\varepsilon _{\mu _{124}\nu _{123}}p_{31}^{\nu _{2}}p_{41}^{\nu
_{3}}\Delta _{4\mu _{3}}^{\nu _{1}}+\varepsilon _{\mu _{123}\nu
_{123}}p_{32}^{\nu _{2}}p_{41}^{\nu _{3}}\Delta _{4\mu _{4}}^{\nu _{1}}%
\right] \\
&&+\frac{8}{3}\varepsilon _{\mu _{1234}\nu _{12}}\left[ \left( k_{1}^{\nu
_{3}}+k_{2}^{\nu _{3}}+k_{3}^{\nu _{3}}\right) p_{32}^{\nu _{2}}-\left(
k_{1}^{\nu _{3}}+k_{2}^{\nu _{3}}+k_{4}^{\nu _{3}}\right) p_{41}^{\nu _{2}}%
\right] \Delta _{4\nu _{3}}^{\nu _{1}},
\end{eqnarray*}%
owing this elements the computations follow in the same vein as the first
version and all the other scenarios we presented.

\subsection{RAGF, Linearity and Uniqueness}

\textbf{RAGF: First Version. }For the contraction associated to the first
vector vertex, the divergent terms obey%
\begin{equation*}
p_{21}^{\mu _{2}}\left( T_{\mu _{1234}}^{AVVV}\right) _{1}=T_{\mu _{1}\mu
_{3}\mu _{4}}^{AVV}\left( 1,3,4\right) -T_{\mu _{1}\mu _{3}\mu
_{4}}^{AVV}\left( 2,3,4\right) ,
\end{equation*}%
and to the finite part, it is necessary to use identities\footnote{%
One exemple is 
\begin{equation*}
\varepsilon _{\mu _{134}\nu _{123}}\left[ p_{31}^{\nu _{2}}p_{41}^{\nu
_{3}}p_{21}^{2}-\left( p_{21}\cdot p_{41}\right) p_{21}^{\nu
_{3}}p_{31}^{\nu _{2}}+\left( p_{21}\cdot p_{31}\right) p_{21}^{\nu
_{3}}p_{41}^{\nu _{2}}\right] J_{4}^{\nu _{1}}=\varepsilon _{\mu _{134}\nu
_{234}}p_{31}^{\nu _{2}}p_{41}^{\nu _{3}}p_{21}^{\nu _{4}}\left( p_{21\nu
_{1}}J_{4}^{\nu _{1}}\right)
\end{equation*}%
} to exchange the position of indexes as done in $4D$ and the results in the
appendix (\ref{AppInt6D}), to determine that it obeys $p_{21}^{\mu
_{2}}\left( T_{\mu _{1234}}^{AVVV}\right) _{1}^{\mathrm{finite}}=0$, and it
is quite direct to verify that the part of the surface terms comply exactly
with 
\begin{eqnarray*}
p_{32}^{\mu _{3}}\left( T_{\mu _{1234}}^{AVVV}\right) _{1} &=&T_{\mu _{1}\mu
_{2}\mu _{4}}^{AVV}\left( 1,2,4\right) -T_{\mu _{1}\mu _{2}\mu
_{4}}^{AVV}\left( 1,3,4\right) \\
p_{43}^{\mu _{4}}\left( T_{\mu _{1234}}^{AVVV}\right) _{1} &=&T_{\mu _{1}\mu
_{2}\mu _{3}}^{AVV}\left( 1,2,3\right) -T_{\mu _{1}\mu _{2}\mu
_{3}}^{AVV}\left( 1,2,4\right) .
\end{eqnarray*}

For the contraction with the momentum corresponding to the axial vertex, we
must find among other things the expression%
\begin{equation}
T_{\mu _{4}\mu _{2}\mu _{3}}^{AVV}\left( 1,2,3\right) -T_{\mu _{2}\mu
_{3}\mu _{4}}^{AVV}\left( 2,3,4\right) =\frac{8}{6}\varepsilon _{\mu
_{234}\nu _{123}}p_{32}^{\nu _{2}}\left[ 2p_{21}^{\nu _{3}}p_{41}^{\nu
_{4}}-p_{41}^{\nu _{3}}\left( P_{23}+P_{24}+P_{43}\right) ^{\nu _{4}}\right]
\Delta _{4\nu _{4}}^{\nu _{1}},
\end{equation}%
but in the contraction of the computed four-point function, as expected, we
must use 
\begin{eqnarray*}
&&p_{21}^{\nu _{2}}p_{31}^{\nu _{3}}p_{41}^{\nu _{4}}\left[ -\varepsilon
_{\mu _{34}\nu _{1234}}\Delta _{4\mu _{2}}^{\nu _{1}}+\varepsilon _{\mu
_{24}\nu _{1234}}\Delta _{4\mu _{3}}^{\nu _{1}}-\varepsilon _{\mu _{23}\nu
_{1234}}\Delta _{4\mu _{4}}^{\nu _{1}}\right] \\
&=&-\varepsilon _{\mu _{234}\nu _{234}}p_{41}^{\nu _{4}}p_{32}^{\nu
_{2}}p_{42}^{\nu _{3}}\Delta _{4\rho }^{\rho }+\varepsilon _{\mu _{234}\nu
_{123}}\left[ -p_{41}^{\nu _{3}}p_{32}^{\nu _{2}}p_{42}^{\nu
_{4}}+p_{41}^{\nu _{3}}p_{42}^{\nu _{2}}p_{32}^{\nu _{4}}+p_{32}^{\nu
_{2}}p_{42}^{\nu _{3}}p_{41}^{\nu _{4}}\right] \Delta _{4\nu _{4}}^{\nu
_{1}},
\end{eqnarray*}%
that after added to the other terms of the expression, with some trivial
algebraic manipulations to find the configuration of $P_{ij}$ in the
expression of the $AVV$ parts, we find out 
\begin{equation*}
p_{41}^{\mu _{1}}\left( T_{\mu _{1234}}^{AVVV}\right) _{1}=T_{\mu _{4}\mu
_{2}\mu _{3}}^{AVV}\left( 1,2,3\right) -T_{\mu _{2}\mu _{3}\mu
_{4}}^{AVV}\left( 2,3,4\right) +\frac{8}{3}\varepsilon _{\mu _{234}\nu
_{234}}p_{41}^{\nu _{4}}p_{32}^{\nu _{2}}p_{42}^{\nu _{3}}\Delta _{4\rho
}^{\rho },
\end{equation*}%
where we see an additional term depending on the external momenta and the
surface's term trace. On the other hand, to the finite part we use using the
identity\footnote{%
The specific substitution necessary is%
\begin{equation*}
\varepsilon _{\mu _{34}\nu _{1234}}J_{4\mu _{2}}^{\nu _{1}}-\varepsilon
_{\mu _{24}\nu _{1234}}J_{4\mu _{3}}^{\nu _{1}}+\varepsilon _{\mu _{23}\nu
_{1234}}J_{4\mu _{4}}^{\nu _{1}}=-\varepsilon _{\mu _{234}\nu _{123}}J_{4\nu
_{4}}^{\nu _{1}}+\varepsilon _{\mu _{234}\nu _{124}}J_{4\nu _{3}}^{\nu
_{1}}-\varepsilon _{\mu _{234}\nu _{134}}J_{4\nu _{2}}^{\nu
_{1}}+\varepsilon _{\mu _{234}\nu _{234}}J_{4\nu _{1}}^{\nu _{1}}
\end{equation*}%
} $\varepsilon _{\lbrack \mu _{34}\nu _{1234}}J_{4\mu _{2}]}^{\nu _{1}}=0$
to exchange indexes in search of contractions with the momenta, in addition
to\footnote{%
When multiplied by $p_{21}^{\nu _{2}}p_{31}^{\nu _{3}}p_{41}^{\nu
_{4}}J_{4}^{\nu _{1}}$ the desired result takes the form of 
\begin{equation*}
\varepsilon _{\mu _{234}\nu _{123}}\left[ p_{21}^{\nu _{2}}p_{31}^{\nu
_{3}}p_{41}^{2}-\left( p_{41}\cdot p_{31}\right) p_{21}^{\nu
_{2}}p_{41}^{\nu _{3}}+\left( p_{41}\cdot p_{21}\right) p_{31}^{\nu
_{2}}p_{41}^{\nu _{3}}\right] J_{4}^{\nu _{1}}=\varepsilon _{\mu _{234}\nu
_{234}}p_{21}^{\nu _{2}}p_{31}^{\nu _{3}}p_{41}^{\nu _{4}}\left[ p_{41\nu
_{1}}J_{4}^{\nu _{1}}\right] .
\end{equation*}%
} $\varepsilon _{\lbrack \mu _{234}\nu _{123}}p_{41\nu _{4}]}J_{4}^{\nu
_{1}} $, and the reductions of the six dimensional basic finite functions,
we reach at the finite part of the contraction. Summing up both parts the
total amplitude obeys%
\begin{eqnarray*}
p_{41}^{\mu _{1}}\left( T_{\mu _{1234}}^{AVVV}\right) _{1} &=&-2mT_{\mu
_{234}}^{PVVV}+T_{\mu _{4}\mu _{2}\mu _{3}}^{AVV}\left( 1,2,3\right) -T_{\mu
_{2}\mu _{3}\mu _{4}}^{AVV}\left( 2,3,4\right) \\
&&+\frac{8}{3}\varepsilon _{\mu _{234}\nu _{123}}p_{21}^{\nu
_{1}}p_{31}^{\nu _{2}}p_{41}^{\nu _{3}}\left[ \Delta _{4\rho }^{\rho
}+i\left( 4\pi \right) ^{-3}\right] .
\end{eqnarray*}

\textbf{RAGF: Second Version}

The pattern is the same, the conditioning term for the satisfaction of the
relations appears in the second vertex, and the three other are found
without restriction, for this reason we only list the conditioned one%
\begin{equation*}
p_{21}^{\mu _{2}}\left( T_{\mu _{1234}}^{AVVV}\right) _{2}=T_{\mu
_{134}}^{AVV}\left( 1,3,4\right) -T_{\mu _{134}}^{AVV}\left( 2,3,4\right) +%
\frac{8}{3}\varepsilon _{\mu _{134}\nu _{123}}p_{21}^{\nu _{1}}p_{31}^{\nu
_{2}}p_{41}^{\nu _{3}}\left[ \Delta _{4\rho }^{\rho }+i\left( 4\pi \right)
^{-3}\right] .
\end{equation*}

For the uniqueness, we subtract the expressions and using the same
manipulations for the RAGFs and get%
\begin{equation*}
\left( T_{\mu _{1234}}^{AVVV}\right) _{1}-\left( T_{\mu
_{1234}}^{AVVV}\right) _{2}=-\frac{8}{3}\varepsilon _{\mu _{1234}\nu
_{12}}p_{32}^{\nu _{1}}p_{43}^{\nu _{2}}\left[ \Delta _{4\rho }^{\rho }+%
\frac{i}{\left( 4\pi \right) ^{3}}\right] .
\end{equation*}%
Showing once more that connection among linearity and uniqueness in the
sense we have posed it, different formulae to the traces do not deliver
identical tensor and their equivalence depends on the value of surface term.
These results are interrelated to a low energy theorem, or the behavior of
the $T_{\mu _{234}}^{PVVV}$ amplitude for all bilinears set to zero.%
\begin{equation*}
2mT_{\mu _{234}}^{PVVV}=\frac{8i}{3\left( 4\pi \right) ^{3}}\varepsilon
_{\mu _{234}\nu _{123}}p_{21}^{\nu _{1}}p_{32}^{\nu _{2}}p_{42}^{\nu
_{3}}\not=0
\end{equation*}

Adopting $q_{1}=q_{2}+q_{3}+q_{4}$, where $q_{2}$, $q_{3}$, $q_{4}$ the
incoming momenta and $q_{1}$ is outgoing, as in $4D$. The tensor casting the
properties of the $AVVV$ box will be given by%
\begin{eqnarray*}
F_{\mu _{1234}} &=&\varepsilon _{\mu _{1234}\nu _{12}}q_{2}^{\nu
_{1}}q_{3}^{\nu _{2}}F_{1}+\varepsilon _{\mu _{1234}\nu _{12}}q_{2}^{\nu
_{1}}q_{4}^{\nu _{2}}F_{2}+\varepsilon _{\mu _{1234}\nu _{12}}q_{3}^{\nu
_{1}}q_{4}^{\nu _{2}}F_{3} \\
&&+\varepsilon _{\mu _{234}\nu _{123}}q_{2}^{\nu _{1}}q_{3}^{\nu
_{2}}q_{4}^{\nu _{3}}\left[ q_{2\mu _{1}}G_{1}+q_{3\mu _{1}}G_{2}+q_{4\mu
_{1}}G_{3}\right] \\
&&+\varepsilon _{\mu _{134}\nu _{123}}q_{2}^{\nu _{1}}q_{3}^{\nu
_{2}}q_{4}^{\nu _{3}}\left[ q_{2\mu _{2}}G_{4}+q_{3\mu _{2}}G_{5}+q_{4\mu
_{2}}G_{6}\right] \\
&&+\varepsilon _{\mu _{124}\nu _{123}}q_{2}^{\nu _{1}}q_{3}^{\nu
_{2}}q_{4}^{\nu _{3}}\left[ q_{2\mu _{3}}G_{7}+q_{3\mu _{3}}G_{8}+q_{4\mu
_{3}}G_{9}\right] \\
&&+\varepsilon _{\mu _{123}\nu _{123}}q_{2}^{\nu _{1}}q_{3}^{\nu
_{2}}q_{4}^{\nu _{3}}\left[ q_{2\mu _{4}}G_{10}+q_{3\mu _{4}}G_{11}+q_{4\mu
_{4}}G_{12}\right] ,
\end{eqnarray*}%
and then the four contractions $q_{i}^{\mu _{i}}F_{\mu _{1234}}=\varepsilon
_{\mu _{1\cdots \hat{\imath}\cdots 4}\nu _{123}}q_{2}^{\nu _{1}}q_{3}^{\nu
_{2}}q_{4}^{\nu _{3}}V_{i}$, the hat means to suppress that index, they
allow to trace the following fact, in the point $q_{i}\cdot q_{j}=0$, it is
straightforward to obtain $V_{1}\left( 0\right) +V_{2}\left( 0\right)
-V_{3}\left( 0\right) +V_{4}\left( 0\right) =0$. Then, what we have
independent of the consistency of the method or manipulations is that if the
WI built out of divergence of the axial current is satisfied or not, we have%
\begin{equation*}
V_{1}\left( 0\right) =\Omega ^{PVVV}\left( 0\right) \not=0,
\end{equation*}%
hence there are vector vertexes where the WIs are not satisfied, and
vice-versa, if the vector currents are conserved the axial must not be due
to 
\begin{equation*}
V_{1}\left( 0\right) +\Omega ^{PVVV}\left( 0\right) +\mathcal{A}=0,
\end{equation*}%
and the series of propositions derived for 4D, in sections (\ref{LE4D} and %
\ref{LED4DSTS}), follow here by analogous constructions, that we see as
unnecessary, it would connect the consequences of kinematical properties of
finite functions to the obstructions, inconsistencies, linearity violation,
lack of uniqueness, and all that. The value of the investigation proposed
here is that this results are the same in all dimensions.

\section{Final Remarks and Perspectives}

\label{finalremarks}In this study, a detailed probe of a significant number
of pseudo-tensor diagrams that correspond to anomalous amplitudes in two,
four, and six dimensions is performed, following a strategy to cope with the
divergences introduced in the thesis of O.A. Battistel \cite%
{PhdBattistel1999}. Using the idea that is possible to separate in two sets
the integrand of an arbitrary perturbative amplitude using systematically an
identity that localizes the divergent parts in a set where the physical
parameters, masses, and momenta, can be factored out of the integrals and
other where only finite integrals reside and are freely integrated.

This procedure is applied to the bubbles, triangles, and box, enabling us to
write down any expression as a sum of standard tensors, that is comprised of
what we have called sign tensors, and another amplitude, of parity even,
formed by vertexes of the same nature the main amplitude is composed. In
four dimensions, for example, we get schematically 
\begin{equation*}
\left( T_{\mu _{123}}^{\Gamma _{1}\Gamma _{2}\Gamma _{3}}\right)
_{i}=4iC_{i\mu _{123}}\pm i\varepsilon _{\mu _{123}\nu _{1}}\left( \text{%
Corresponding sub-amplitude}\right) ^{\nu _{1}}.
\end{equation*}%
Thus, after splitting off and organizing the divergent parts, without
further action, the finite ones are integrated. In this point, summing up
these two parts, the scalar objects $I_{\log }^{\left( 2n\right) }$ exactly
cancel in all cases, letting the final result as a sum of finite tensors and
surface terms, $\Delta _{n+1;\mu _{12}}^{2n}$, as defined along this work.
Such recipe crucially relies on the principle that it is possible to write
the integral of a sum as the sum of their integrals the linearity of
integration.

The role of this aspect as a fundamental element in this discussion then
emerges, writing in 4D three equations $q_{i}^{\mu _{i}}T_{\mu
_{123}}^{\Gamma _{123}}=T_{\left( -\right) i}^{AV}+\varepsilon _{\mu
_{ab}\nu _{12}}q_{2}^{\nu _{1}}q_{3}^{\nu _{2}}\Omega _{i}$ for the RAGFs,
it follows that, if the vanishing of the $AV$ or their difference, in that
equations were a possibility, then that would allow the vector and partial
axial symmetry to be true. Such hope could be based on the fact that if
these structures were only functions of routing differences, then using the
charge conjugation matrix, $C\gamma _{\mu }C^{-1}=-\gamma _{\mu }^{T}$, the
properties of the spinor-propagators, and traces, it would be possible to
prove that $T_{\mu \nu }^{AV}\left( p\right) =-T_{\mu \nu }^{AV}\left(
p\right) $, or that they are vanishing. To have this property means to have
translational invariance or translational invariance in momentum-space, but
computations reveal that such structures, in principle, depend on the
unphysical and arbitrary sum of routings and are proportional to a surface
term, $T^{AV}\sim \varepsilon _{\alpha \mu \nu _{1}\nu _{2}}p_{ij}^{\nu
_{1}}P_{ji}^{\nu _{3}}\Delta _{\nu _{3}}^{\nu _{2}}$, that violates the
mentioned symmetry in momentum space. Well a partial solution is to make the
surface term zero, and we are back in the symmetric scenario, routing
invariance.

Nevertheless, as demonstrated in section (\ref{LE4D}), regarding low-energy
theorems, a tensor with the characteristics of $AVV$, for example, a
function of the differences among the routings related by contraction to the
well-defined tensor $PVV$, must satisfy, in this case, $\left. p_{31}^{\mu
_{1}}T_{\mu _{123}}^{AVV}\right\vert _{0}=0\not=\left. -2mT_{\mu
_{23}}^{PVV}\right\vert _{0}$, what is impossible. At this point, anyone
should notice that the satisfaction of the RAGF, or linearity, can not be
satisfied for any value of the surface term, in particular, not for
vanishing value.

Therefore, the next step is to consider that the undetermined content of all
the tensors we investigated are always combinations of routings, surface
terms, and the $\varepsilon $-tensor. With these assumptions in mind, in the
form of hypotheses, plus the known RHS of the relations, the $AV$-part, we
lay down, in definitive: it is impossible without additional conditions to
satisfy all the RAGF, in other words, they are not valid for any value of
the surface term, a result developed in the section (\ref{LED4DSTS}). Beyond
this, the obligation to satisfy all of them, the RAGFs, makes the
kinematical property, at zero, of the $PVV$, the value and the reason why
the surface term can not be vanishing, see eq. (\ref{ME}).For this reason,
translational invariance in momentum space, and linearity are incompatible
properties for these perturbative amplitudes. Moreover, if one could rule
out any role for the internal momenta by adopting $\Delta _{3\mu \nu }=0$
and have linearity, one would be wrestling with the fact that three-point
amplitudes related by contraction do not vanish in zero,%
\begin{equation*}
2\Delta _{3\alpha }^{\alpha }=i\left[ \Omega _{3}\left( 0\right) -\Omega
_{2}\left( 0\right) -\Omega _{1}\left( 0\right) \right] =-i/\left( 2\pi
\right) ^{2},
\end{equation*}%
and, by necessity, when choosing the linear scenario, we can transform by
use of linear combinations the internal momenta in terms of external ones.
Because there are two independent external variables, we have two parameters
available, that is not enough to keep all WIs. As expected, due
non-ambiguous kinematical reasons.

The explicitly finite values of $\Omega _{i}$ always satisfies the last
equality above, for arbitrary masses as well. Nevertheless, such a result is
irreconcilable with that coming from a tensor with the properties of $T_{\mu
_{123}}^{\Gamma _{123}}$, in which case the result should be$\ $equal to
zero, but then, there must be some way to understand why this does not
happen in any, minimally consistent, manipulation. After writing the
internal momenta in terms of the external ones, assuming all arbitrary
violations of the RAGF taking the surface term as an arbitrary quantity, we
reach a tensor under the hypotheses stated in section (\ref{LE4D}). Follows
that, all violations are encompassed by $V_{i}=\Omega _{i}+\mathcal{A}_{i}$
and obey $\mathcal{A}_{3}-\mathcal{A}_{2}-\mathcal{A}_{1}=\left( 2\pi
\right) ^{-2}$, demonstrated through the eq.'s (\ref{genViol}). In it reside
the straightforward fact that when two WI are satisfied, the third is
violated by a unique amount independent of any consistent computational
philosophy because there is no ambiguity in the values of finite amplitudes.

In what concerns the consequence of the Dirac traces, surface terms, and
Schouten identities, in all these amplitudes arises the trace of $2n+2$
Dirac matrices and an odd number of the chiral matrices, schematically%
\begin{equation*}
\mathrm{tr}(\gamma _{\ast }\gamma _{\mu _{1}}\cdots \gamma _{\mu
_{2n+2}})\sim \sum \pm g_{\mu _{i}\mu _{j}}\varepsilon _{\mu _{1}\cdots \hat{%
\mu}_{i}\cdots \hat{\mu}_{j}\cdots \mu _{2n}}.
\end{equation*}%
Applying the definition of $\gamma _{\ast }=$ $\varepsilon _{\nu
_{1234}}\gamma ^{\nu _{1234}}/24$ or using the identity $\gamma _{\ast
}\gamma _{\mu _{i}}=\varepsilon _{\mu _{i}\nu _{123}}\gamma ^{\nu _{123}}/6$
in the adjacent position of the matrix $\gamma _{\mu _{i}}$, we have shown
the tensors calculated to correspond to the versions defined as the main
ingredients of the investigation, that a priori are not equivalent for $%
i\not=j$. If the surface term is adopted zero, the obtained set of tensors
have the property they violate the RAGF around that vertex, that means, to
the vertex corresponding to $\gamma _{\mu _{i}}$, at the level of the
diagram, the WI gets violated in the same vertex. These specific types of
substitutions deliver different expressions in the number of monomials, but
their difference, after integration, is a combination of finite and null
integrals. Their linear combination, of the versions, forms the building
block of any other identity, and all the common substitutions are a subset
of these possibilities. Arbitrary combinations of these building blocks can
be used to explain any result obtained in scenarios without internal
momentum ambiguities.

Adopting $\Delta _{3\mu \nu }=0$ makes the amplitudes to depend on the
traces used, the Schouten identity inside the integral that connects the
integrands ceases to make it in the final integrated results. At the end of
the day this is what breaks the linearity of integration and violates the
RAGFs. Different formulae for the traces do not deliver identical tensors,
and the equivalence depends on the value of the surface term.

For the term uniqueness that we have employed, some definition is necessary
for it to work as a concrete criterion. A criterion that makes the
amplitudes unique in a universal sense is impossible since they are
divergent quantities. After renormalized, they become dependent on an
arbitrary mass scale, and this is beginning of renormalization group
equations.

However once an expression is attributed by a regularization, there is no
other way to get another result of the same procedure in even tensors. On
the other hand, to our amplitudes, adopting the same interpretation to the
surface terms led to various different tensors. In this narrow sense we have
defined uniqueness: if the stance on the divergent quantities is the same,
uniqueness implies only one answer. Through this definition, all
mathematical manipulations leads to one result. Apart the manipulations used
in this work, when bilinears present in the expressions are not reduced or
taking traces after the integrals manipulated, the result is the same. Here
comes the point of having a narrow definition the unique answer is a
function of the routings $k_{1},k_{2},\cdots k_{i}\cdots $ taken as
independent variables. The consequence is that choices that break
momentum-space homogeneity must be adopted and establishes that one does not
have a unique function of the external momenta. The mathematically unique
answer, that does not depend on any sequence of algebraic operations, is not
unique in another form, in a way that preserves homogeneity, that is,
freedom of the origin of integration.

As of rule, there is the attractive option of making the surface term zero
as done in even amplitudes and by a convenient choice of trace to obtain the
symmetry content with the condition that not all symmetries can be present
and not all distribution of anomalies is possible. In this scenario, there
is a myriad of tensors to represent some amplitude, but it is a choice that
can be made. Notwithstanding, there is one keeping linearity of integration,
turning amplitudes unique functions of their routings, violating
momentum-space homogeneity, and then to make physical interpretation is
necessary to write the routings as combinations of the physical momenta. The
conclusions about the symmetries are the same, but now in a different
context: The integral of the sum is the sum of the integrals.

\appendix

\section{Traces of a String of Six Gamma and the Chiral Matrix}

\label{Tr6G4D}

The a way to insert a Levi-Civita tensor in the traces with the chiral
matrix come from the use of%
\begin{equation*}
\gamma _{\ast }\gamma _{\left[ \mu _{1}\cdots \mu _{r}\right] }=\frac{%
i^{n-1+r\left( r+1\right) }}{\left( 2n-r\right) !}\varepsilon _{\mu
_{1}\cdots \mu _{r}\nu _{r+1}\cdots \nu _{2n}}\gamma ^{\left[ \nu
_{r+1}\cdots \nu _{2n}\right] }
\end{equation*}%
in $2n=4$ dimensions they are the identities with $0,1,2,3,4$
antisymmetrized products, giving rise in traces of a string of six gamma
matrices to $\left( 15,10,7,6,7\right) $ monomials respectively.

\textbf{Trace Using }$\gamma _{\ast }=i\varepsilon _{\nu _{1}\nu _{2}\nu
_{3}\nu _{4}}\gamma ^{\nu _{1}\nu _{2}\nu _{3}\nu _{4}}/4!$\textbf{\
(Definition)}

The three main positions to deploy the definition of the chiral matrix is
around the gamma matrices present in the vertexes $\Gamma _{1},\Gamma _{2},$
and $\Gamma _{3}$, in the left or the right they return the same integrated
results.

Distinct positions of the chiral matrix in the trace. First one

\begin{eqnarray*}
t_{1} &=&\mathrm{tr}\left( \gamma _{\ast }\gamma _{\mu _{1}\nu _{1}\mu
_{2}\nu _{2}\mu _{3}\nu _{3}}\right) =i\varepsilon ^{\alpha _{1}\alpha
_{2}\alpha _{3}\alpha _{4}}\mathrm{tr}\left( \gamma _{\alpha _{1}\alpha
_{2}\alpha _{3}\alpha _{4}}\gamma _{\mu _{1}\nu _{1}\mu _{2}\nu _{2}\mu
_{3}\nu _{3}}\right) /4! \\
&=&+g_{\mu _{1}\nu _{1}}\varepsilon _{\mu _{2}\nu _{2}\mu _{3}\nu
_{3}}-g_{\mu _{1}\mu _{2}}\varepsilon _{\nu _{1}\nu _{2}\mu _{3}\nu
_{3}}+g_{\mu _{1}\nu _{2}}\varepsilon _{\nu _{1}\mu _{2}\mu _{3}\nu
_{3}}-g_{\mu _{1}\mu _{3}}\varepsilon _{\nu _{1}\mu _{2}\nu _{2}\nu
_{3}}+g_{\mu _{1}\nu _{3}}\varepsilon _{\nu _{1}\mu _{2}\nu _{2}\mu _{3}} \\
&&+g_{\nu _{1}\mu _{2}}\varepsilon _{\mu _{1}\nu _{2}\mu _{3}\nu
_{3}}-g_{\nu _{1}\nu _{2}}\varepsilon _{\mu _{1}\mu _{2}\mu _{3}\nu
_{3}}+g_{\nu _{1}\mu _{3}}\varepsilon _{\mu _{1}\mu _{2}\nu _{2}\nu
_{3}}-g_{\nu _{1}\nu _{3}}\varepsilon _{\mu _{1}\mu _{2}\nu _{2}\mu
_{3}}+g_{\mu _{2}\nu _{2}}\varepsilon _{\mu _{1}\nu _{1}\mu _{3}\nu _{3}} \\
&&-g_{\mu _{2}\mu _{3}}\varepsilon _{\mu _{1}\nu _{1}\nu _{2}\nu
_{3}}+g_{\mu _{2}\nu _{3}}\varepsilon _{\mu _{1}\nu _{1}\nu _{2}\mu
_{3}}+g_{\nu _{2}\mu _{3}}\varepsilon _{\mu _{1}\nu _{1}\mu _{2}\nu
_{3}}-g_{\nu _{2}\nu _{3}}\varepsilon _{\mu _{1}\nu _{1}\mu _{2}\mu
_{3}}+g_{\mu _{3}\nu _{3}}\varepsilon _{\mu _{1}\nu _{1}\mu _{2}\nu _{2}}.
\end{eqnarray*}%
Second one%
\begin{eqnarray*}
t_{2} &=&\mathrm{tr}\left( \gamma _{\mu _{1}\nu _{1}}\gamma _{\ast }\gamma
_{\mu _{2}\nu _{2}\mu _{3}\nu _{3}}\right) =i\varepsilon ^{\alpha _{1}\alpha
_{2}\alpha _{3}\alpha _{4}}\mathrm{tr}\left( \gamma _{\mu _{1}\nu
_{1}}\gamma _{\alpha _{1}\alpha _{2}\alpha _{3}\alpha _{4}}\gamma _{\mu
_{2}\nu _{2}\mu _{3}\nu _{3}}\right) /4! \\
&=&+g_{\mu _{1}\nu _{1}}\varepsilon _{\mu _{2}\nu _{2}\mu _{3}\nu
_{3}}+g_{\mu _{1}\mu _{2}}\varepsilon _{\nu _{1}\nu _{2}\mu _{3}\nu
_{3}}-g_{\mu _{1}\nu _{2}}\varepsilon _{\mu _{2}\nu _{2}\mu _{3}\nu
_{3}}+g_{\mu _{1}\mu _{3}}\varepsilon _{\nu _{1}\mu _{2}\nu _{2}\nu
_{3}}-g_{\mu _{1}\nu _{3}}\varepsilon _{\nu _{1}\mu _{2}\nu _{2}\mu _{3}} \\
&&-g_{\nu _{1}\mu _{2}}\varepsilon _{\mu _{1}\nu _{2}\mu _{3}\nu
_{3}}+g_{\nu _{1}\nu _{2}}\varepsilon _{\mu _{1}\mu _{2}\mu _{3}\nu
_{3}}-g_{\nu _{1}\mu _{3}}\varepsilon _{\mu _{1}\mu _{2}\nu _{2}\nu
_{3}}+g_{\nu _{1}\nu _{3}}\varepsilon _{\mu _{1}\mu _{2}\nu _{2}\mu
_{3}}+g_{\mu _{2}\nu _{2}}\varepsilon _{\mu _{1}\nu _{1}\mu _{3}\nu _{3}} \\
&&-g_{\mu _{2}\mu _{3}}\varepsilon _{\mu _{1}\nu _{1}\nu _{2}\nu
_{3}}+g_{\mu _{2}\nu _{3}}\varepsilon _{\mu _{1}\nu _{1}\nu _{2}\mu
_{3}}+g_{\nu _{2}\mu _{3}}\varepsilon _{\mu _{1}\nu _{1}\mu _{2}\nu
_{3}}-g_{\nu _{2}\nu _{3}}\varepsilon _{\mu _{1}\nu _{1}\mu _{2}\mu
_{3}}+g_{\mu _{3}\nu _{3}}\varepsilon _{\mu _{1}\nu _{1}\mu _{2}\nu _{2}}.
\end{eqnarray*}%
Third one%
\begin{eqnarray*}
t_{3} &=&\mathrm{tr}\left( \gamma _{\mu _{1}\nu _{1}\mu _{2}\nu _{2}}\gamma
_{\ast }\gamma _{\mu _{3}\nu _{3}}\right) =i\varepsilon ^{\alpha _{1}\alpha
_{2}\alpha _{3}\alpha _{4}}\mathrm{tr}\left( \gamma _{\mu _{1}\nu _{1}\mu
_{2}\nu _{2}}\gamma _{\alpha _{1}\alpha _{2}\alpha _{3}\alpha _{4}}\gamma
_{\mu _{3}\nu _{3}}\right) /4! \\
&=&+g_{\mu _{1}\nu _{1}}\varepsilon _{\mu _{2}\nu _{2}\mu _{3}\nu
_{3}}-g_{\mu _{1}\mu _{2}}\varepsilon _{\nu _{1}\nu _{2}\mu _{3}\nu
_{3}}+g_{\mu _{1}\nu _{2}}\varepsilon _{\mu _{2}\nu _{2}\mu _{3}\nu
_{3}}+g_{\mu _{1}\mu _{3}}\varepsilon _{\nu _{1}\mu _{2}\nu _{2}\nu
_{3}}-g_{\mu _{1}\nu _{3}}\varepsilon _{\nu _{1}\mu _{2}\nu _{2}\mu _{3}} \\
&&+g_{\nu _{1}\mu _{2}}\varepsilon _{\mu _{1}\nu _{2}\mu _{3}\nu
_{3}}-g_{\nu _{1}\nu _{2}}\varepsilon _{\mu _{1}\mu _{2}\mu _{3}\nu
_{3}}-g_{\nu _{1}\mu _{3}}\varepsilon _{\mu _{1}\mu _{2}\nu _{2}\nu
_{3}}+g_{\nu _{1}\nu _{3}}\varepsilon _{\mu _{1}\mu _{2}\nu _{2}\mu
_{3}}+g_{\mu _{2}\nu _{2}}\varepsilon _{\mu _{1}\nu _{1}\mu _{3}\nu _{3}} \\
&&+g_{\mu _{2}\mu _{3}}\varepsilon _{\mu _{1}\nu _{1}\nu _{2}\nu
_{3}}-g_{\mu _{2}\nu _{3}}\varepsilon _{\mu _{1}\nu _{1}\nu _{2}\mu
_{3}}-g_{\nu _{2}\mu _{3}}\varepsilon _{\mu _{1}\nu _{1}\mu _{2}\nu
_{3}}+g_{\nu _{2}\nu _{3}}\varepsilon _{\mu _{1}\nu _{1}\mu _{2}\mu
_{3}}+g_{\mu _{3}\nu _{3}}\varepsilon _{\mu _{1}\nu _{1}\mu _{2}\nu _{2}},
\end{eqnarray*}%
we omit the imaginary unit. Now, these three expressions cast all the
indexes of the tensor and they have fifteen terms each, in a narrow sense
they could be called symmetric and considered to be as respecting all the
symmetry among the indexes, see, by example, the appendix of the ref. \cite%
{AguilaVictoria1998} or the refs. \cite{Wu2006}\cite{Viglioni2016}, we do
not focus on such adjectives, but on the fact they are enough to obtain any
other result by a careful analysis, and encompassing any possible
manipulations with these structures.

First things first, the sign differences are the unique distinguishing
factor in that traces, they effectively sample the indexes among finite and
surface terms in the real calculations. The aim is to demonstrate that any
expression to the triangles investigated are just linear combinations of the
ones we have detailed in the main body of this work.

Making the combinations, only using sums and not Schouten identities%
\begin{equation*}
t_{ij}=\frac{1}{2}\left( t_{i}+t_{j}\right) ,
\end{equation*}%
we will have%
\begin{eqnarray*}
t_{12} &=&-g_{\mu _{1}\nu _{1}}\varepsilon _{\mu _{2}\mu _{3}\nu _{2}\nu
_{3}}-g_{\mu _{2}\nu _{2}}\varepsilon _{\mu _{1}\mu _{3}\nu _{1}\nu
_{3}}+g_{\mu _{2}\nu _{3}}\varepsilon _{\mu _{1}\mu _{3}\nu _{1}\nu _{2}} \\
&&-g_{\nu _{2}\mu _{3}}\varepsilon _{\mu _{1}\mu _{2}\nu _{1}\nu
_{3}}-g_{\mu _{3}\nu _{3}}\varepsilon _{\mu _{1}\mu _{2}\nu _{1}\nu
_{2}}-g_{\mu _{2}\mu _{3}}\varepsilon _{\mu _{1}\nu _{1}\nu _{2}\nu
_{3}}-g_{\nu _{2}\nu _{3}}\varepsilon _{\mu _{1}\mu _{2}\mu _{3}\nu _{1}},
\end{eqnarray*}%
\begin{eqnarray*}
t_{13} &=&-g_{\mu _{3}\nu _{3}}\varepsilon _{\mu _{1}\mu _{2}\nu _{1}\nu
_{2}}-g_{\mu _{1}\nu _{1}}\varepsilon _{\mu _{2}\mu _{3}\nu _{2}\nu
_{3}}+g_{\mu _{1}\nu _{2}}\varepsilon _{\mu _{2}\mu _{3}\nu _{1}\nu _{3}} \\
&&-g_{\nu _{1}\mu _{2}}\varepsilon _{\mu _{1}\mu _{3}\nu _{2}\nu
_{3}}-g_{\mu _{2}\nu _{2}}\varepsilon _{\mu _{1}\mu _{3}\nu _{1}\nu
_{3}}-g_{\mu _{1}\mu _{2}}\varepsilon _{\mu _{3}\nu _{1}\nu _{2}\nu
_{3}}-g_{\nu _{1}\nu _{2}}\varepsilon _{\mu _{1}\mu _{2}\mu _{3}\nu _{3}},
\end{eqnarray*}%
\begin{eqnarray*}
t_{23} &=&-g_{\mu _{2}\nu _{2}}\varepsilon _{\mu _{1}\mu _{3}\nu _{1}\nu
_{3}}-g_{\mu _{1}\nu _{1}}\varepsilon _{\mu _{2}\mu _{3}\nu _{2}\nu
_{3}}+g_{\mu _{1}\nu _{3}}\varepsilon _{\mu _{2}\mu _{3}\nu _{1}\nu _{2}} \\
&&-g_{\nu _{1}\mu _{3}}\varepsilon _{\mu _{1}\mu _{2}\nu _{2}\nu
_{3}}-g_{\mu _{3}\nu _{3}}\varepsilon _{\mu _{1}\mu _{2}\nu _{1}\nu
_{2}}-g_{\mu _{1}\mu _{3}}\varepsilon _{\mu _{2}\nu _{1}\nu _{2}\nu
_{3}}-g_{\nu _{1}\nu _{3}}\varepsilon _{\mu _{1}\mu _{2}\mu _{3}\nu _{2}},
\end{eqnarray*}%
and we can employ the identities involving the antisymmetrized products to
compute the same trace as well obtaining other formulas.

\textbf{Trace Using} $\gamma _{\ast }\gamma _{\mu _{1}}=-i\varepsilon _{\mu
_{1}\nu _{1}\nu _{2}\nu _{3}}\gamma ^{\nu _{1}\nu _{2}\nu _{3}}/3!$

The straightforward application%
\begin{eqnarray*}
\eta _{1}\left( a\right) &=&\mathrm{tr}\left( \gamma _{\ast }\gamma
_{abcdef}\right) =g_{\text{$b$}c}\varepsilon _{\text{$a$}def}-g_{\text{$b$}%
d}\varepsilon _{\text{$a$}cef}+g_{\text{$b$}e}\varepsilon _{\text{$a$}%
cdf}-g_{\text{$b$}f}\varepsilon _{\text{$a$}cde} \\
&&+g_{\text{$c$}d}\varepsilon _{\text{$a$}bef}-g_{\text{$c$}e}\varepsilon _{%
\text{$a$}bdf}+g_{\text{$c$}f}\varepsilon _{\text{$a$}bde}+g_{de}\varepsilon
_{abcf}+g_{ef}\varepsilon _{abcd}-g_{df}\varepsilon _{abce}
\end{eqnarray*}%
the notation means that it uses a product with one Dirac matrix with index $%
a $ in the substitution of $\gamma _{\ast }\gamma _{a}$%
\begin{equation*}
\eta _{1}\left( a\right) =-i\varepsilon _{a}^{\quad \nu _{1}\nu _{2}\nu _{3}}%
\mathrm{tr}\left( \gamma _{\nu _{1}\nu _{2}\nu _{3}}\gamma _{bcdef}\right)
/6.
\end{equation*}

\textbf{The Trace Using} $\gamma _{\ast }\gamma _{\left[ ab\right]
}=-i\varepsilon _{ab\nu _{1}\nu _{2}}\gamma ^{\nu _{1}\nu _{2}}/2!$

The application of this one requires to express the ordinary product in
terms of the antisymmetrized one%
\begin{equation*}
\gamma _{\ast }\gamma _{ab}=-\frac{i}{2}\varepsilon _{ab\nu _{1}\nu
_{2}}\gamma ^{\nu _{1}\nu _{1}}+g_{ab}\gamma _{\ast },
\end{equation*}%
thereby follows 
\begin{eqnarray*}
\eta _{2}\left( ab\right) &=&\mathrm{tr}\left( \gamma _{\ast }\gamma
_{abcdef}\right) =g_{ab}\varepsilon _{cdef}+g_{cd}\varepsilon
_{abef}-g_{ce}\varepsilon _{abdf}+g_{cf}\varepsilon _{abde} \\
&&+g_{d\text{$e$}}\varepsilon _{abcf}-g_{df}\varepsilon
_{abce}+g_{ef}\varepsilon _{abcd}
\end{eqnarray*}

\textbf{The Trace Using} $\gamma _{\ast }\gamma _{\left[ abc\right]
}=i\varepsilon _{abc\nu }\gamma ^{\nu }$

Expressing the antisymmetric product as common products we get%
\begin{equation*}
\gamma _{\ast }\gamma _{abc}=i\varepsilon _{abc\nu }\gamma ^{\nu }+\gamma
_{\ast }\left( g_{bc}\gamma _{a}-g_{ac}\gamma _{b}+g_{ab}\gamma _{c}\right)
\end{equation*}%
with arbitrary indexes we get%
\begin{equation*}
\eta _{3}\left( abc\right) =\mathrm{tr}\left( \gamma _{\ast }\gamma
_{abcdef}\right) =g_{ab}\varepsilon _{cdef}-g_{ac}\varepsilon
_{bdef}+g_{bc}\varepsilon _{adef}+g_{de}\varepsilon
_{abcf}-g_{df}\varepsilon _{abce}+g_{ef}\varepsilon _{abcd}
\end{equation*}%
where the notation means that we absorb the indexes $a$, $b$ and $c$, with
the identity and compute the resulting trace, it can be used to apply the
substitution in any place desired. The use of this identity is a common
choice on computation of this type of diagrams. In them, and all other
possible results, after integration, we get some of the results obtained
through the linear combinations $t_{12}$, $t_{13}$, and $t_{23}$.

\textbf{The Trace Using} $\gamma _{\ast }\gamma _{\left[ abcd\right]
}=i\varepsilon _{abcd}$

With the help of%
\begin{eqnarray*}
\gamma _{\ast }\gamma _{abcd} &=&i\varepsilon _{abcd}\mathbf{1}+g_{ab}\gamma
_{\ast }\gamma _{\left[ cd\right] }-g_{ac}\gamma _{\ast }\gamma _{\left[ bd%
\right] }+g_{ad}\gamma _{\ast }\gamma _{\left[ bc\right] } \\
&&+g_{bc}\gamma _{\ast }\gamma _{\left[ ad\right] }-g_{bd}\gamma _{\ast
}\gamma _{\left[ ac\right] }+g_{cd}\gamma _{\ast }\gamma _{\left[ ab\right]
}+\left( g_{ab}g_{cd}-g_{ac}g_{bd}+g_{ad}g_{bc}\right) \gamma _{\ast }
\end{eqnarray*}%
under the trace and with arbitrary indexes, we get%
\begin{eqnarray*}
\eta _{4}\left( abcd\right) &=&\mathrm{tr}\left( \gamma _{\ast }\gamma
_{abcdef}\right) =+g_{ab}\varepsilon _{cdef}-g_{ac}\varepsilon
_{bdef}+g_{ad}\varepsilon _{bcef} \\
&&+g_{bc}\varepsilon _{adef}-g_{bd}\varepsilon _{acef}+g_{cd}\varepsilon
_{abef}+g_{ef}\varepsilon _{abcd}.
\end{eqnarray*}

\textbf{The Interconnection Among the Formulas: }The difference on the
integrated amplitudes either will identically vanishing as the integrand are
exactly equal as in%
\begin{equation*}
\left[ t_{12}-\eta _{2}\left( \mu _{1}\nu _{1}\right) \right] =0;\qquad %
\left[ t_{13}-\eta _{4}\left( \mu _{1}\nu _{1}\mu _{2}\nu _{2}\right) \right]
=0,
\end{equation*}%
that is why is needless to say anything more, or will vanish because the
difference inside a explicit computation always corresponds, when
integrated, to finite null integrals, see%
\begin{eqnarray*}
\left[ t_{1}-\eta _{1}\left( \mu _{1}\right) \right] \frac{K_{123}^{\nu
_{123}}}{D_{123}} &=&\varepsilon _{\mu _{2}\mu _{3}\nu _{1}\nu _{2}}t_{\mu
_{1}}^{\left( -+\right) \nu _{12}}+g_{\mu _{1}\mu _{2}}t_{\mu
_{3}}^{ASS}-g_{\mu _{1}\mu _{3}}t_{\mu _{2}}^{ASS} \\
\left[ t_{12}+\eta _{2}\left( \nu _{1}\mu _{2}\right) \right] \frac{%
K_{123}^{\nu _{123}}}{D_{123}} &=&-\varepsilon _{\mu _{2}\mu _{3}\nu _{1}\nu
_{2}}t_{\mu _{1}}^{\left( -+\right) \nu _{12}}+\varepsilon _{\mu _{1}\mu
_{3}\nu _{1}\nu _{2}}t_{\mu _{2}}^{\left( -+\right) \nu _{12}}-g_{\mu
_{2}\mu _{3}}t_{\mu _{1}}^{ASS}+g_{\mu _{1}\mu _{3}}t_{\mu _{2}}^{ASS} \\
\left[ t_{12}-\eta _{3}\left( \mu _{1}\nu _{1}\mu _{2}\right) \right] \frac{%
K_{123}^{\nu _{123}}}{D_{123}} &=&+\varepsilon _{\mu _{1}\mu _{3}\nu _{1}\nu
_{2}}t_{\mu _{2}}^{\left( -+\right) \nu _{12}}+g_{\mu _{1}\mu _{2}}t_{\mu
_{3}}^{ASS}-g_{\mu _{2}\mu _{3}}t_{\mu _{1}}^{ASS} \\
\left[ t_{13}+\eta _{4}\left( \nu _{1}\mu _{2}\nu _{2}\mu _{3}\right) \right]
\frac{K_{123}^{\nu _{123}}}{D_{123}} &=&-\varepsilon _{\mu _{1}\mu _{2}\nu
_{1}\nu _{2}}t_{\mu _{3}}^{\left( -+\right) \nu _{12}}-\varepsilon _{\mu
_{2}\mu _{3}\nu _{1}\nu _{2}}t_{\mu _{1}}^{\left( -+\right) \nu
_{12}}+g_{\mu _{2}\mu _{3}}t_{\mu _{1}}^{ASS}-g_{\mu _{1}\mu _{2}}t_{\mu
_{3}}^{ASS}
\end{eqnarray*}%
and as was showed in the text the well defined integrals corresponding to $%
\varepsilon _{\alpha \beta \nu _{1}\nu _{2}}t_{\rho }^{\left( -+\right) \nu
_{12}}$ eq. (\ref{T-+}) and $t_{\mu }^{ASS}$ in eq. (\ref{ASS}),are null%
\begin{equation*}
\varepsilon _{\alpha \beta \nu _{1}\nu _{2}}T_{\rho }^{\left( -+\right) \nu
_{12}}=T_{\mu }^{ASS}=0
\end{equation*}%
delivering the conclusion that any form of substitution or manipulation is
accounted by the linear combination of the version one, two or three
replacing the definition of $\gamma _{\ast }$ left or right of the matrices $%
\gamma _{\mu _{1}}$, $\gamma _{\mu _{2}}$, and $\gamma _{\mu _{3}}$. Whose
consequence is that it is enough to unfold any feature of such calculations
with the basic versions we described. What we showed here is the forms that
identically correspond, not that all differences are finite and vanishing.
The form obtained from $t_{12}$ is not identical without conditions to any $%
t_{i}$, for example.

\section{The Integrals in Two Dimensions}

\label{AppInt2D}

\begin{equation*}
\bar{J}_{1}\left( k_{i}\right) =I_{\log }^{\left( 2\right) };\quad \bar{J}%
_{1}^{\mu }\left( k_{i}\right) =-k_{i}^{\nu }\Delta _{2\nu }^{\left(
2\right) \mu }
\end{equation*}%
\begin{eqnarray}
J_{2} &=&i\left( 4\pi \right) ^{-1}\left[ Z_{0}^{\left( -1\right) }\left(
p^{2},m^{2}\right) \right]  \label{2DJ2} \\
J_{2}^{\mu _{1}} &=&i\left( 4\pi \right) ^{-1}\left[ -p^{\mu
_{1}}Z_{1}^{\left( -1\right) }\right]  \label{2DJ2mu1} \\
J_{2}^{\mu _{1}\mu _{2}} &=&i\left( 4\pi \right) ^{-1}\left[ -\frac{1}{2}%
g^{\mu _{1}\mu _{2}}Z_{0}^{\left( 0\right) }+p^{\mu _{1}}p^{\mu
_{2}}Z_{2}^{\left( -1\right) }\right]  \notag \\
\bar{J}_{2}^{\mu _{1}\mu _{2}} &=&J_{2}^{\mu _{1}\mu _{2}}+\frac{1}{2}%
(\Delta _{2}^{\left( 2\right) \mu _{1}\mu _{2}}+g^{\mu _{1}\mu _{2}}I_{\log
}^{\left( 2\right) })  \notag
\end{eqnarray}%
Reductions%
\begin{eqnarray*}
\left( n+1\right) p^{2}Z_{n+2}^{\left( -1\right) } &=&\left( n+1\right)
p^{2}Z_{n+1}^{\left( -1\right) }-\left( n+1\right) m^{2}Z_{n}^{\left(
-1\right) }-1 \\
\left( n+2\right) p^{2}Z_{n+1}^{\left( 0\right) } &=&\left( n+1\right)
p^{2}Z_{n}^{\left( 0\right) }-m^{2}nZ_{n-1}^{\left( 0\right) }-\frac{n}{%
\left( n+1\right) \left( n+2\right) }p^{2};\quad n=0,1,3,\cdots \\
Z_{0}^{\left( 0\right) } &=&2p^{2}Z_{2}^{\left( -1\right)
}-p^{2}Z_{1}^{\left( -1\right) }
\end{eqnarray*}%
They imply%
\begin{eqnarray*}
J_{2}^{\mu _{1}} &=&-\frac{1}{2}p^{\mu _{1}}J_{2};\quad p_{\mu
_{1}}J_{2}^{\mu _{1}}=-\frac{1}{2}p^{2}J_{2} \\
p_{\mu _{1}}J_{2}^{\mu _{1}\mu _{2}} &=&-\frac{1}{2}p^{2}J_{2}^{\mu
_{2}};\quad g_{\mu _{1}\mu _{2}}J_{2}^{\mu _{1}\mu _{2}}=m^{2}J_{2}+\frac{i}{%
4\pi }
\end{eqnarray*}

\section{The Integrals in Four Dimensions}

\label{AppInt4D}\textbf{Two Point: }%
\begin{equation*}
J_{2}\left( p_{ij}\right) =-\frac{i}{\left( 4\pi \right) ^{2}}Z_{0}^{\left(
0\right) }\left( p_{ij}^{2},m^{2}\right) ;\text{ and\qquad }J_{2\mu }\left(
p_{ij}\right) =\frac{i}{\left( 4\pi \right) ^{2}}p_{ij\mu }Z_{1}^{\left(
0\right) }\left( p_{ij}^{2},m^{2}\right)
\end{equation*}%
\begin{eqnarray*}
\bar{J}_{2} &=&I_{\log }^{\left( 4\right) }+J_{2} \\
\bar{J}_{2\mu } &=&\int \frac{\mathrm{d}^{4}k}{\left( 2\pi \right) ^{4}}%
\frac{K_{i\mu }}{D_{ij}}=-\frac{1}{2}(P_{ij}^{\nu }\Delta _{3\mu \nu
}^{\left( 4\right) }+p_{ji\mu }I_{\log }^{\left( 4\right) })+J_{2\mu }
\end{eqnarray*}

\textbf{Three Point: }%
\begin{eqnarray*}
J_{3} &=&i\left( 4\pi \right) ^{-2}\left[ Z_{00}^{\left( -1\right) }\left(
p,q\right) \right] \\
J_{3\mu } &=&i\left( 4\pi \right) ^{-2}\left[ -p_{\mu }Z_{10}^{\left(
-1\right) }-q_{\mu }Z_{01}^{\left( -1\right) }\right] \\
J_{3\mu _{1}\mu _{2}} &=&i\left( 4\pi \right) ^{-2}\left[ p_{\mu _{1}}p_{\mu
_{2}}Z_{20}^{\left( -1\right) }+q_{\mu _{1}}q_{\mu _{2}}Z_{02}^{\left(
-1\right) }+\left( p_{\mu _{1}}q_{\mu _{2}}+p_{\mu _{2}}q_{\mu _{1}}\right)
Z_{11}^{\left( -1\right) }-\frac{1}{2}g_{\mu _{1}\mu _{2}}Z_{00}^{\left(
0\right) }\right]
\end{eqnarray*}%
\begin{equation*}
\bar{J}_{3\mu _{1}\mu _{2}}=J_{3\mu _{1}\mu _{2}}+\frac{1}{4}\left( \Delta
_{3\mu _{1}\mu _{2}}^{\left( 4\right) }+g_{\mu _{1}\mu _{2}}I_{\log
}^{\left( 4\right) }\right) ,
\end{equation*}%
it is worth mention that the arguments $p$ and $q$ are only general
variables that tag the entries of the functions, they must be carefully
substitute for the ones that appear in a particular part of the
investigation. \textbf{Reductions }of the basic functions. The two point
basic function that appear satisfy a simple relation $2Z_{1}^{\left(
0\right) }=Z_{0}^{\left( 0\right) }$, as the three point obey%
\begin{eqnarray*}
2\left[ p^{2}Z_{10}^{\left( -1\right) }+\left( p\cdot q\right)
Z_{01}^{\left( -1\right) }\right] &=&p^{2}Z_{00}^{\left( -1\right) }+\left[
Z_{0}^{\left( 0\right) }\left( q\right) -Z_{0}^{\left( 0\right) }\left(
q-p\right) \right] \\
2\left[ q^{2}Z_{01}^{\left( -1\right) }+\left( p\cdot q\right)
Z_{10}^{\left( -1\right) }\right] &=&q^{2}Z_{00}^{\left( -1\right) }+\left[
Z_{0}^{\left( 0\right) }\left( p\right) -Z_{0}^{\left( 0\right) }\left(
q-p\right) \right] \\
2\left[ p^{2}Z_{20}^{\left( -1\right) }+\left( p\cdot q\right)
Z_{11}^{\left( -1\right) }\right] &=&p^{2}Z_{10}^{\left( -1\right)
}+Z_{00}^{\left( 0\right) }-Z_{1}^{\left( 0\right) }\left( q-p\right) \\
2\left[ q^{2}Z_{02}^{\left( -1\right) }+\left( p\cdot q\right)
Z_{11}^{\left( -1\right) }\right] &=&q^{2}Z_{01}^{\left( -1\right)
}+Z_{00}^{\left( 0\right) }-Z_{1}^{\left( 0\right) }\left( q-p\right) \\
2\left[ p^{2}Z_{11}^{\left( -1\right) }+\left( p\cdot q\right)
Z_{02}^{\left( -1\right) }\right] &=&p^{2}Z_{01}^{\left( -1\right) }+\left[
Z_{1}^{\left( 0\right) }\left( q\right) -Z_{1}^{\left( 0\right) }\left(
q-p\right) \right] \\
2\left[ q^{2}Z_{11}^{\left( -1\right) }+\left( p\cdot q\right)
Z_{20}^{\left( -1\right) }\right] &=&q^{2}Z_{10}^{\left( -1\right) }+\left[
Z_{1}^{\left( 0\right) }\left( p\right) -Z_{1}^{\left( 0\right) }\left(
q-p\right) \right]
\end{eqnarray*}%
\begin{equation}
2Z_{00}^{\left( 0\right) }=\left[ p^{2}Z_{10}^{\left( -1\right)
}+q^{2}Z_{01}^{\left( -1\right) }\right] -2m^{2}Z_{00}^{\left( -1\right)
}-1+2Z_{1}^{\left( 0\right) }\left( q-p\right)  \label{Z00^0}
\end{equation}%
therefore it is possible to show that the tensors $J$ satisfy%
\begin{eqnarray*}
p^{\mu _{1}}J_{3\mu _{1}} &=&-\frac{1}{2}p^{2}J_{3}+\frac{1}{2}\left[
J_{2}\left( q\right) -J_{2}\left( q-p\right) \right] \\
q^{\mu _{1}}J_{3\mu _{1}} &=&-\frac{1}{2}q^{2}J_{3}+\frac{1}{2}\left[
J_{2}\left( p\right) -J_{2}\left( q-p\right) \right]
\end{eqnarray*}%
\begin{eqnarray*}
p^{\mu _{1}}J_{3\mu _{1}\mu _{2}} &=&-\frac{1}{2}p^{2}J_{3\mu _{2}}+\frac{1}{%
2}\left[ J_{2\mu _{2}}\left( q\right) +J_{2\mu _{2}}\left( q-p\right)
+q_{\mu _{2}}J_{2}\left( q-p\right) \right] \\
q^{\mu _{1}}J_{3\mu _{1}\mu _{2}} &=&-\frac{1}{2}q^{2}J_{3\mu _{2}}+\frac{1}{%
2}\left[ J_{2\mu _{2}}\left( p\right) +J_{2\mu _{2}}\left( q-p\right)
+q_{\mu _{2}}J_{2}\left( q-p\right) \right]
\end{eqnarray*}%
\begin{equation}
g^{\mu _{1}\mu _{2}}J_{3\mu _{1}\mu _{2}}=m^{2}J_{3}+\frac{i}{2\left( 4\pi
\right) ^{2}}+J_{2}\left( q-p\right)  \label{trJ3}
\end{equation}

\section{The Integrals in Six Dimensions}

\label{AppInt6D}\textbf{Three Point Functions}%
\begin{equation*}
J_{3}\left( p,q\right) =i\left( 4\pi \right) ^{-3}\left[ -Z_{00}^{\left(
0\right) }\left( p,q\right) \right] ;\qquad \bar{J}_{3}=I_{\log }^{\left(
6\right) }+J_{3}
\end{equation*}%
\begin{eqnarray*}
J_{3\mu _{1}}\left( k_{1},k_{2},k_{3}\right) &=&i\left( 4\pi \right) ^{-3} 
\left[ p_{21\mu _{1}}Z_{10}^{\left( 0\right) }+p_{31\mu _{1}}Z_{01}^{\left(
0\right) }\right] \\
\bar{J}_{3}^{\mu _{1}}\left( k_{1},k_{2},k_{3}\right) &=&-\frac{1}{3}\left(
k_{1}^{\nu _{1}}+k_{2}^{\nu _{1}}+k_{3}^{\nu _{1}}\right) \Delta _{4\nu
_{1}}^{\left( 6\right) \mu _{1}}-\frac{1}{3}\left( p_{21}^{\mu
_{1}}+p_{31}^{\mu _{1}}\right) I_{\log }^{\left( 6\right) }+J_{3}^{\mu _{1}},
\end{eqnarray*}%
\textbf{Four Point Functions}%
\begin{eqnarray*}
J_{4} &=&i\left( 4\pi \right) ^{-3}\left[ Z_{000}^{\left( -1\right) }\left(
p,q,r\right) \right] \\
J_{4\mu _{1}} &=&i\left( 4\pi \right) ^{-3}\left[ -p_{\mu
_{1}}Z_{100}^{\left( -1\right) }-q_{\mu _{1}}Z_{010}^{\left( -1\right)
}-r_{\mu _{1}}Z_{001}^{\left( -1\right) }\right]
\end{eqnarray*}%
\begin{eqnarray*}
&&J_{4\mu _{1}\mu _{2}}=i\left( 4\pi \right) ^{-3}\left[ -\frac{1}{2}g_{\mu
_{1}\mu _{2}}Z_{000}^{\left( 0\right) }+p_{\mu _{1}}p_{\mu
_{2}}Z_{200}^{\left( -1\right) }+q_{\mu _{1}}q_{\mu _{2}}Z_{020}^{\left(
-1\right) }+r_{\mu _{1}}r_{\mu _{2}}Z_{002}^{\left( -1\right) }\right] \\
&&+i\left( 4\pi \right) ^{-3}\left[ \left( p_{\mu _{1}}q_{\mu _{2}}+p_{\mu
_{2}}q_{\mu _{1}}\right) Z_{110}^{\left( -1\right) }+\left( p_{\mu
_{1}}r_{\mu _{2}}+p_{\mu _{2}}r_{\mu _{1}}\right) Z_{101}^{\left( -1\right)
}+\left( q_{\mu _{1}}r_{\mu _{2}}+r_{\mu _{1}}q_{\mu _{2}}\right)
Z_{011}^{\left( -1\right) }\right] \\
\bar{J}_{4\mu _{1}\mu _{2}} &=&J_{4\mu _{1}\mu _{2}}+\frac{1}{6}(\Delta
_{4\mu _{1}\mu _{2}}^{\left( 6\right) }+g_{\mu _{1}\mu _{2}}I_{\log
}^{\left( 6\right) })
\end{eqnarray*}%
\textbf{Reductions}%
\begin{eqnarray*}
2\left[ p^{2}Z_{100}^{\left( -1\right) }+\left( p\cdot q\right)
Z_{010}^{\left( -1\right) }+\left( p\cdot r\right) Z_{001}^{\left( -1\right)
}\right] &=&p^{2}Z_{000}^{\left( -1\right) }+Z_{00}^{\left( 0\right) }\left(
q,r\right) -Z_{00}^{\left( 0\right) }\left( p_{42},p_{43}\right) \\
2\left[ q^{2}Z_{010}^{\left( -1\right) }+\left( p\cdot q\right)
Z_{100}^{\left( -1\right) }+\left( q\cdot r\right) Z_{001}^{\left( -1\right)
}\right] &=&q^{2}Z_{000}^{\left( -1\right) }+Z_{00}^{\left( 0\right) }\left(
p,r\right) -Z_{00}^{\left( 0\right) }\left( p_{42},p_{43}\right) \\
2\left[ r^{2}Z_{001}^{\left( -1\right) }+\left( p\cdot r\right)
Z_{100}^{\left( -1\right) }+\left( q\cdot r\right) Z_{010}^{\left( -1\right)
}\right] &=&r^{2}Z_{000}^{\left( -1\right) }+Z_{00}^{\left( 0\right) }\left(
p,q\right) -Z_{00}^{\left( 0\right) }\left( p_{42},p_{43}\right)
\end{eqnarray*}%
\begin{eqnarray*}
2\left[ p^{2}Z_{200}^{\left( -1\right) }+\left( p\cdot q\right)
Z_{110}^{\left( -1\right) }+\left( p\cdot r\right) Z_{101}^{\left( -1\right)
}\right] &=&p^{2}Z_{100}^{\left( -1\right) }+Z_{000}^{\left( 0\right)
}-Z_{10}^{\left( 0\right) }\left( p_{42},p_{43}\right) \\
2\left[ p^{2}Z_{110}^{\left( -1\right) }+\left( p\cdot q\right)
Z_{020}^{\left( -1\right) }+\left( p\cdot r\right) Z_{011}^{\left( -1\right)
}\right] &=&p^{2}Z_{010}^{\left( -1\right) }+Z_{10}^{\left( 0\right) }\left(
p_{31},p_{41}\right) -Z_{01}^{\left( 0\right) }\left( p_{42},p_{43}\right) \\
2\left[ p^{2}Z_{101}^{\left( -1\right) }+\left( p\cdot q\right)
Z_{011}^{\left( -1\right) }+\left( p\cdot r\right) Z_{002}^{\left( -1\right)
}\right] &=&p^{2}Z_{001}^{\left( -1\right) }+Z_{01}^{\left( 0\right) }\left(
q,r\right) \\
&&-\left[ Z_{00}^{\left( 0\right) }-Z_{10}^{\left( 0\right) }-Z_{01}^{\left(
0\right) }\right] \left( p_{42},p_{43}\right)
\end{eqnarray*}%
\begin{eqnarray*}
2\left[ q^{2}Z_{020}^{\left( -1\right) }+\left( p\cdot q\right)
Z_{110}^{\left( -1\right) }+\left( q\cdot r\right) Z_{011}^{\left( -1\right)
}\right] &=&q^{2}Z_{010}^{\left( -1\right) }+Z_{000}^{\left( 0\right)
}-Z_{01}^{\left( 0\right) }\left( p_{42},p_{43}\right) \\
2\left[ q^{2}Z_{110}^{\left( -1\right) }+\left( p\cdot q\right)
Z_{200}^{\left( -1\right) }+\left( q\cdot r\right) Z_{101}^{\left( -1\right)
}\right] &=&q^{2}Z_{100}^{\left( -1\right) }+Z_{10}^{\left( 0\right) }\left(
p,r\right) -Z_{10}^{\left( 0\right) }\left( p_{42},p_{43}\right) \\
2\left[ q^{2}Z_{011}^{\left( -1\right) }+\left( p\cdot q\right)
Z_{101}^{\left( -1\right) }+\left( q\cdot r\right) Z_{002}^{\left( -1\right)
}\right] &=&q^{2}Z_{001}^{\left( -1\right) }+Z_{01}^{\left( 0\right) }\left(
p,r\right) \\
&&-\left[ Z_{00}^{\left( 0\right) }-Z_{10}^{\left( 0\right) }-Z_{01}^{\left(
0\right) }\right] \left( p_{42},p_{43}\right)
\end{eqnarray*}%
\begin{eqnarray*}
2\left[ r^{2}Z_{002}^{\left( -1\right) }+\left( q\cdot r\right)
Z_{011}^{\left( -1\right) }+\left( p\cdot r\right) Z_{101}^{\left( -1\right)
}\right] &=&r^{2}Z_{001}^{\left( -1\right) }+Z_{000}^{\left( 0\right) } \\
&&-\left[ Z_{00}^{\left( 0\right) }-Z_{10}^{\left( 0\right) }-Z_{01}^{\left(
0\right) }\right] \left( p_{42},p_{43}\right) \\
2\left[ r^{2}Z_{011}^{\left( -1\right) }+\left( p\cdot r\right)
Z_{110}^{\left( -1\right) }+\left( q\cdot r\right) Z_{020}^{\left( -1\right)
}\right] &=&r^{2}Z_{010}^{\left( -1\right) }+Z_{01}^{\left( 0\right) }\left(
p,q\right) -Z_{01}^{\left( 0\right) }\left( p_{42},p_{43}\right) \\
2\left[ r^{2}Z_{101}^{\left( -1\right) }+\left( p\cdot r\right)
Z_{200}^{\left( -1\right) }+\left( q\cdot r\right) Z_{110}^{\left( -1\right)
}\right] &=&r^{2}Z_{100}^{\left( -1\right) }+Z_{10}^{\left( 0\right) }\left(
p,q\right) -Z_{10}^{\left( 0\right) }\left( p_{42},p_{43}\right)
\end{eqnarray*}%
\begin{equation*}
-3Z_{000}^{\left( 0\right) }=+\frac{1}{3}+2m^{2}Z_{000}^{\left( -1\right) }-%
\left[ p^{2}Z_{100}^{\left( -1\right) }+q^{2}Z_{010}^{\left( -1\right)
}+r^{2}Z_{001}^{\left( -1\right) }\right] -Z_{00}^{\left( 0\right) }\left(
p_{42},p_{43}\right)
\end{equation*}%
they imply the relations%
\begin{eqnarray*}
2p^{\mu _{1}}J_{4\mu _{1}} &=&-p^{2}J_{4}+J_{3}\left( q,r\right)
-J_{3}\left( r-p,r-q\right) \\
2q^{\mu _{1}}J_{4\mu _{1}} &=&-q^{2}J_{4}+J_{3}\left( p,r\right)
-J_{3}\left( r-p,r-q\right) \\
2r^{\mu _{1}}J_{4\mu _{1}} &=&-r^{2}J_{4}+J_{3}\left( p,q\right)
-J_{3}\left( r-p,r-q\right)
\end{eqnarray*}%
\begin{eqnarray*}
2p^{\mu _{1}}J_{4\mu _{1}\mu _{2}} &=&-p^{2}J_{4\mu _{2}}+J_{3\mu
_{2}}\left( p_{42},p_{43}\right) +J_{3\mu _{2}}\left( p_{31},p_{41}\right)
+p_{41\mu _{2}}J_{3}\left( p_{42},p_{43}\right) \\
2q^{\mu _{1}}J_{4\mu _{1}\mu _{2}} &=&-q^{2}J_{4\mu _{2}}+J_{3\mu
_{2}}\left( p_{42},p_{43}\right) +J_{3\mu _{2}}\left( p_{21},p_{41}\right)
+p_{41\mu _{2}}J_{3}\left( p_{42},p_{43}\right) \\
2r^{\mu _{1}}J_{4\mu _{1}\mu _{2}} &=&-r^{2}J_{4\mu _{2}}+J_{3\mu
_{2}}\left( p_{42},p_{43}\right) +J_{3\mu _{2}}\left( p_{21},p_{31}\right)
+p_{41\mu _{2}}J_{3}\left( p_{42},p_{43}\right)
\end{eqnarray*}%
\begin{eqnarray*}
2g^{\mu _{12}}J_{4\mu _{1}\mu _{2}} &=&\frac{i}{3\left( 4\pi \right) ^{3}}%
+2m^{2}J_{4}+2J_{3}\left( p_{42},p_{43}\right) \\
p &=&p_{21};\quad q=p_{31};\quad r=p_{41},\text{ in the computations }
\end{eqnarray*}

\section{Subamplitudes}

\label{AppSub}\textbf{The }$SAP$\textbf{\ subamplitude}

Integrand%
\begin{equation*}
S_{ij}=\left( K_{i}\cdot K_{j}-m^{2}\right)
\end{equation*}%
\begin{equation*}
\left( t^{SAP}\right) ^{\nu _{1}}=4\left[ K_{1}^{\nu _{1}}S_{23}+K_{2}^{\nu
_{1}}\left( S_{13}+2m^{2}\right) -K_{3}^{\nu _{1}}\left(
S_{12}+2m^{2}\right) \right] \frac{1}{D_{123}}
\end{equation*}

\begin{eqnarray*}
\left( T^{SAP}\right) ^{\nu _{1}} &=&-4\left( p_{32}\cdot p_{31}\right)
J_{3}^{\nu _{1}}+2\left[ \left( p_{31}^{\nu _{1}}p_{21}^{2}-p_{21}^{\nu
_{1}}p_{31}^{2}-4m^{2}p_{32}^{\nu _{1}}\right) J_{3}-p_{32}^{\nu
_{1}}J_{2}\left( p_{32}\right) -p_{31}^{\nu _{1}}J_{2}\left( p_{31}\right) %
\right] \\
&&-2\left[ P_{21}^{\nu _{2}}\Delta _{3\nu _{2}}^{\nu _{1}}+\left(
p_{32}^{\nu _{1}}+p_{31}^{\nu _{1}}\right) I_{\log }\right]
\end{eqnarray*}%
\textbf{The }$SPA$\textbf{\ subamplitude}%
\begin{eqnarray*}
\left( T^{SPA}\right) ^{\nu _{1}} &=&+4\left( p_{21}\cdot p_{31}\right)
J_{3}^{\nu _{1}}+2\left[ \left( p_{31}^{\nu _{1}}p_{21}^{2}+p_{21}^{\nu
_{1}}p_{31}^{2}-4m^{2}p_{21}^{\nu _{1}}\right) J_{3}-p_{21}^{\nu
_{1}}J_{2}\left( p_{21}\right) -p_{31}^{\nu _{1}}J_{2}\left( p_{31}\right) %
\right] \\
&&+2\left[ P_{32}^{\nu _{2}}\Delta _{\nu _{2}}^{\nu _{1}}-\left( p_{21}^{\nu
_{1}}+p_{31}^{\nu _{1}}\right) I_{\log }\right]
\end{eqnarray*}%
\textbf{The }$ASP$\textbf{\ subamplitude}%
\begin{eqnarray*}
\left( T^{ASP}\right) ^{\nu _{1}} &=&-4\left( p_{21}\cdot p_{32}\right)
J_{3}^{\nu _{1}}+2\left[ \left( p_{31}^{\nu _{1}}p_{21}^{2}-p_{21}^{\nu
_{1}}p_{31}^{2}-4m^{2}p_{32}^{\nu _{1}}\right) J_{3}+p_{21}^{\nu
_{1}}J_{2}\left( p_{21}\right) -p_{32}^{\nu _{1}}J_{2}\left( p_{32}\right) %
\right] \\
&&+2\left[ P_{31}^{\nu _{2}}\Delta _{\nu _{2}}^{\nu _{1}}+\left( p_{21}^{\nu
_{1}}-p_{32}^{\nu _{1}}\right) I_{\log }\right]
\end{eqnarray*}%
\textbf{The }$PVP$\textbf{\ subamplitude}%
\begin{eqnarray*}
\left( T^{PVP}\right) ^{\nu _{1}} &=&+4\left( p_{32}\cdot p_{31}\right)
J_{3}^{\nu _{1}}+2\left[ \left( p_{21}^{\nu _{1}}p_{31}^{2}-p_{31}^{\nu
_{1}}p_{21}^{2}\right) J_{3}+p_{32}^{\nu _{1}}J_{2}\left( p_{32}\right)
+p_{31}^{\nu _{1}}J_{2}\left( p_{31}\right) \right] \\
&&+2\left[ P_{21}^{\nu _{2}}\Delta _{3\nu _{2}}^{\nu _{1}}+\left(
p_{32}^{\nu _{1}}+p_{31}^{\nu _{1}}\right) I_{\log }\right]
\end{eqnarray*}%
\textbf{The }$PSA$\textbf{\ subamplitude}%
\begin{eqnarray*}
\left( T^{PSA}\right) ^{\nu _{1}} &=&-4\left( p_{21}\cdot p_{31}\right)
J_{3}^{\nu _{1}}-2\left[ \left( p_{21}^{\nu _{1}}p_{31}^{2}+p_{31}^{\nu
_{1}}p_{21}^{2}-4m^{2}p_{31}^{\nu _{1}}\right) J_{3}-p_{21}^{\nu
_{1}}J_{2}\left( p_{21}\right) -p_{31}^{\nu _{1}}J_{2}\left( p_{31}\right) %
\right] \\
&&+2\left[ -P_{32}^{\nu _{2}}\Delta _{3\nu _{2}}^{\nu _{1}}+\left(
p_{21}^{\nu _{1}}+p_{31}^{\nu _{1}}\right) I_{\log }\right]
\end{eqnarray*}%
\textbf{The }$APS$\textbf{\ subamplitude}%
\begin{eqnarray*}
\left( T^{APS}\right) ^{\nu _{1}} &=&+4\left( p_{21}\cdot p_{32}\right)
J_{3}^{\nu _{1}}+2\left[ \left( p_{21}^{\nu _{1}}p_{31}^{2}-p_{31}^{\nu
_{1}}p_{21}^{2}-4m^{2}p_{21}^{\nu _{1}}\right) J_{3}+p_{32}^{\nu
_{1}}J_{2}\left( p_{32}\right) -p_{21}^{\nu _{1}}J_{2}\left( p_{21}\right) %
\right] \\
&&-2\left[ P_{31}^{\nu _{2}}\Delta _{3\nu _{2}}^{\nu _{1}}-\left(
p_{32}^{\nu _{1}}-p_{21}^{\nu _{1}}\right) I_{\log }\right]
\end{eqnarray*}%
\textbf{The }$PAS$\textbf{\ subamplitude}%
\begin{eqnarray*}
\left( T^{PAS}\right) ^{\nu _{1}} &=&+4\left( p_{32}\cdot p_{31}\right)
J_{3}^{\nu _{1}}+2\left[ \left( p_{21}^{\nu _{1}}p_{31}^{2}-p_{31}^{\nu
_{1}}p_{21}^{2}+4m^{2}p_{31}^{\nu _{1}}\right) J_{3}+p_{32}^{\nu
_{1}}J_{2}\left( p_{32}\right) +p_{31}^{\nu _{1}}J_{2}\left( p_{31}\right) %
\right] \\
&&+2\left[ P_{21}^{\nu _{2}}\Delta _{3\nu _{2}}^{\nu _{1}}+\left(
p_{32}^{\nu _{1}}+p_{31}^{\nu _{1}}\right) I_{\log }\right]
\end{eqnarray*}%
\textbf{The }$PPV$\textbf{\ subamplitude}%
\begin{eqnarray*}
\left( T^{PPV}\right) ^{\nu _{1}} &=&+4\left( p_{21}\cdot p_{31}\right)
J_{3}^{\nu _{1}}+2\left[ \left( p_{31}^{\nu _{1}}p_{21}^{2}+p_{21}^{\nu
_{1}}p_{31}^{2}\right) J_{3}-\left[ p_{21}^{\nu _{1}}J_{2}\left(
p_{21}\right) +p_{31}^{\nu _{1}}J_{2}\left( p_{31}\right) \right] \right] \\
&&+2\left[ P_{32}^{\nu _{2}}\Delta _{\nu _{2}}^{\nu _{1}}-\left( p_{21}^{\nu
_{1}}+p_{31}^{\nu _{1}}\right) I_{\log }\right]
\end{eqnarray*}%
\textbf{The }$VSS$\textbf{\ subamplitude}%
\begin{eqnarray*}
\left( T^{VSS}\right) ^{\nu _{1}} &=&4\left[ \left( p_{21}\cdot
p_{32}\right) +4m^{2}\right] J_{3}^{\nu _{1}}+2\left( p_{21}^{\nu
_{1}}p_{31}^{2}-p_{31}^{\nu _{1}}p_{21}^{2}+4m^{2}p_{31}^{\nu _{1}}\right)
J_{3} \\
&&+2\left[ p_{32}^{\nu _{1}}J_{2}\left( p_{32}\right) -p_{21}^{\nu
_{1}}J_{2}\left( p_{21}\right) \right] -2\left[ P_{31}^{\nu _{2}}\Delta
_{\nu _{2}}^{\nu _{1}}-\left( p_{32}^{\nu _{1}}-p_{21}^{\nu _{1}}\right)
I_{\log }\right]
\end{eqnarray*}%
\textbf{The }$SVS$\textbf{\ subamplitude}%
\begin{eqnarray*}
\left( T^{SVS}\right) ^{\nu _{1}} &=&4\left[ -\left( p_{32}\cdot
p_{31}\right) +4m^{2}\right] J_{3}^{\nu _{1}}+2\left( p_{31}^{\nu
_{1}}p_{21}^{2}-p_{21}^{\nu _{1}}p_{31}^{2}+4m^{2}p_{21}^{\nu _{1}}\right)
J_{3} \\
&&-2\left[ p_{32}^{\nu _{1}}J_{2}\left( p_{32}\right) +p_{31}^{\nu
_{1}}J_{2}\left( p_{31}\right) \right] -2\left[ P_{21}^{\nu _{2}}\Delta
_{\nu _{2}}^{\nu _{1}}+\left( p_{32}^{\nu _{1}}+p_{31}^{\nu _{1}}\right)
I_{\log }\right]
\end{eqnarray*}%
\textbf{The }$SSV$\textbf{\ subamplitude}%
\begin{eqnarray*}
\left( T^{SSV}\right) ^{\nu _{1}} &=&+4\left[ -\left( p_{21}\cdot
p_{31}\right) +4m^{2}\right] J_{3}^{\nu _{1}}+2\left[ \left( -p_{31}^{\nu
_{1}}p_{21}^{2}-p_{21}^{\nu _{1}}p_{31}^{2}+4m^{2}\left( p_{21}^{\nu
_{1}}+p_{31}^{\nu _{1}}\right) \right) J_{3}\right] \\
&&+2\left[ p_{21}^{\nu _{1}}J_{2}\left( p_{21}\right) +p_{31}^{\nu
_{1}}J_{2}\left( p_{31}\right) \right] -2\left[ P_{32}^{\nu _{2}}\Delta
_{\nu _{2}}^{\nu _{1}}-\left( p_{21}^{\nu _{1}}+p_{31}^{\nu _{1}}\right)
I_{\log }\right]
\end{eqnarray*}

\begin{acknowledgments}
The authors would like to thank Sebasti\~{A}\pounds o A. Dias for helpful
suggestions and support in this investigation. L. Ebani, T. J. Girardi and
J.F. Thuorst, acknowledge the financial support of CAPES and CNPQ.
\end{acknowledgments}

\end{document}